\documentclass[referee]{aa}  

\usepackage{graphicx}
\usepackage{txfonts}
\begin{document}

   \title{Relativistic Coulomb Screening in Pulsational Pair Instability Supernovae}
   \titlerunning{Screening in PPISN}
   \authorrunning{Famiano et al.}


\author{M.A. Famiano
    \inst{1-3}\fnmsep\thanks{Corresponding Author, \email{michael.famiano@wmich.edu}}
    \and
    K. Mori
    \inst{4}\fnmsep\thanks{\email{kanji.mori@fukuoka-u.ac.jp}}
    \and
    A.B. Balantekin
    \inst{2,5}\fnmsep\thanks{\email{baha@physics.wisc.edu}}
    \and
    T. Kajino
    \inst{2,6,7}\fnmsep\thanks{\email{kajino@buaa.edu.cn}}
    \and
    M. Kusakabe
    \inst{7}\fnmsep\thanks{\email{kusakabe@buaa.edu.cn}}
    \and
    G. Mathews
    \inst{3,8}\fnmsep\thanks{ \email{gmathews@nd.edu}}
}

\institute{Department of Physics, Western Michigan University, Kalamazoo, MI 49008 USA
    \and
    National Astronomical Observatory of Japan, 2-21-1 Osawa, Mitaka, Tokyo 181-8588 Japan
    \and
    Joint Institute for Nuclear Astrophysics - Center for the Evolution of the Elements, USA
    \and
    Research Institute of Stellar Explosive Phenomena, Fukuoka University,
    8-19-1 Nanakuma, Jonan-ku, Fukuoka-shi, Fukuoka 814-0180, Japan
    \and
    Department of Physics, University of Wisconsin-Madison, Madison, Wisconsin 53706 USA
    \and
    Graduate School of Science, The University of Tokyo, 7-3-1 Hongo, Bunkyo-ku, Tokyo, 113-0033 Japan
    \and
    School of Physics, Beihang University, 37 Xueyuan Road, Haidian-qu, Beijing 100083, China
    \and
    Center for Astrophysics, Department of Physics, University of Notre Dame, 225 Nieuwland Science Hall, Notre Dame, IN 46556, USA
}

   \date{Received XXXX; accepted XXXX}

 
  \abstract
   {Pulsational pair-instabilitye supernovae (PPISNe) and pair instability supernovae (PISNe) are the result of a thermonuclear runaway in the presence of a background electron-positron pair plasma.  As such, their evolution and resultant black hole (BH) masses could possibly be affected by screening corrections due to the electron pair plasma.}
   {Sensitivity of PISNe and PPISNe to relativistic weak screening has been explored.  }
   {In this paper a  weak screening model that includes effects from relativistic pair production has been developed and applied  at temperatures
approaching and exceeding the threshold for pair production.  This screening model replaces ``classical'' screening commonly used in astrophysics. 
Modifications to the weak screening electron Debye length are incorporated in a computationally tractable analytic
form with.}
   {In PPISNe the BH masses were found to increase somewhat at high temperatures,
though this increase is small. The BH collapse is also 
found to occur at earlier times, and the pulsational morphology also changes.  In addition to the 
resultant BH mass, the sensitivity to the screening model of the pulsational period, the pulse structure,
the PPISN-to-PISN transition, and the shift in the BH mass gap has been analyzed. The 
dependence of the composition of the ejected mass was also examined.}
   {}

\maketitle
%

\section{Introduction}
\label{sec:intro}
Currently, there is considerable interest in the black hole mass gap (BHMG).  That is, black 
hole masses in the range of $\sim 50-120$ M$_\odot$ are not expected to exist 
\citep{Woosley07,Belczynski16,woosley17,Woosley19} because progenitor stars in this mass range 
are disrupted by pair-instability supernovae (PISNe) or pulsational pair-instability supernovae
(PPISNe).   The existence of this mass gap, however, has been brought into question by the 
observed LIGO/VIRGO gravitational wave event GW190521 \citep{Abbot20} from which two black 
holes of masses $m_1 = 85^{+21}_{-14}$ M$_\sun$ and $m_2 = 66^{+17}_{-18}$ M$_\sun$ were 
deduced.  Since both of these merging black holes are well within the putative black-hole mass 
gap, a re-examination of the constraints on black-hole masses from the pair instability and 
PPISNe is warranted \citep[e.g.,][]{farmer19,marchant19,abbott20b,croon07,VanSon20,Woosley21}.
Indeed, the existence of the BHMG and the associated  progenitor evolution is sensitive to a 
host of astrophysical parameters as noted in \citet{farmer19, sakstein20,Woosley21}.  In this 
paper we consider one additional sensitivity, i.e. the effects of relativistic screening of the
thermonuclear reaction rates in the background associated pair plasma.  Although, we find that the
effects on the BHMG are small, we point out a number of interesting features of the evolution
of the progenitors that depend upon the nature of the electron screening.  In what follows we
briefly summarize  the physics of PISNe and PPISNe in \S 1.1.  We describe a model for electron
screening in a relativistic pair plasma in \S 1.2.  The computations are described in \S 2, 
and results given in \S 3.  \S 4 contains the conclusions of this work.

\subsection{Pulsational Pair Instability Supernovae and Pair Instability Supernovae}
PISNe are caused  by the production of electron-positron pairs in the cores of massive (M$_{ZAMS}\gtrsim$ 80 M$_\odot$) stars at the end of their lifetimes~\citep{leung19}. At temperatures $^>_\sim 10^9$ K, electron pairs are produced by the thermal plasma, e.g. via $\gamma + \gamma \Leftrightarrow e^- + e^+$.   As the photons are absorbed into pair production, the radiation pressure support of the core diminishes.  Subsequently, the core contracts  and the  temperature rises.   This leads to  explosive  thermonuclear oxygen burning.  The release of this thermonuclear energy can be comparable to the binding energy of the star and lead to substantial mass ejection.

The details of the explosion and associated nucleosynthesis, however,  depend upon the mass and metallicity of the progenitor star. At low metallicity, $Z \sim 10^{-3}$, stars in the mass range of $\sim 50-90$ M$_\odot$  can experience  a sequence of contractions and explosions.  This is referred to as a PPISN. Eventually, these stars can return to hydrostatic equilibrium and then collapse.
However, the resultant black holes are significantly less massive than they would have been without the onset of  the pair instability. 

On the other hand, more massive progenitors with M$\ge 90$ M$_\odot$ experience such  violent explosions that no black-hole remnant remains.  These have been dubbed  as PISNe. 
The most massive black hole that can be formed before mass loss from the PPISN  becomes significant defines the lower limit of the black hole mass gap. In the heaviest  progenitors for which M$ ^>_\sim  240$ M$_\odot$, the pair instability is quenched because the interior temperatures are high enough to photodisintegrate heavy elements.   This precludes the onset of runaway thermonuclear burning. The lightest black hole formed in this way [$\sim 120$ M$_\odot$ after mass loss \citep{heger02}] defines the upper edge of the black hole mass gap.
\subsection{Coulomb Screening in Nuclear Reactions}
Astrophysical reaction rates in stellar plasmas can be very sensitive to the 
tails of the Coulomb potentials between two reacting charged nuclei.  This is because 
charged nuclei in astrophysical environments have center-of-mass energies $\sim kT$ well below
the height of the Coulomb barriers.
For an environment at a certain temperature, the average thermonuclear reaction rates
between charged particles are determined by integrating the energy-dependent cross
section times the Mawell-Boltzmann energy distribution for the reactants involved
to obtain 
the average thermonuclear reaction rate (TRR), $\left\langle\sigma v\right\rangle$ \citep{boyd08,iliadis}.  
Resonances at specific energies are similarly determined with a term in the cross section
which defines the resonance.

However, one must also account for ``screening'' between the two reacting, nuclei due to the background
charges and (fully or partially) ionized nuclei.  Coulomb screening results in a reduction in the
effective Coulomb barrier between the two reacting nuclei and an increase in
the WKB penetrability of the barrier, leading to an increase in the overall rate.  This effect has been well studied and is 
incorporated into multiple existing astrophysics calculations and codes~\citep{wu17,liu16,spitaleri16,yakovlev14,kravchuk14,potekhin13,quarati07,
	shaviv00,adelberger98,shalybkov87,wang83,
	www82,itoh77,jancovici77,graboske73,dewitt73,salpeter69,salpeter54}. 

When evaluating the effects of electron screening on the Coulomb potential, even a small shift in the potential can result in significant changes in
the classical potential turning points, resulting in an increase in the reaction rate.  For positively charged nuclei the overall reaction rates increase because charges are re-distributed non-uniformly in the presence of the central 
nuclear potential. 
Despite the extensive development of nuclear screening effects on thermonuclear reaction rates, relativistic plasma effects have not yet been fully addressed.  If the temperature is high enough, electron-positron pairs are created, and these add additional charged particles
to the plasma.  Recalling that any charge added to the plasma results in a net reduction in the Coulomb potential, pair production
can have an additional effect on screening. 
In addition, if the plasma is magnetized, the electron-positron momentum distributions are quantized into Landau levels, further changing the
overall energy distribution, resultant numbers of electrons and positrons, and the overall screening characteristics of the plasma~\citep{famiano20}.
Applying these effects to massive stars \citep{kozyreva17, woosley17, spera17, takahashi18} or to neutrino cooling mechanisms~\citep{itoh96} is potentially very interesting.

For a hot, ionized plasma, the Coulomb potential between two reacting nuclei is reduced by the presence of background charges.  Assuming charge neutrality, the electrostatic potential, $\phi$ of a particle with charge $Ze$ immersed in a set of background charge distributions is
determined via the Poisson-Boltzmann equation:
\begin{equation}
\label{PB}
\nabla^2\phi(r) = -4\pi Ze^2\delta(\mathbf{r}^3) -4\pi\sum_{z\ge-1} ze n_z\exp\left[-\frac{Zze^2\phi(r)}{kT}\right]~~.
\end{equation}
Here, the summation in the last term is over all charges $ze$ in the plasma having number density $n_z$.  Note that
this sum includes electrons, with $z=-1$.
The results of this calculation are widely used in astrophysical calculations, despite the assumption of a classical (Boltzmann) 
energy distribution for all of the particles, given by the second term in the above equations.
With this, evaluations of the electron degeneracy must be explicitly determined to obtain the electron number density and resulting
energy distribution.  

In the case of a hot, unmagnetized plasma, the energy distribution must be replaced with the corresponding distribution
assuming Fermi-Dirac statistics.  
The Poisson-Boltzmann
equation must be replaced with the equivalent equation assuming Fermi statistics and chemical potential, $\mu$~\citep{famiano16,famiano20}:
\begin{eqnarray}
\label{PF}
\nabla^2\phi_r &= -4\pi Ze\delta(\mathbf{r}^3) -4\pi\sum_{z>0} ze n_z\exp{\left[-\frac{Zze^2\phi_r}{kT}\right]}
\\\nonumber
&-\frac{4e}{\pi}\int\limits_0^\infty{p^2dp \left[\frac{1}{\exp(E-\mu- e\phi_r)/T+1}-\frac{1}{\exp(E+\mu+e\phi_r)/T+1}\right]}~~.
\end{eqnarray}
Relativistic effects come from
the high thermal energy. (Natural units are used: $k=\hbar=c=1$.) 
This relationship can also be deduced from a solution to the Schwinger-Dyson equation for
the photon propagator \citep{kapusta06}.
Here, the much heavier nuclei can be safely treated with Boltzmann statistics.  

If the temperature is high enough, such that the average Coulomb energy between two interacting particles is much less 
than the thermal energy of the plasma, $E_C/kT\ll1$, ``weak screening'' can apply.  In such a plasma, the electrons are mostly
non-degenerate.  Equations \ref{PB} and \ref{PF} can be expanded to first order in the  potential.  This expansion to $\mathcal{O}(\phi)$
is known as the Debye-H\"uckel approximation, and the usual $1/r$ Coulomb potential 
takes on a Yukawa form, $\phi(r)\propto e^{-r/\lambda_D}/r$ where the characteristic length $\lambda_D$ is
known as the Debye length.  A similar Thomas-Fermi screening length can be derived using the density of states
at the electron Fermi surface~\citep{ichimaru93} with a screening length, $\lambda_{TF}\propto\partial n/\partial\mu$.   

At the other extreme in the temperature-density relationship is ``strong screening.'' At high densities, the average distance between
nuclei is small enough such that the average Coulomb energy energy is much larger than the thermal energy, $E_C/kT\gg 1$.
Here, the linear approximation for Debye-H\"uckel or Thomas-Fermi screening is inadequate.  Likewise, the electron chemical potential
becomes larger and must be accounted for in Equation \ref{PF}.  In this regime, generally the ``ion sphere'' screening model is used, in
which the potential is modified by the difference in Coulomb energy between the reactants and products of a reaction~\citep{clayton,salpeter69, salpeter54}.

Between weak and strong screening, where $E_C\sim kT$, there is still much work to be done, and computational methods can be tricky. 
While the electron degeneracy and Coulomb potentials can be numerically calculated~\citep{graboske73}, in order to maintain computational efficiency,  approximate methods are often adopted to treat screening in this region.  Numerical fits and tables have also been used
to determine reaction-rate enhancements due to screening.  Other commonly-used tools rely on interpolation or averaging schemes to 
determine the screening enhancement in this region~\citep{libnucnet,www82,salpeter69,mesa11}.
	
Ultimately, the increase in nuclear reaction rates is expressed by the ``screening enhancement factor'' (SEF), $f$ which relates the unscreened 
and screened rates, $\langle \sigma v\rangle_{uns}$ and $\langle \sigma v\rangle_{scr}$ respectively.  In the Salpeter approximation~\citep{salpeter54}, the
screened rate is then $\langle\sigma v\rangle_{scr}=f\langle\sigma v\rangle_{uns}$.  This results from
assuming that the shift in the screening potential is much less than the height of the Coulomb potential and 
can be treated to first order in the WKB approximation.  The value of $f$ is then expressed as
$f=e^H$ \citep{graboske73,jancovici77,salpeter54,salpeter69,www82}.  
The value of $H$ is unitless and is determined based upon the screening model used~\citep{das16,kravchuk14,itoh77,alastuey78,dewitt73,quarati07}.
In the  case of intermediate screening discussed briefly above, $H_I$ commonly results from an interpolation between
strong and weak screening or a type of geometric mean~\citep{mesa11, mesa15, mesa18, libnucnet}.  For example, \citet{www82} 
sets $H_I = H_SH_W/\sqrt{H_S^2+H_W^2}$ while the \texttt{MESA} code~\citep{mesa11,mesa15,mesa18} relies on a type of
linear interpolation based upon the effective screening parameter discussed in \S\ref{screen_model}.

In this paper, effects of the inclusion of Fermi-Dirac statistics in the electron screening on PPISN/PISN models
are studied.  Various characteristics and results from several PPISN/PISN simulations are studied using a screening
model in which the weak screening factor has been replaced with one developed using the Poisson equation shown in Equation \ref{PF}. 
\section{Methods}
\subsection{Computational Model}
\label{computer_model}
We have used the \texttt{MESA} code~\citep{mesa} {\it v.11123} to model the  collapse and SN explosions of
PPISN and PISN respectively.  Here, we follow the computations of \citet{marchant19}.  While we summarize the calculation
briefly here, we refer to the original paper
 for details~\citep{marchant19}.
While it is noted that differences exist in various models
and computations, we adopt this model for a consistent comparison between the two screening formulations.  In this model, we treat stellar evolution beginning with He
cores, assuming the H envelope has since been ejected by the wind~\citep{leung19}.  We concentrate primarily on non-rotating models with a metallicity, 
$Z=Z_\odot/10$, where $Z_\odot = 0.0142$.  Aside from changing
the screening, nearly all of the parameters used here are 
kept the same as in \citet{marchant19}.  This includes the assumption of
hydrostatic equilibrium in early phases of the evolution, with
the HLLC solver~\citep{toro94} used when the weighted adiabatic exponent
$\langle \Gamma_1\rangle < (4/3 + 0.01)$ and the central temperature exceeds
10$^9$ K.  Additionally, a somewhat higher Fe core infall limit of
8$\times$10$^8$ cm s$^-1$ was used to avoid interrupting collapse from the
pair-instability.

The \texttt{approx21} network 
was used at the onset of the pulsational phase.  Some uncertainties have
been 
associated with the use of this network~\citep{marchant19}, particularly for 
more massive stars (e.g., M$\sim$ 200M$_\odot$).  However, at lower masses
within the PPISN region, the uncertainties in this choice of network have been
found to be relatively small.

There are three possible results of this model for the He core progenitors
studied.  One possibility is a direct collapse in which the Fe core 
velocity exceeds the infall limit without undergoing a pulse, forming
a black hole (BH).  Here, we adopt the enclosed 
baryonic mass prior to direct core
collapse as a measure of the resultant black-hole gravitational mass \citep{fryer99}, although we note that the BH gravitational mass can be significantly less than the baryonic mass.  We also note that recent studies have indicated that a slightly larger mass may result
even for massive cores due to fallback in subsequent explosions~\citep{chan18,kuroda18}.  This is expected to occur
for lower-mass models with M$\lesssim$45 M$_\odot$ and very high mass
models with M$\gtrsim$ 240 M$_\odot$~\citep{farmer19}.

For intermediate masses with 45$\lesssim$M/M$_\odot\lesssim$90, the star
is expected to undergo a PPISN, in which it ejects a possibly significant
amount of mass in the wind and pulses before finally collapsing into 
a BH. The number of pulses that the progenitor undergoes varies with each
model, and we study here the dependence on the screening model.

For progenitors with 90$\lesssim$M/M$_\odot\lesssim$240, the 
star explodes in a PISN, determined by the velocity of each mass element
in the model exceeding the stellar escape velocity.

For all of these outcomes, we examine the resultant BH mass, the time
to final outcome, the pulse morphology, and the resultant nucleosynthesis - 
particularly the composition of the ejected mass.
\subsection{Screening Model}
\label{screen_model}
The screening model described in \citet{famiano16} was used to determine
and effective Debye screening length at high temperatures, and the SEF was calculated in the Salpeter approximation.  A specialized screening subroutine was written and adapted into \texttt{MESA} and implemented with an appropriate 
inlist. 
For weak screening, two screening modes were explored. The `extended' screening mode implemented in
existing versions of \texttt{MESA} and based upon the
formulation of \citet{graboske73} was used as a default 
model for comparison.  The new `relativistic' 
mode incorporated the intermediate and strong extended screening modes while replacing the 
weak SEF with the relativistic factor.  Here,
the screened reaction rate, $\Gamma_{scr}$, between two nuclei of charge $Z_1$
and $Z_2$ is enhanced over the unscreened rate, $\Gamma_0$, according to:
\begin{equation}
\label{SEF}
    \Gamma_{scr} = \Gamma_0 e^{H_w}~~,
\end{equation}
where the weak screening exponent varies between the two models.  For extended screening, the
weak screening exponent is defined in the literature~\citep{dewitt73,alastuey78,itoh79}.  For both relativistic screening, the
exponent is defined using the Salpeter approximation:
\begin{equation}
\label{weak_h}
    H_w \equiv \frac{Z_1Z_2e^2}{\lambda T}~~,
\end{equation}
where $\lambda$ is the Debye screening length derived in each
model.  The screening length must also include that of the background ions, $\lambda_I$ so the total
screening length is:
\begin{equation}
    \label{total_length}
    \frac{1}{\lambda} = \left[\frac{1}{\lambda_I^2} + \frac{1}{\lambda_e^2}\right]^{1/2}~~.
\end{equation}
For the classical electron screening length, the total Debye length can be shown to match that
of \citet{dewitt73}, which is currently used in the \texttt{MESA} extended screening model.

The electron Debye length is computed for two different regimes characterized by the environmental 
temperature.  At high temperatures, $kT\sim m_e$,  relativistic effects become more significant.  At lower
temperatures, $kT\lesssim m_e$,  a classical approximation is sufficient.  Thus, two different approximations for
the screening length are used.  For $kT<$ 150 keV, the non-relativistic regime is assumed, and \texttt{MESA}'s default classical
screening length, which corresponds to the Salpeter screening length~\citep{salpeter54} is adopted.
For $kT>$ 150 keV, a relativistic approximation of the Thomas-Fermi screening length, described below, is used.  In the following we discuss the accuracy of this approximation compared to the numerically computed Thomas-Fermi length.  We also highlight the
agreement between the different means to describe screening enhancement factors at the boundary between the relativistic and classical 
regimes.

We develop a new application of the electron screening length in which relativistic effects from
a hot plasma are accounted for.  This is distinguished from the commonly used screening length 
extracted from the linear approximation of the Poisson-Boltzmann equation.  For temperatures that are 
high-enough, pair production can change the total free charges in the plasma, resulting in enhanced screening.
At lower temperatures, the classical screening length suffices.  We thus define the screening length
used in this paper as:
\begin{equation}
\lambda_e = \left\{
{
\begin{array}{cl}
\lambda_c & \mbox{for } T < 150 \mbox{ keV}\\
\lambda_r & \mbox{for } T\ge 150 \mbox{ keV}
\end{array}
}
\right.
\end{equation}
where the usual classical screening length is defined as:
\begin{equation}
    \lambda_c \equiv \left[
        \frac{T}{4\pi e^2Y_e\rho N_A}
    \right]^{1/2}
\end{equation}
and the relativistic electron screening length is described in a compact form below.

The relativistic electron screening length, $\lambda_r$, can be derived following \citet{famiano16}.  For computational speed, we
utilize the following approximation:
\begin{eqnarray}
\label{debye_exp}
\frac{1}{\lambda_r^2}\approx
\frac{4e^2}{\pi}T^2A\left[1-f\right]
\end{eqnarray}
where 
\begin{eqnarray}
    \label{a_define}
    A&\equiv \tilde{\mu}^2+\frac{\pi^2}{3}>3.28 \\\nonumber
    f&\equiv \frac{\tilde{m}^2}{2A}-\frac{\tilde{m}^3}{12A}\frac{1}{\cosh^2\tilde{\mu}/2}
\end{eqnarray}
where terms expressed with a tilde are divided by the temperature, $\tilde{X}\equiv X/T$.

In the weak screening regime, the screening coefficient is 
given by Equation \ref{weak_h}.  For strong screening, the same 
evaluation was used in both models.  In this case, 
the evaluation of \citet{alastuey78} was used with 
plasma parameters from \citet{itoh79}.  This is the default strong
screening treatment for the \texttt{MESA} `extended' screening model.
For intermediate screening, the screening
coefficient in Equation \ref{SEF} is replaced by an interpolation between
the weak and strong screening coefficients as prescribed by the \texttt{MESA} default
screening scheme in which the intermediate screening is a weighted sum of 
the strong and weak screening coefficients, $H_s$ and $H_w$. The 
weighting factor $H$ is an average given by the relative difference between the effective screening parameter
$\Gamma_{eff}$ and the boundaries between the strong and intermediate plus intermediate and weak
screening regimes:
\begin{eqnarray}
\label{interp}
H &=& H_w\left(\frac{\Gamma_s - \Gamma_{eff}}{\Gamma_s-\Gamma_w}\right) + H_s\left(\frac{\Gamma_{eff} - \Gamma_w}{\Gamma_s-\Gamma_w}\right)\\\nonumber
\Gamma_s &>& 0.8\\\nonumber
\Gamma_w &<& 0.3
\end{eqnarray}
Where the effective screening parameter, $\Gamma_{eff}$, is adapted from
\citet{alastuey78} in \citet{www82}. 

\begin{figure*}
    \includegraphics[width=0.5\textwidth]{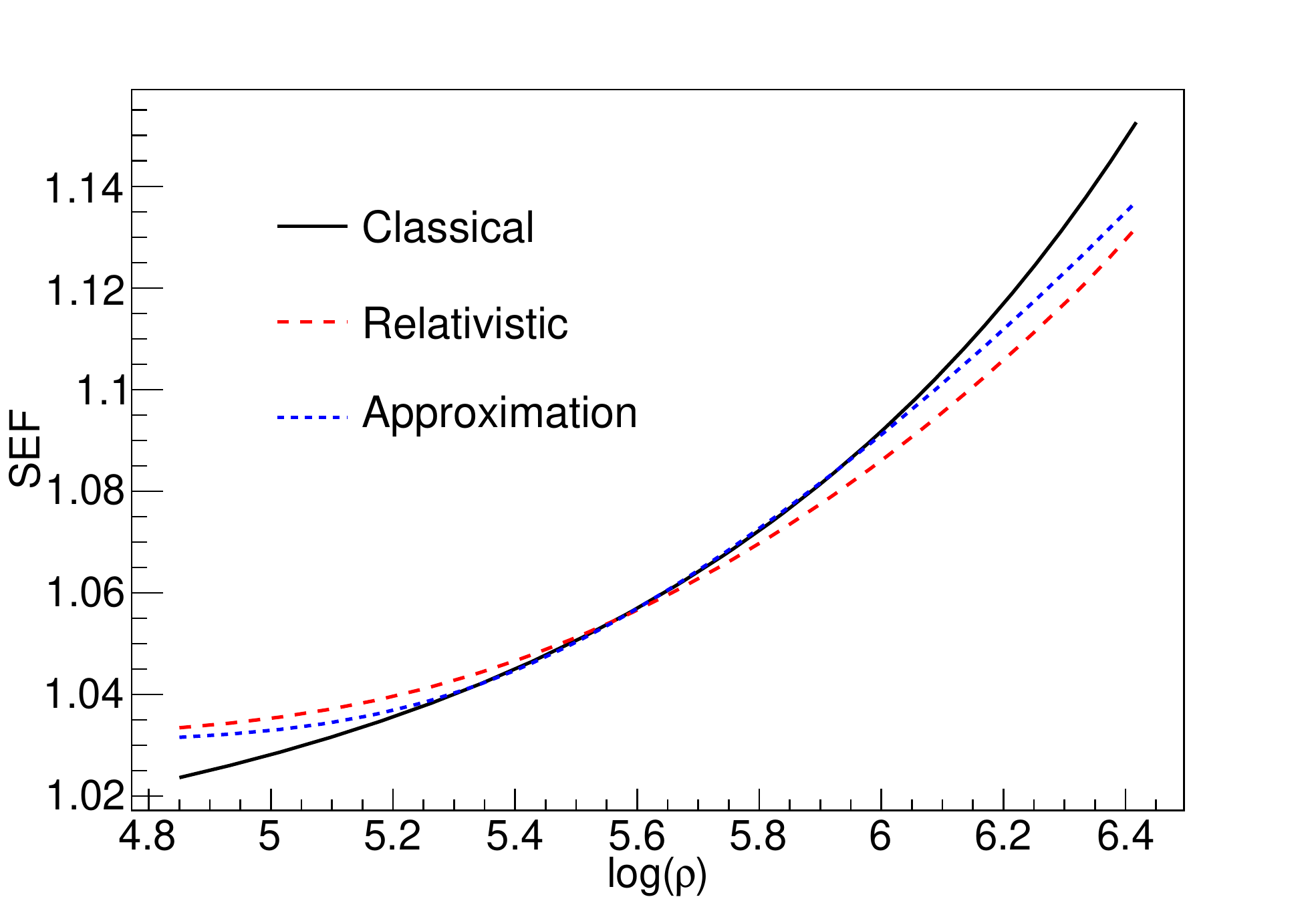}
    \includegraphics[width=0.5\textwidth]{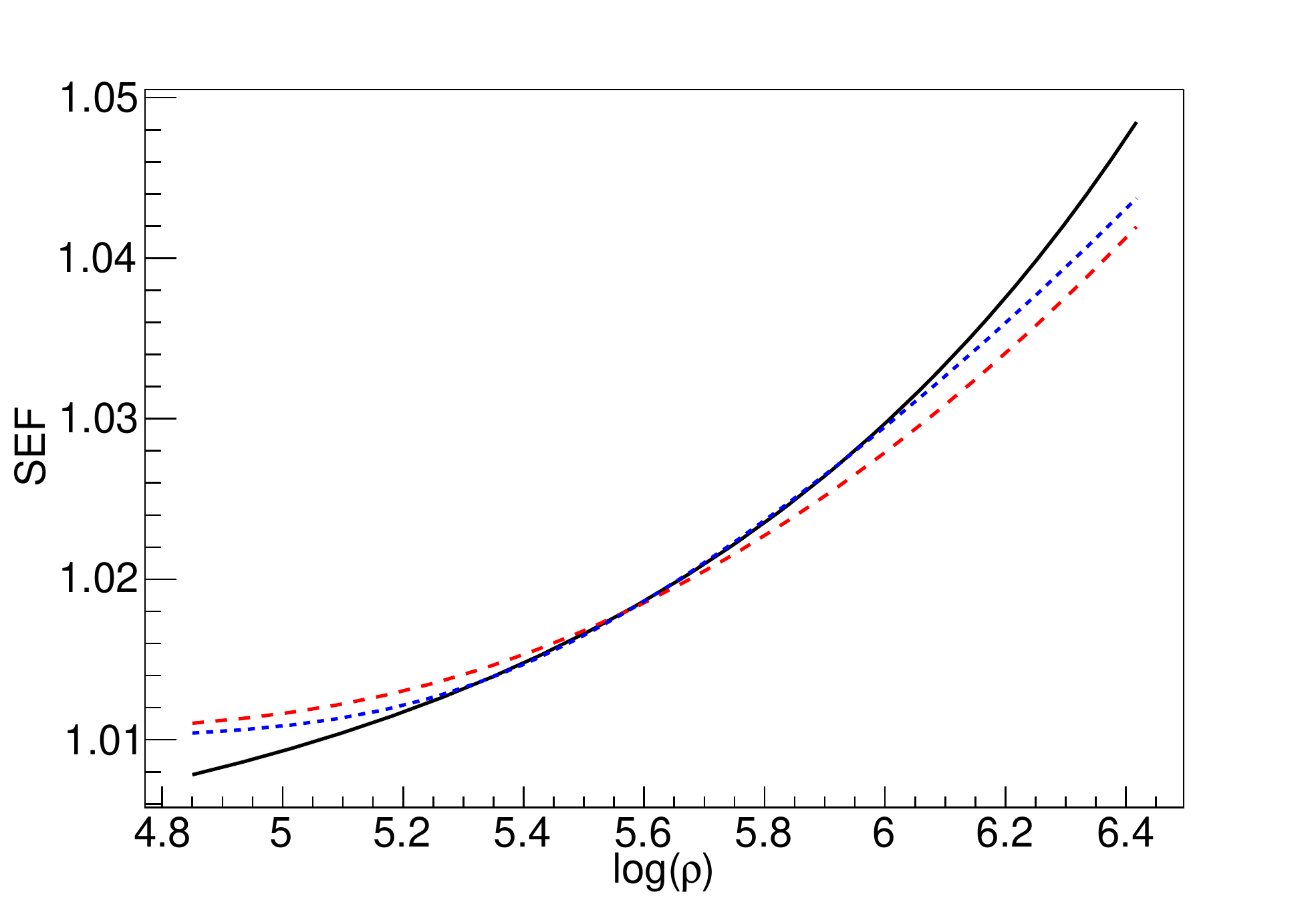}
    \includegraphics[width=0.5\textwidth]{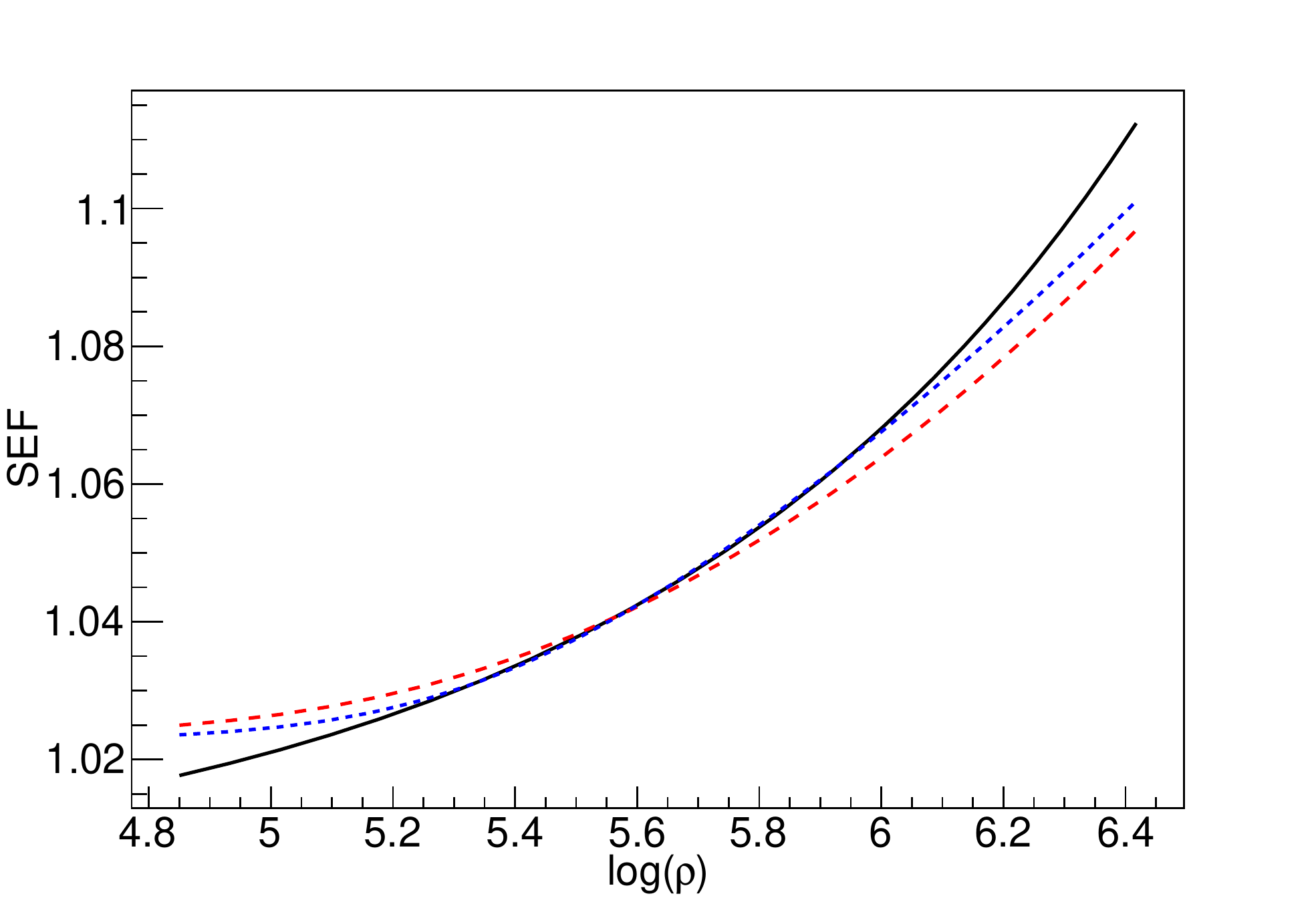}
    \includegraphics[width=0.5\textwidth]{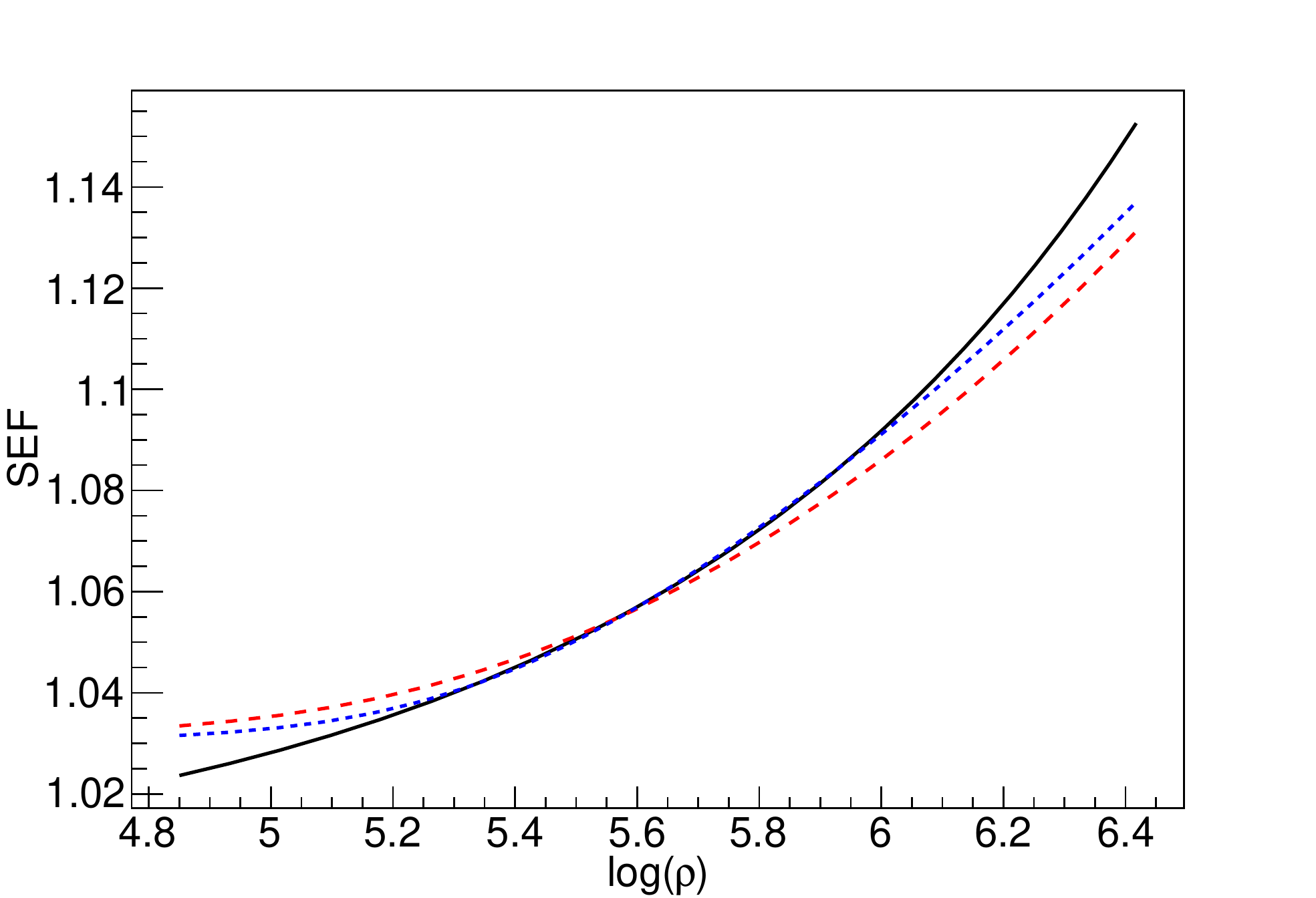}
    \caption{Screening enhancement factors for four major reactant pairs as a function of
    density at $kT$=150 keV and Y$_e$=0.5.  The density at which all SEFs are equal is temperature dependent. (\textit{Top left:} $^{12}$C+$\alpha$, \textit{Top right:}$^{16}$O+$\alpha$,
    \textit{Bottom left:}$^{12}$C+$^{12}$C, \textit{Bottom right:}$^{16}$O+$^{16}$O.)}
    \label{SEF_compare}
\end{figure*}

Thus, the modification to the \texttt{MESA} extended
screening scheme is to replace the weak screening coefficient, $H_w$, in Equation 
\ref{SEF} and \ref{interp} with
that of Equation \ref{weak_h} utilizing the relativistic screening length of Equation \ref{debye_exp}. Further, we restrict
this scheme to temperatures above $kT = 150$ keV ($T_9\equiv1.74$).  
For temperatures below
$kT = 150$ keV, the classical screening length was used.  This has been found to be a 
reasonable demarcation because the difference between the relativistic
and the classical SEFs is  
small at lower temperatures \citep{famiano20}.  This is shown in Figure
\ref{SEF_compare}, in which the SEFs for classical screening, relativistic screening,
and the relativistic approximation are shown for various major reactions as a function of density at $kT$=150 keV.
A density range of 4.8$<$log($\rho$)$<$6.4 is shown. This is chosen because for all models, 
the density when $kT$=150 keV falls within this range.  It can be seen that the SEFs 
at this temperature are relatively low, compared to SEFs at higher $T$ or $\rho$, and that the numerical 
relativistic, the relativistic approximation, and the classical screening lengths only vary by $\sim$ 1\%.
This also 
provides a limitation at very high density to prevent $k$ from becoming
less than zero in the approximation of Equation \ref{debye_exp}.

While computationally efficient, there is some uncertainty in this approximation 
at low temperatures and high densities.  

The $T=150$ keV boundary between the use of the classical screening length and the relativistic screening
length is somewhat arbitrary.  Also, we explore the magnitude of the error induced by using the approximation and the agreement
between the classical and the relativistic screening lengths at the boundary.  In particular, 
any uncertainty must be evaluated for the astrophysical site to which it is applicable.  Because we
are evaluating screening effects in PPISN and PISN, we have explored the errors induced for
all the trajectories calculated in a \texttt{MESA} simulation of representative PPISNs and PISNs.

\begin{figure*}
            \includegraphics[width=0.5\textwidth]{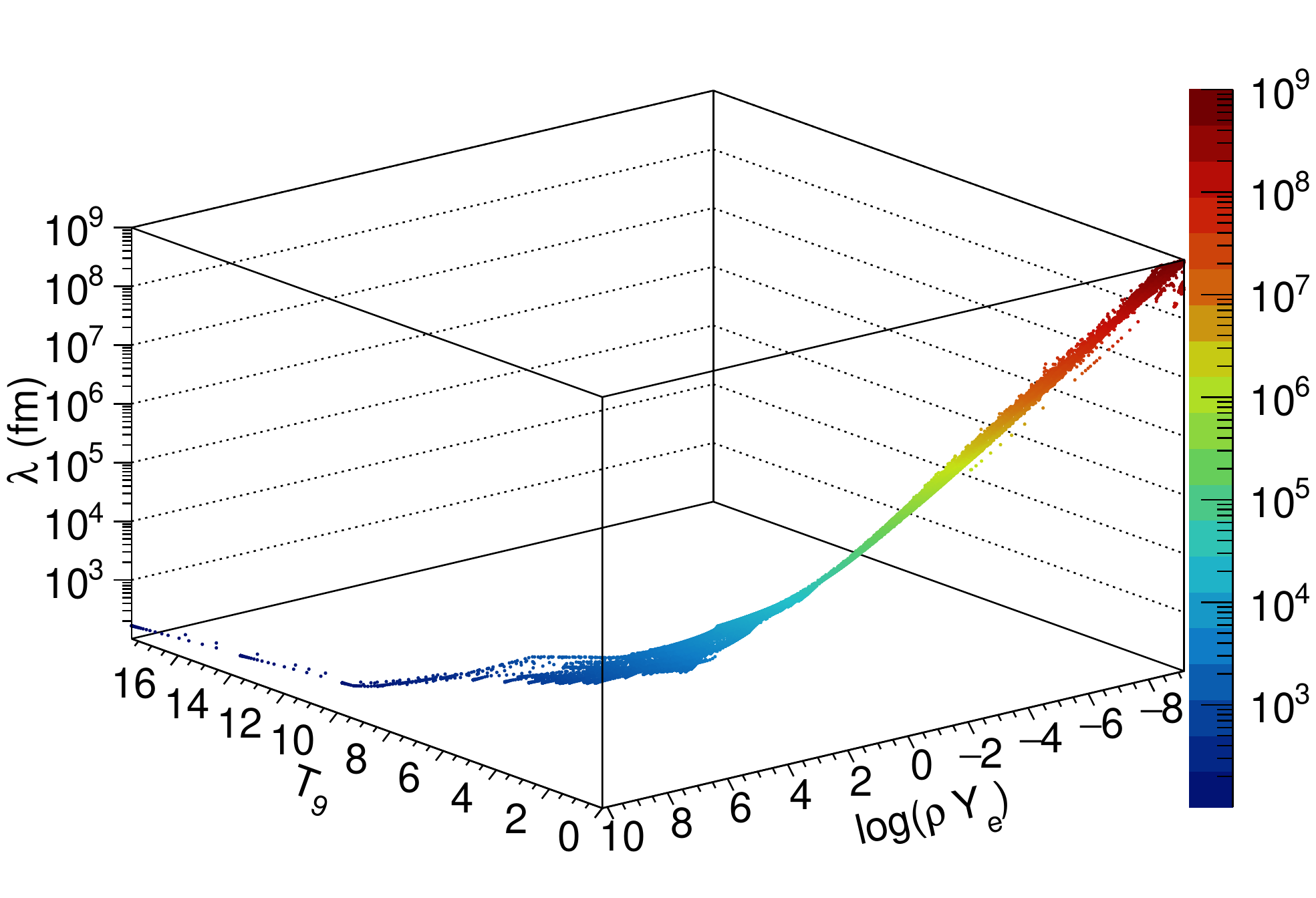}
            \includegraphics[width=0.5\textwidth]{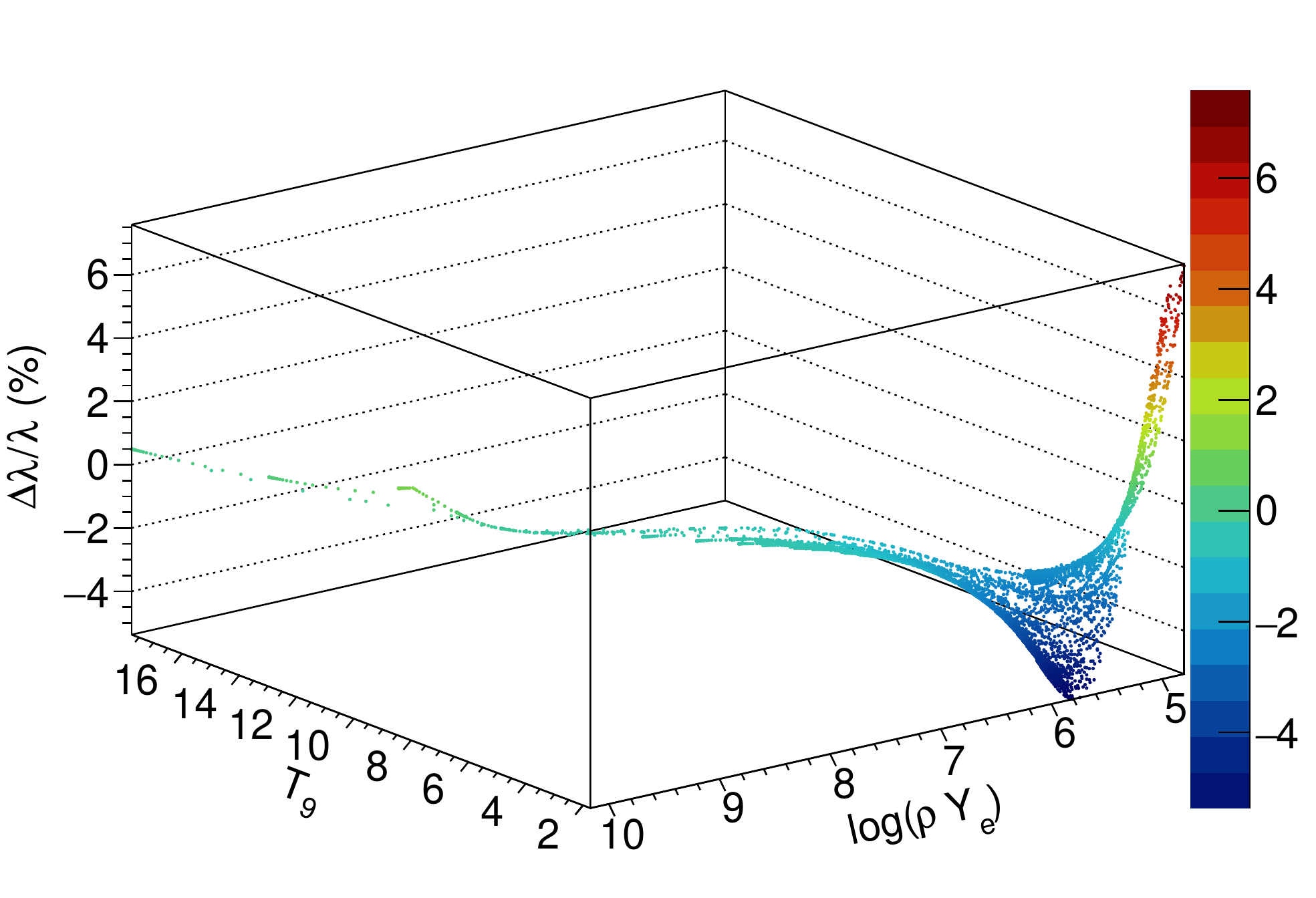}
            \includegraphics[width=0.5\textwidth]{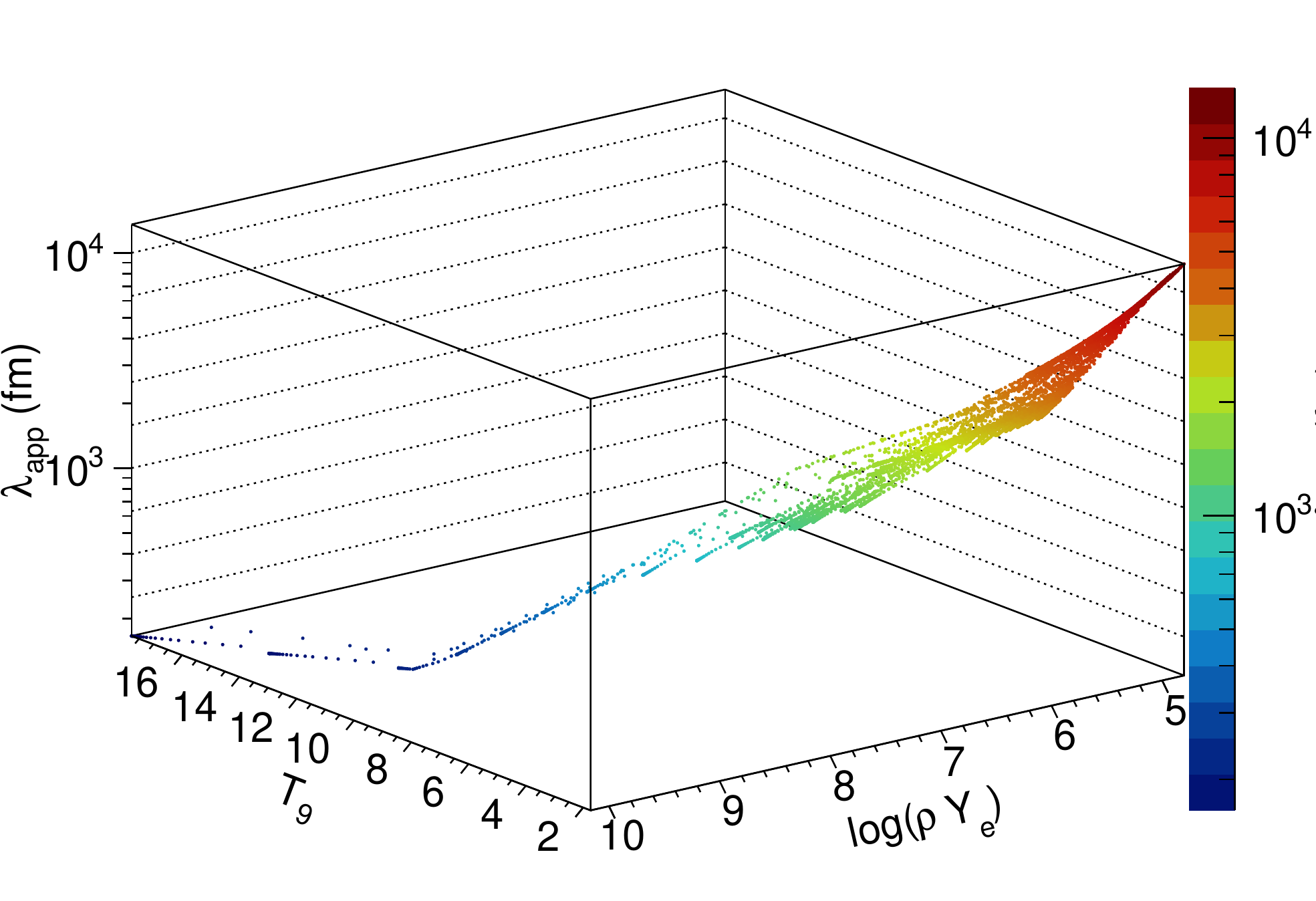}
            \includegraphics[width=0.5\textwidth]{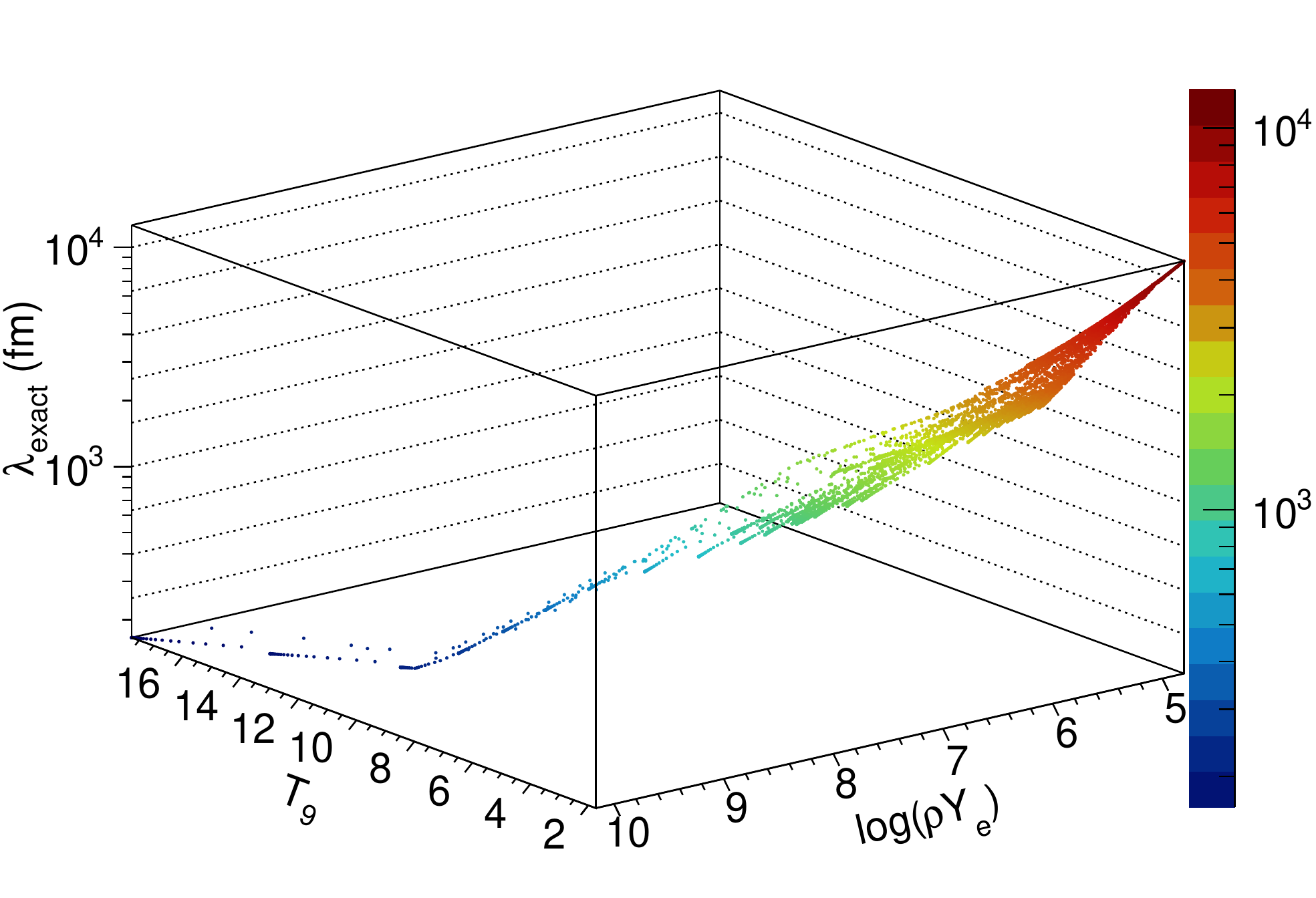}
    \caption{
    Evaluation of the error and applicability of various screening lengths for a 70 M$_\odot$ He core progenitor.  Each point in the figures corresponds to an individual mass-time coordinate in the overall evolution of this model from the start of the simulation to the final collapse.  (\textit{top left}) Calculated screening length for the entire evolution of the star.  The transition from relativistic screening to classical screening occurs at a temperature of 150 keV.  The points mesh seamlessly here.
    (\textit{top right}) The relative uncertainty between the approximation of Equation \ref{debye_exp} and the exact numerically calculated screening length.
    (\textit{bottom left}) The screening length computed using the approximation of Equation \ref{debye_exp} for kT$>$150 keV.
    (\textit{bottom right}) The exact numerically calculated screening length for kT$>$150 keV.
    }
    \label{error_figs}
\end{figure*}

As a representative case, we present calculations of the screening length for temperatures and densities within
a 70 M$_\odot$ He core progenitor in Figure \ref{error_figs}.  Here $T-\rho Y_e$ coordinates are presented for
the evolution of this progenitor.  Every point in this figure corresponds to a unique mass-time coordinate 
in the evolution of the star for zones in which the screening length is applicable ($\Gamma_{eff}<0.8$
\cite{www82}).  Figure \ref{error_figs} (\textit{top left}) shows the screening length at all points in the evolution.  it can 
be seen that the screening lengths match seamlessly at kT=150 keV, where the boundary between relativistic and
classical screening is set.  

The relative error induced in the use of the approximation is shown for all points in the evolution where 
relativistic screening is used in Figure
\ref{error_figs} (\textit{top right}).  While the error at low temperatures and densities is $\sim$5\% or less, the errors
at higher temperatures is quite small.  This is important to note as the effects of relativistic screening
are expected to be more pronounced at higher temperature.

For completeness, the relativistic screening length evaluations are shown in  the bottom row of Figure \ref{error_figs}. These compare
 the adopted approximation (\textit{bottom left}) and the exact numerical computation (\textit{bottom right}).  These evaluations are shown for all points in the evolution for which T$>$150 keV.  These
figures appear nearly identical.

The same evaluation was done for all models explored in this work, and the errors induced in
using the approximation were found to be nearly the same and quite tolerable in every case.

\begin{figure*}
            \includegraphics[width=0.5\textwidth]{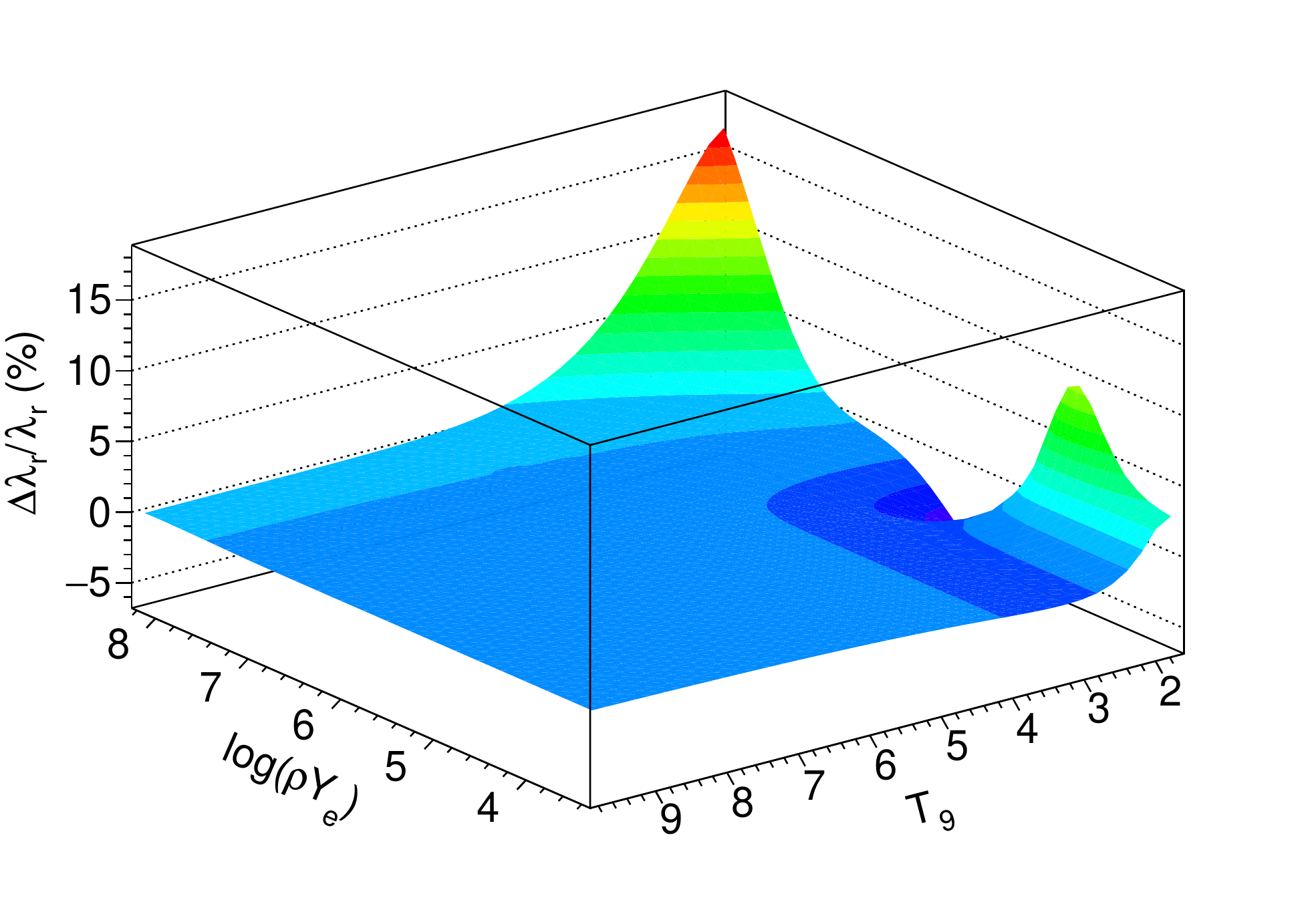}
            \includegraphics[width=0.5\textwidth]{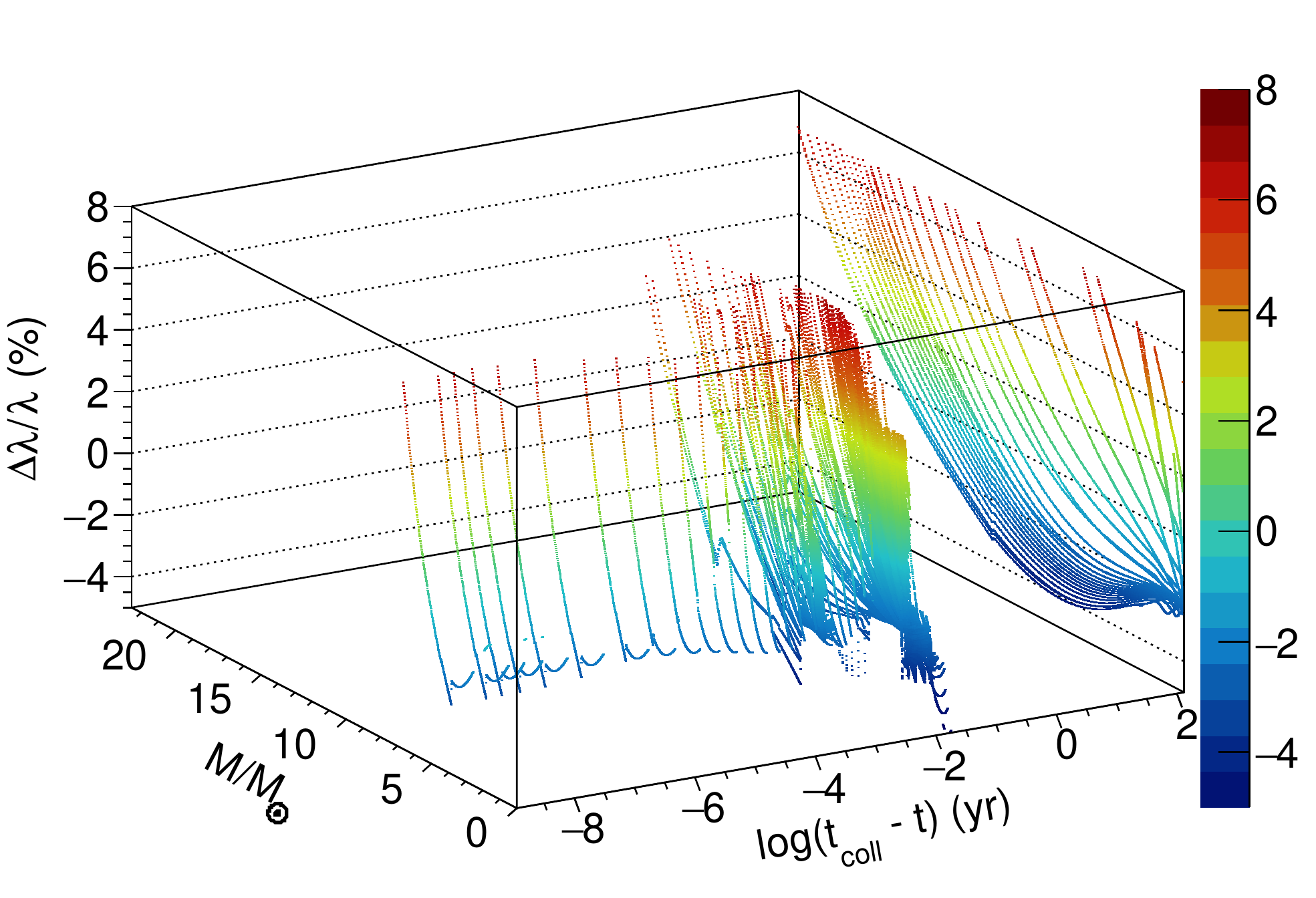}
    \caption{(\textit{left}) Relative error in relativistic Debye length
    using the approximation of Equation \ref{debye_exp}. 
    (\textit{right}) Relative error in relativistic Debye length for
    individual zones in the MESA evolution for a 70 M$_\odot$
    He progenitor as a function of time prior to collapse.  Each dot in the figure is a mass element at a specific time step.
    Only times at which $T>$ 150 keV and mass elements for which weak or intermediate relativistic screening is appropriate are shown.
    }
    \label{debye_compare}
\end{figure*}

The uncertainty in this evaluation is further explored in Figure \ref{debye_compare}, where
the relative difference in the numerically evaluated screening length, $\lambda_r$ and the
approximation of Equation \ref{debye_exp}, $\lambda_{exp}$, is shown as a function of temperature
and density times electron fraction, $\rho Y_e$.  In the left panel of this figure the
relative error of the approximation is shown where
$\Delta\lambda_r\equiv\lambda_{exp}-\lambda_{r}$.  There is some deviation at the lowest temperatures and densities, where the screening length is small.  Also, at low temperatures
and very high densities, the uncertainty is larger.  However,
this temperature-density combination does not occur in any of 
the simulations presented here as seen in Figure \ref{error_figs} and can thus be ignored in this model.  Further, this region
is more likely to be in the strong-screening regime, where
weak screening is invalid in any evaluation, including the commonly used Salpeter approximation.

\begin{figure*}

        \includegraphics[width=0.5\textwidth]{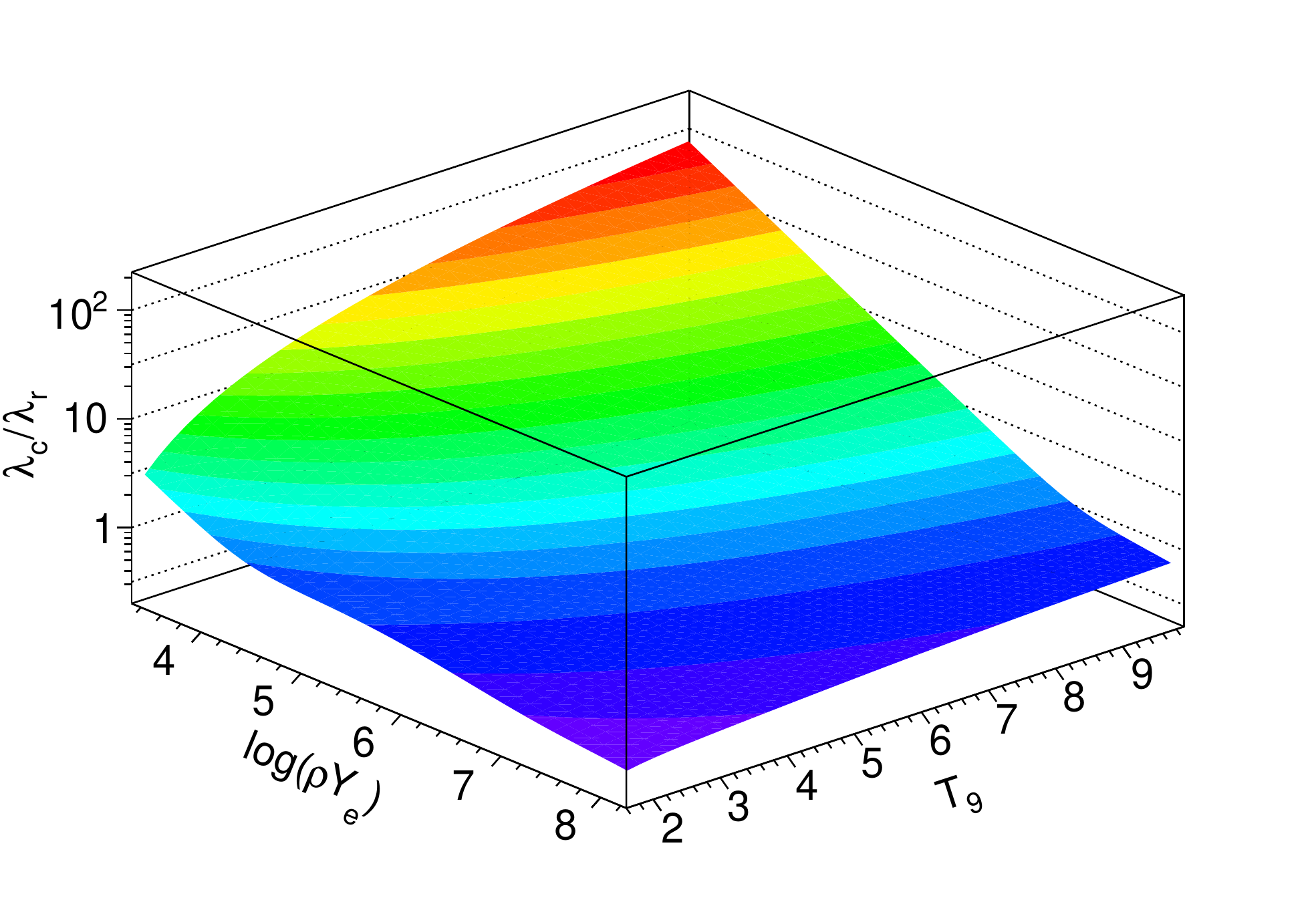}
            \includegraphics[width=0.5\textwidth]{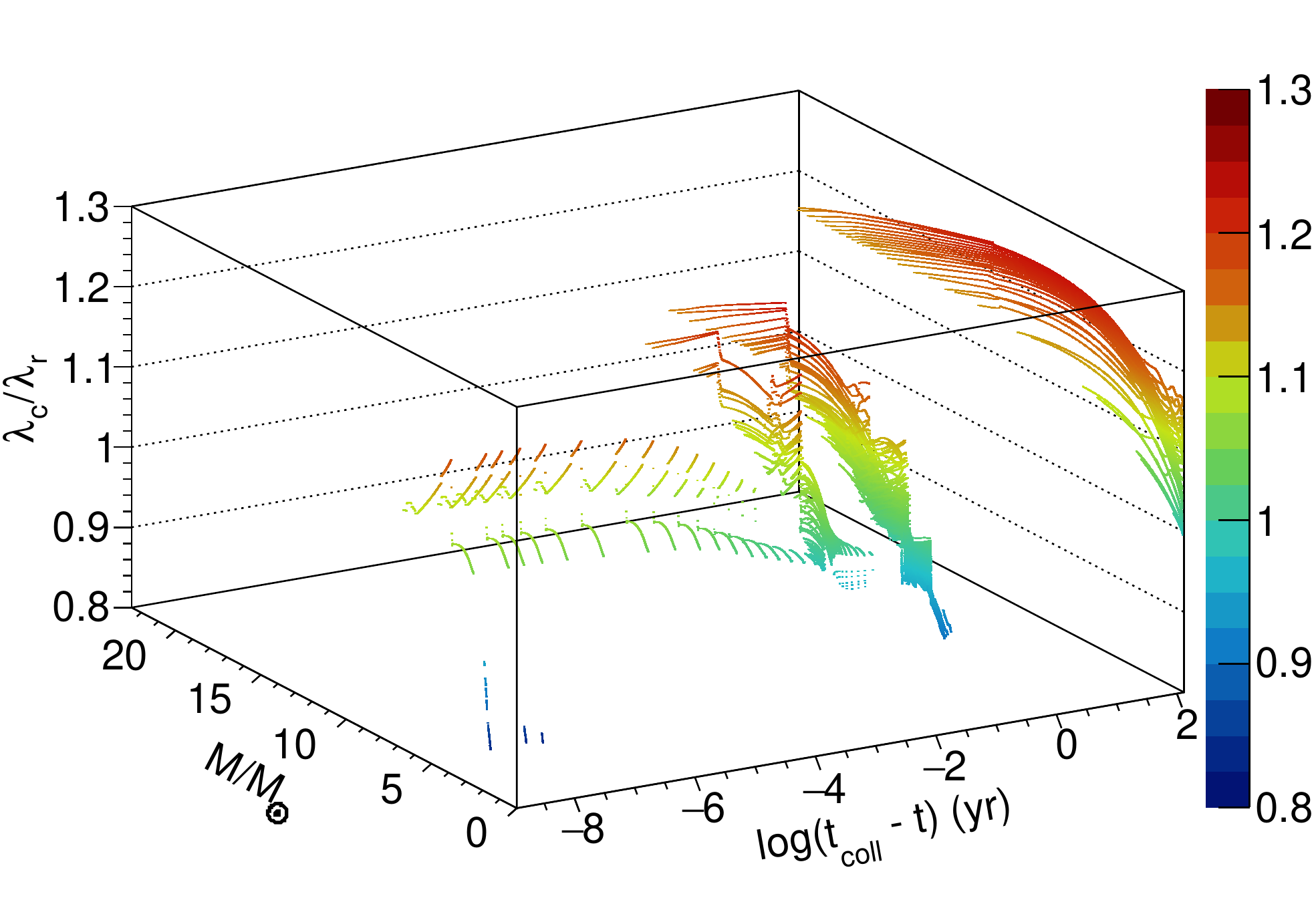}
    \caption{(\textit{left}) Ratio of classical to relativistic electron screening lengths for 
    the weak and intermediate screening as a function of temperature and electron density.  (\textit{right}) Ratio of classical to electron screening lengths for a 70 M$_\odot$ model evolution.  The time is shown as time before final collapse.  Each dot is a mass element and a time step. In the figure, values less
    than unity are blue, while those greater than unity are green.}
    \label{debye_rel_default_compare}
\end{figure*}

The uncertainty is explored further in the right panel of Figure \ref{debye_compare}, 
which shows the error in the relativistic screening length as a function of 
stellar mass and time coordinate for a 70 M$_\odot$ He core progenitor. (For 
Figures \ref{debye_compare} and \ref{debye_rel_default_compare}, only the 
electron screening lengths $\lambda_e$ are compared for clarity.) In 
this figure, the time coordinate is the logarithm of the time before final collapse,
and the mass coordinate is measured from the center of the star.  Each dot in the 
figure represents a mass-time point in the evolution.  The clustering
around pulses near log($t_{coll}-t)\approx 2$ and at log($t_{coll}-t)\approx -2$
correspond to small time steps as the temperature gradients are steep for the pulses. 
Here, only
coordinates for which relativistic weak or intermediate screening is relevant
(i.e., $\Gamma_{eff}<0.8$ and $kT>150$ keV) are displayed 
as these are the only
coordinates for which the relativistic weak screening length is
employed.  Thus, while it appears that the mass-time points
extend above the range of the plot, they actually do not.  
This is because only the relativistic approximation is being evaluated in this figure.  The apparent cutoff in points at the top of the plot is the region where the temperature of the
mass zones drops below 150 keV (farther from the stellar core).  Beyond these points, $kT<$ 150 keV,
so that  the default Salpeter screening length is used. 
These correspond
to the central portions of the star as it heats up; one sees that the
maximum mass coordinate is 20 M$_\odot$ even though there are cooler mass
coordinates in the outer layers of the star.  The outer layers of the star
are not sufficiently hot for a relativistic treatment.  As the central and and outer
layers of the star get hotter, the mass coordinates for which relativistic screening
becomes relevant increases towards the outer edge of the remnant.  It is seen
that the error in the screening length is within 10\% of the true value for all time and mass elements in
this treatment.

The relative difference between the  classical and relativistic Debye lengths $(\lambda_c - \lambda_r)/\lambda_c$ for the same site is shown in Figure \ref{debye_rel_default_compare}.  In the left panel, the ratio of the classical to the 
relativistic Debye length is shown as a function
of $\rho Y_e$ and temperature, $T_9$.  It is seen that the classical Debye length is larger than the 
relativistic Debye length for most of the area of this plot, except at low temperatures and higher densities,
where the classical Debye length is smaller than the relativistic Debye length.   We note that, at constant
density,
the classical Debye length increases monotonically with temperature, while the relativistic Debye length
decreases.   In typical astrophysical conditions, the points where $\lambda_c<\lambda_r$ are 
in the strong screening regime, and weak screening is not relevant anyway.  This is shown in the right
panel for the same mass-time elements as in Figure \ref{debye_compare}.  This shows that the early pulses (that are closer to the surface where the density is lower) have
a shorter $\lambda_r$.  Except for a very few high-density points near the core, the 
relativistic screening length is shorter than the classical screening length.
Naively, it is expected that the relativistic Debye screening always results in increased 
rates, and
we can see
that this is true overall except for a few very short periods of time in a very small region 
near the high-density core. 
\section{Results}
Multiple PPISN and PISN simulations were run varying by the initial mass and metallicity.  For the bulk of the 
simulations, the initial metallicity was set to $Z=Z_\odot/10$.  We examined ejected mass, final BH mass, pulse morphology, ejection time,
and nucleosynthesis in each model.  Representative results are presented here.
\subsection{Pulsational Time}
\begin{figure*}
        \includegraphics[width=0.5\textwidth]{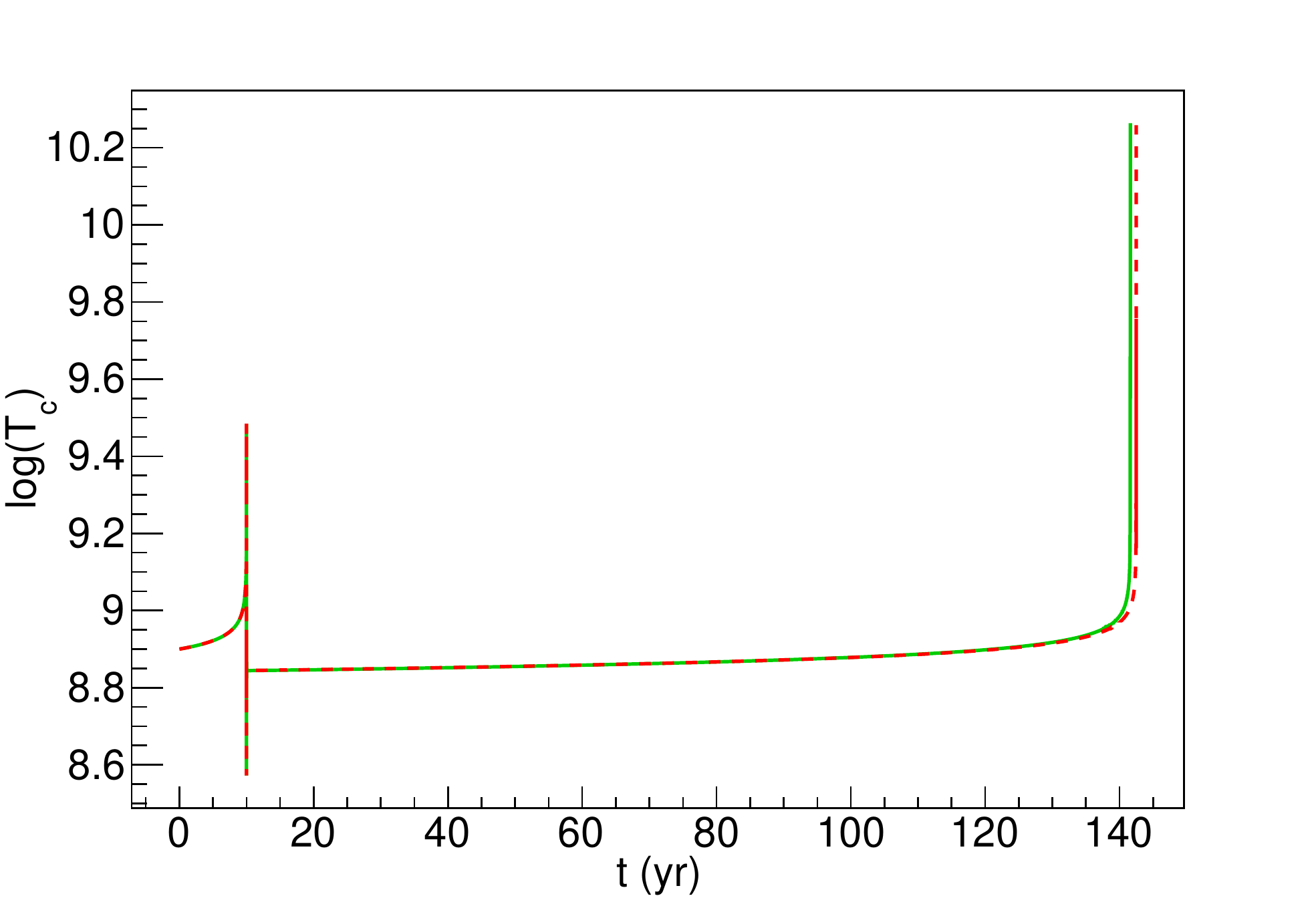}
        \includegraphics[width=0.5\textwidth]{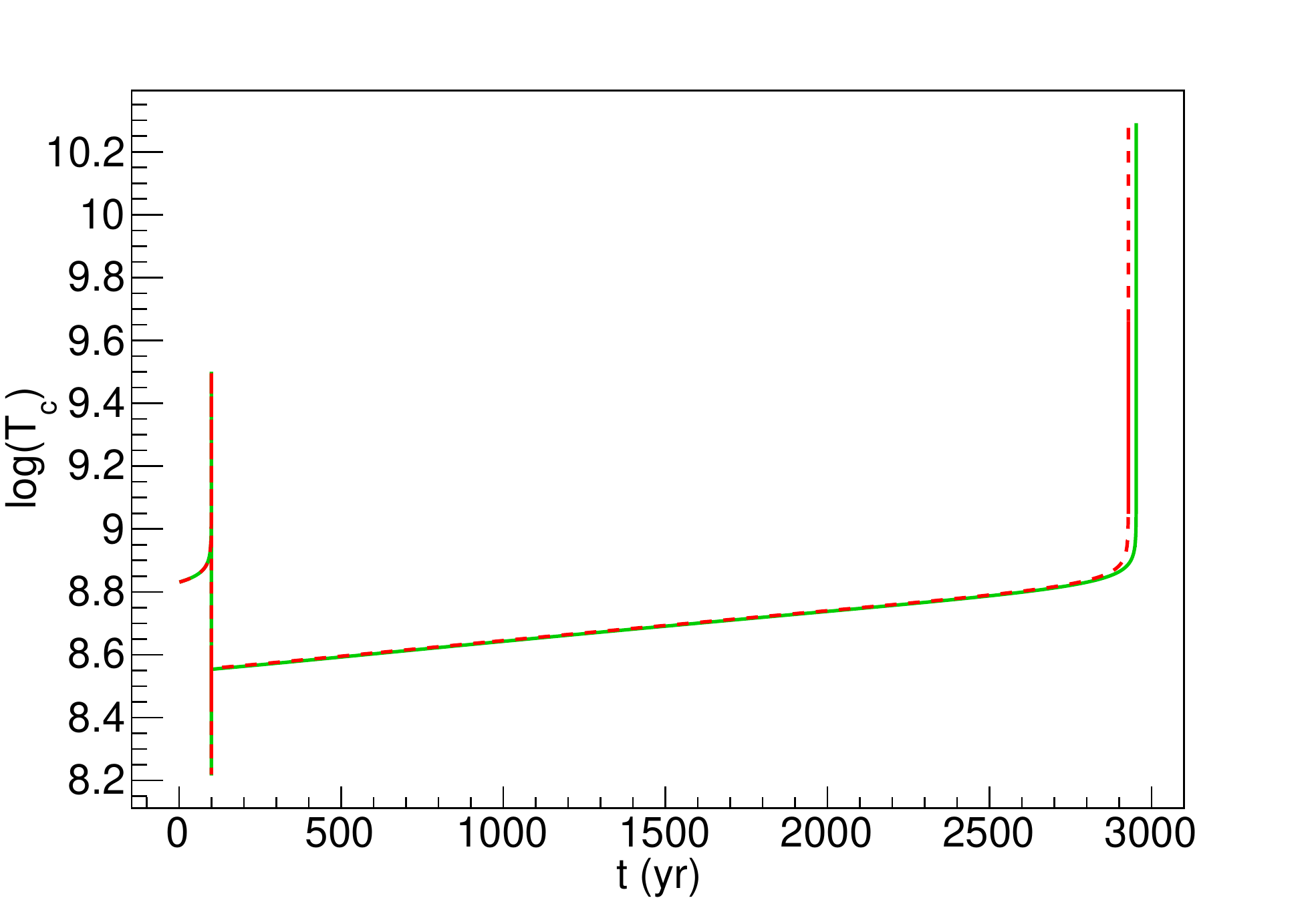}
        \includegraphics[width=0.5\textwidth]{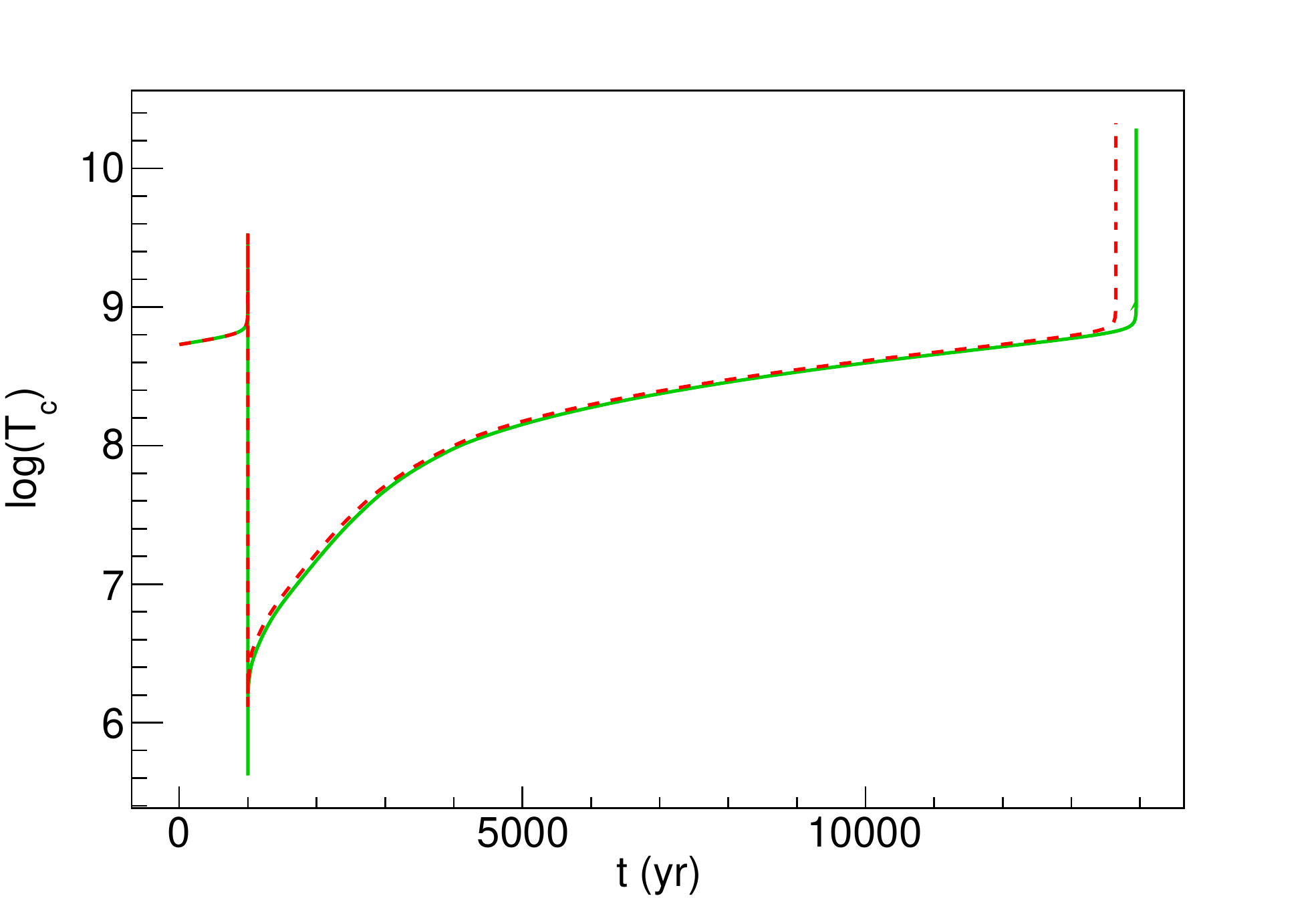}
        \includegraphics[width=0.5\textwidth]{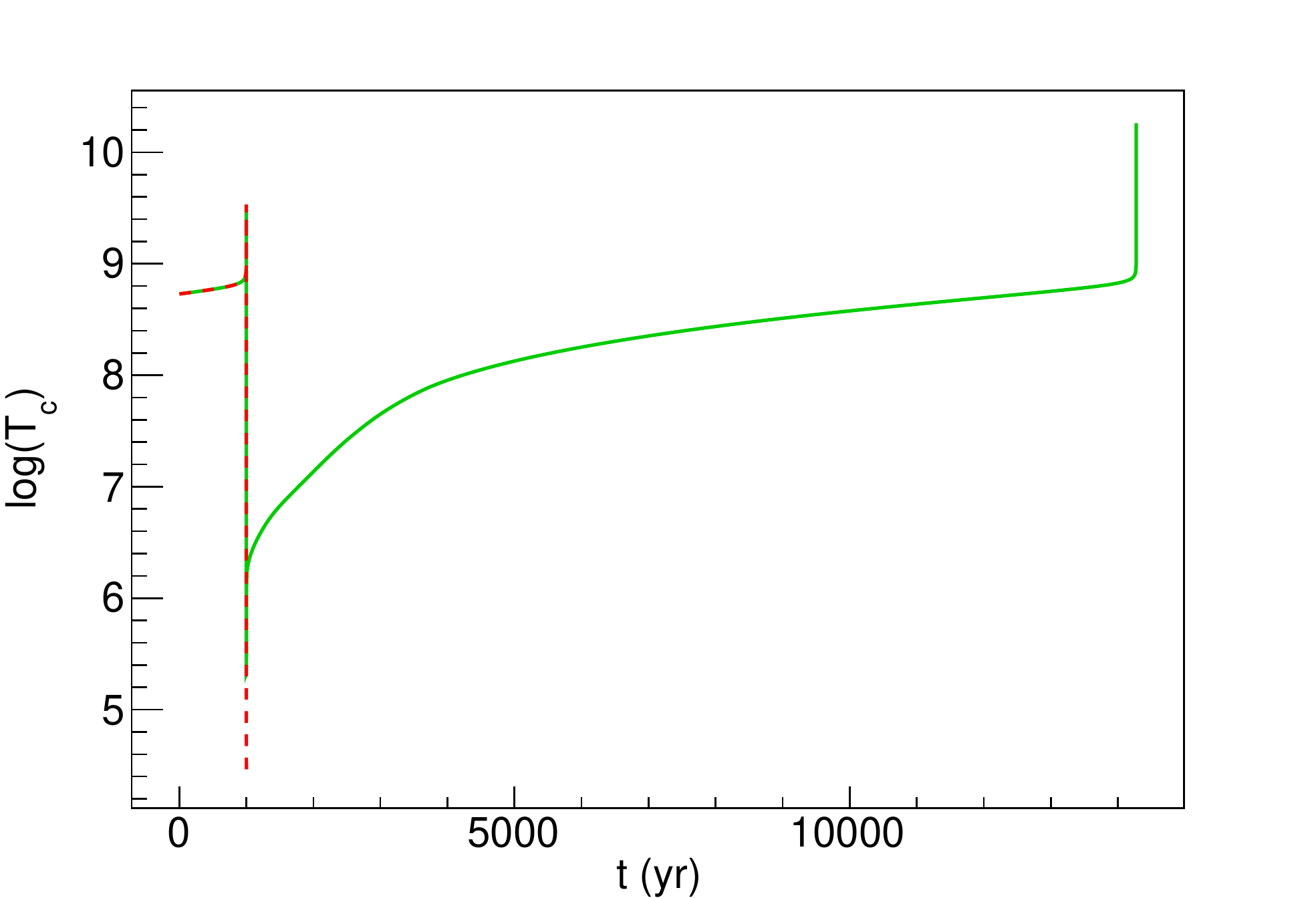}
    \caption{
    Logarithm of central temperature, $\log{T_c}$ versus time for various progenitor models.  The solid green lines correspond to the default screening model, and the
    dashed red lines correspond to the relativistic screening model. (\textit{top left}) 70 M$_\odot$, (\textit{top right}) 76 M$_\odot$, (\textit{bottom  left}) 89 M $_\odot$, and (\textit{bottom right}) 89.02 M$_\odot$}
    \label{temp_v_time}
\end{figure*}

\begin{figure*}
        \includegraphics[width=0.5\textwidth]{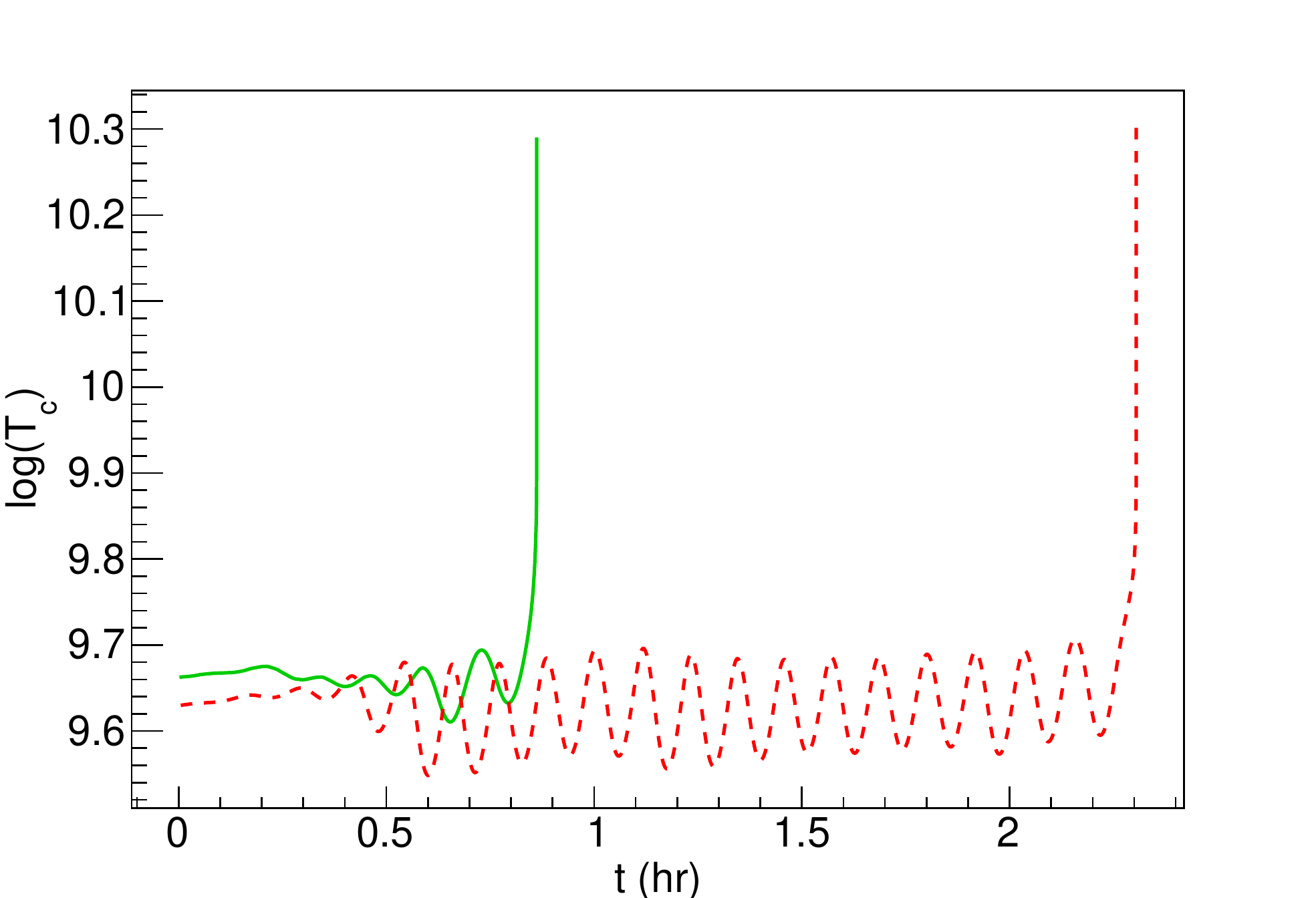}         \includegraphics[width=0.5\textwidth]{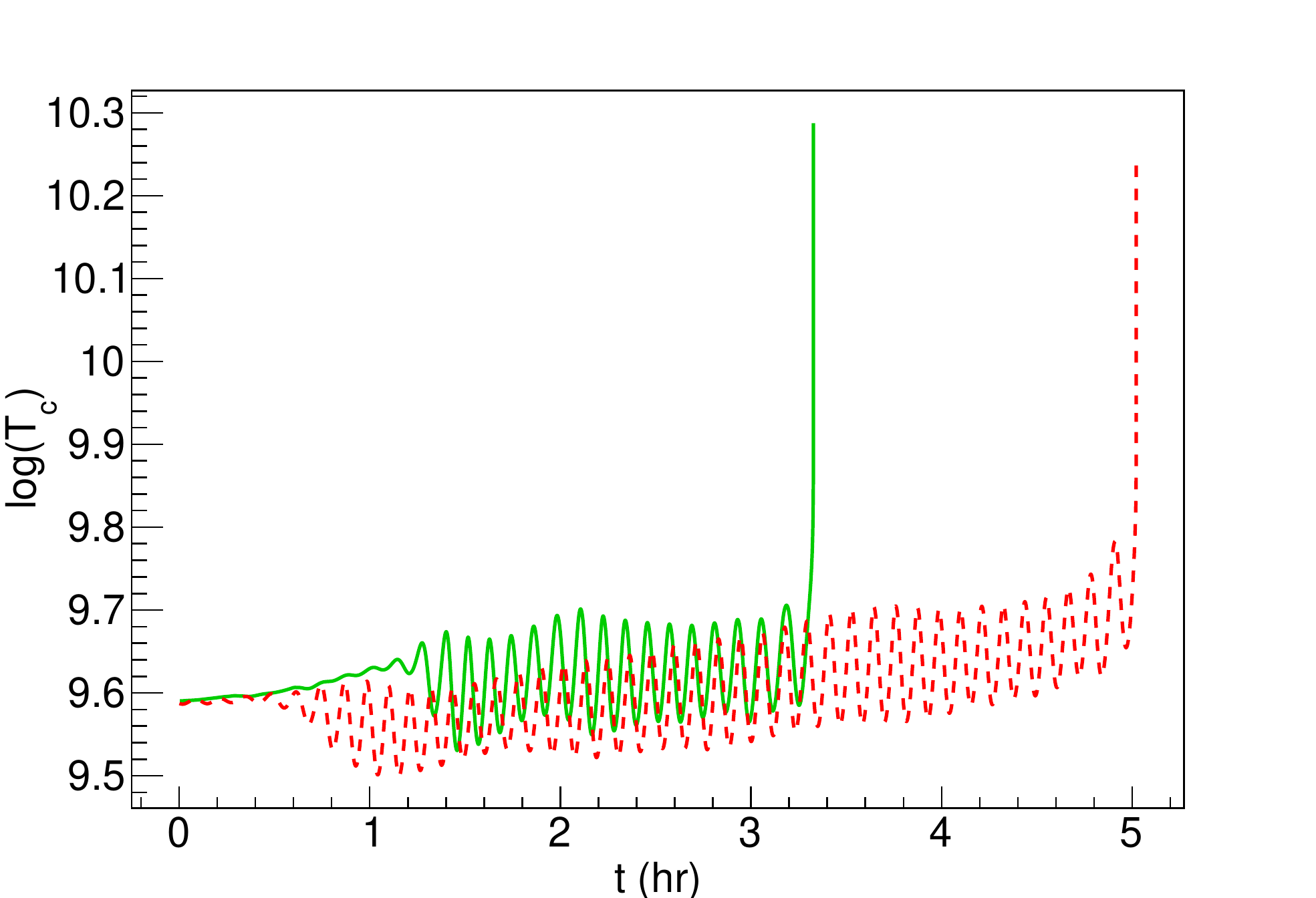}
        \includegraphics[width=0.5\textwidth]{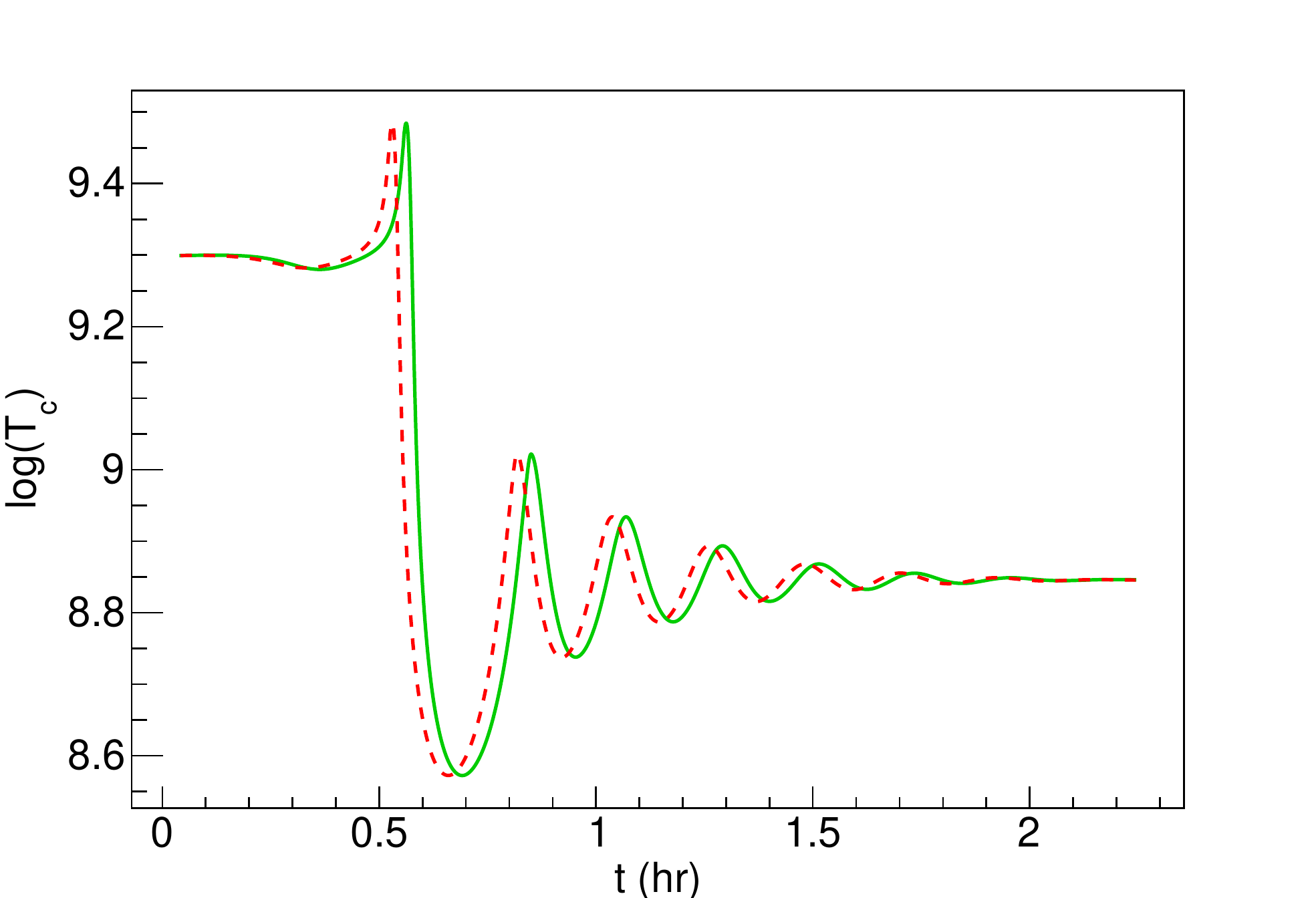}
         \includegraphics[width=0.5\textwidth]{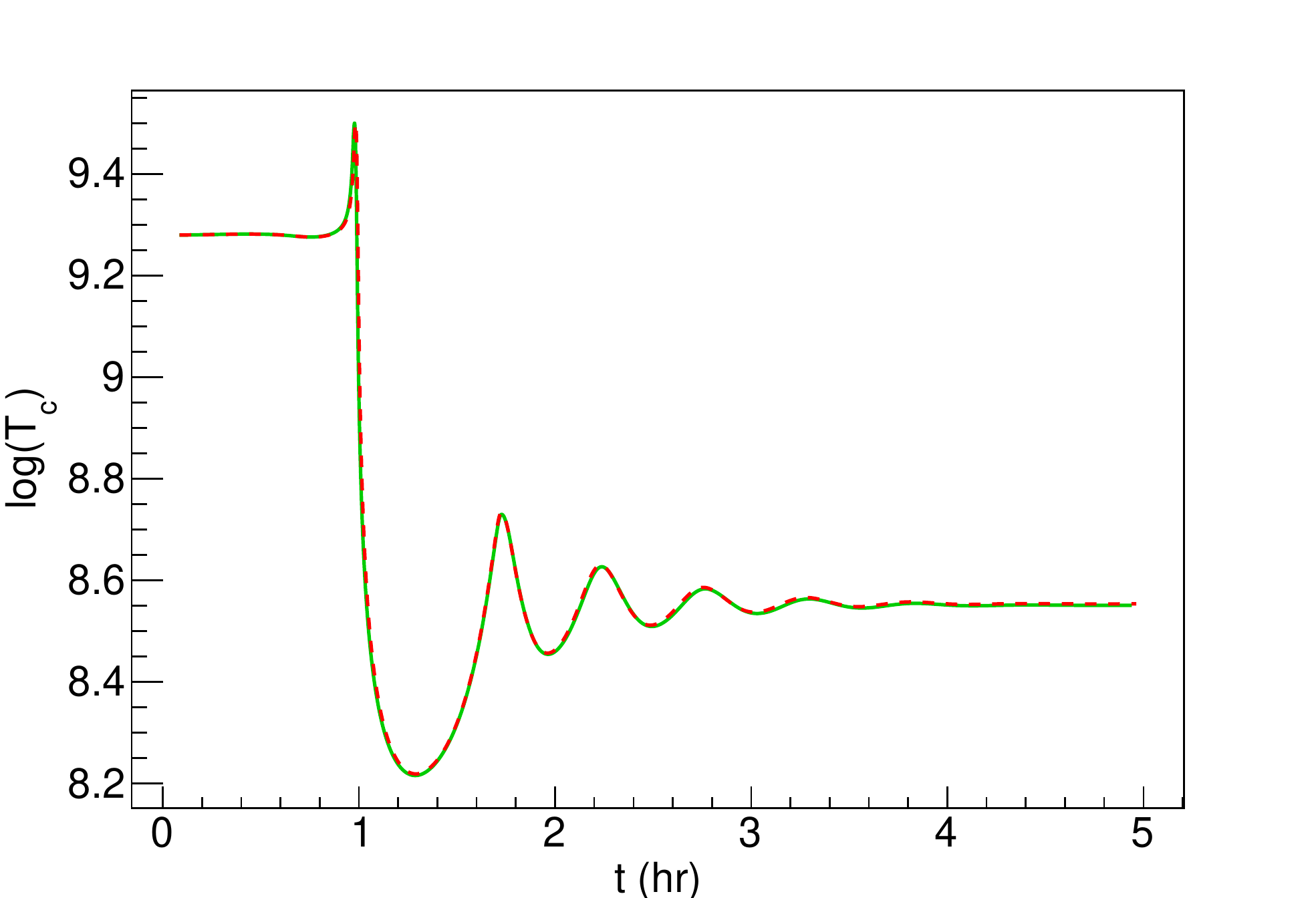}
        \includegraphics[width=0.5\textwidth]{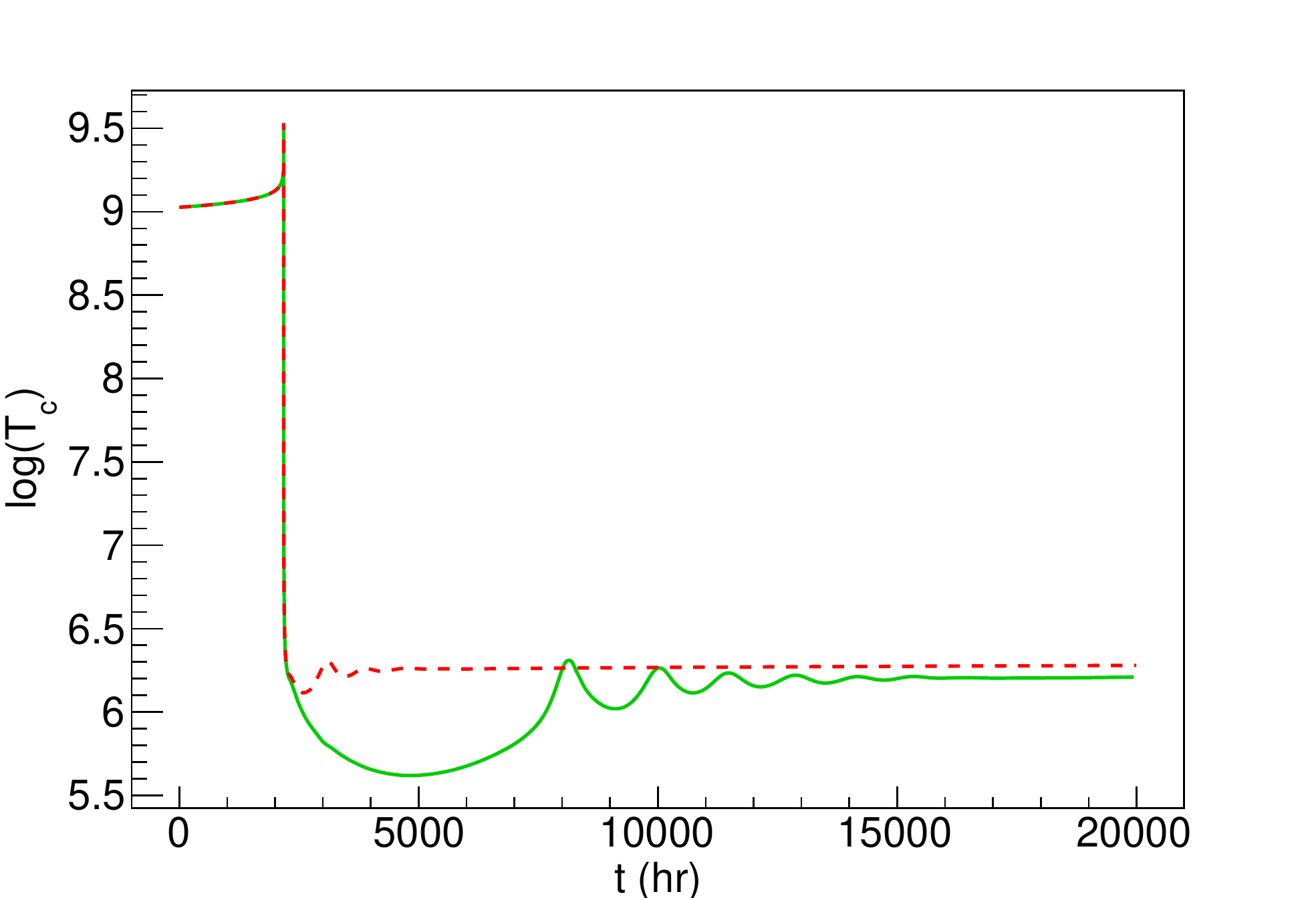}
         \includegraphics[width=0.5\textwidth]{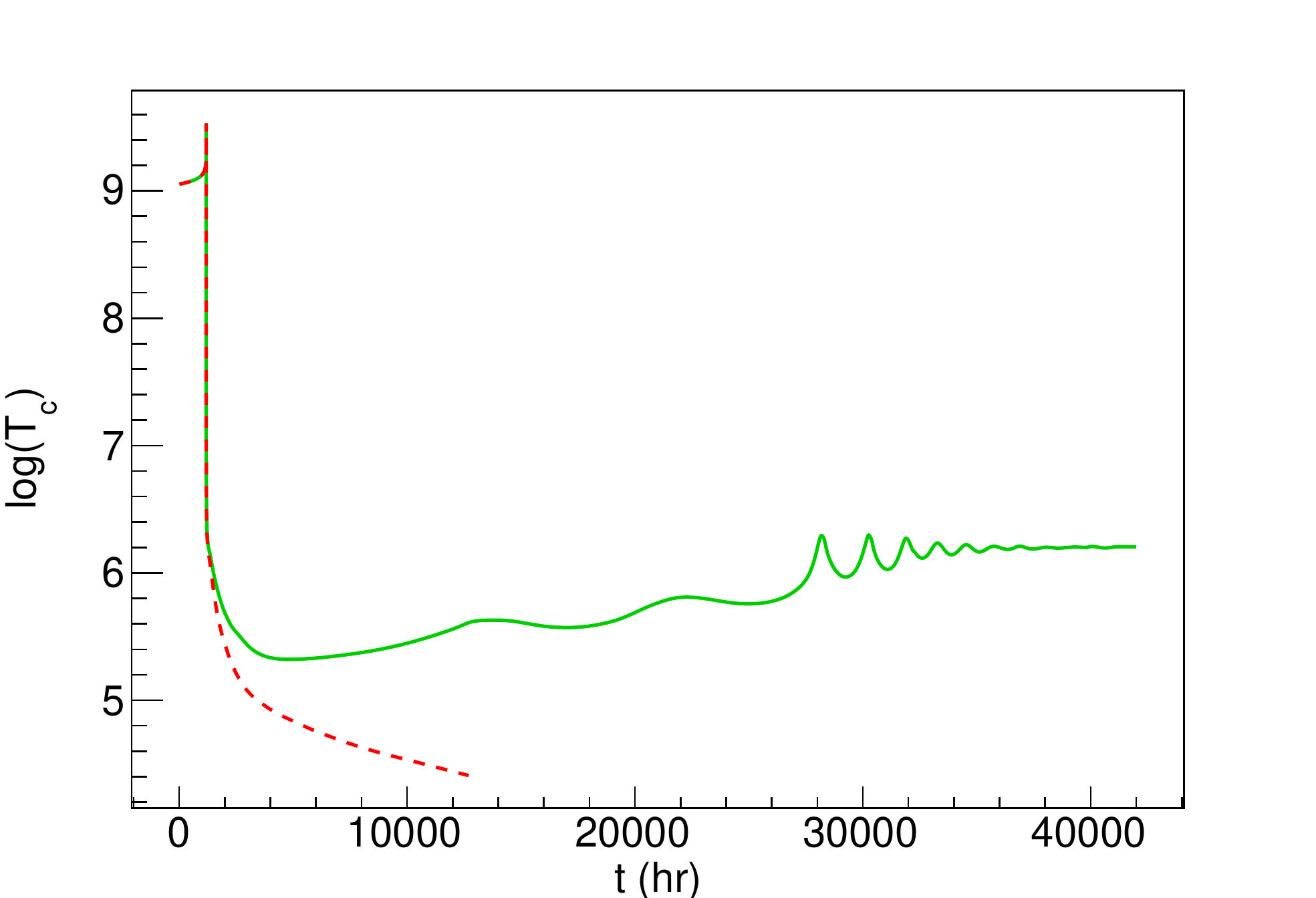}
         \caption{
    Logarithm of central temperature showing detail of the first pulse for several progenitors. The lines are the
    same as in Figure \ref{temp_v_time}.
    (\textit{top left}) 44.5 M$_\odot$,
        (\textit{top right}) 45 M$_\odot$,
            (\textit{middle left}) 70 M$_\odot$,
                (\textit{middle right}) 76 M$_\odot$,
                    (\textit{bottom left}) 89 M$_\odot$, and
                        (\textit{bottom right}) 89.02 M$_\odot$.}
    \label{temp_v_time_first}
\end{figure*}

\begin{figure*}
        \includegraphics[width=0.5\textwidth]{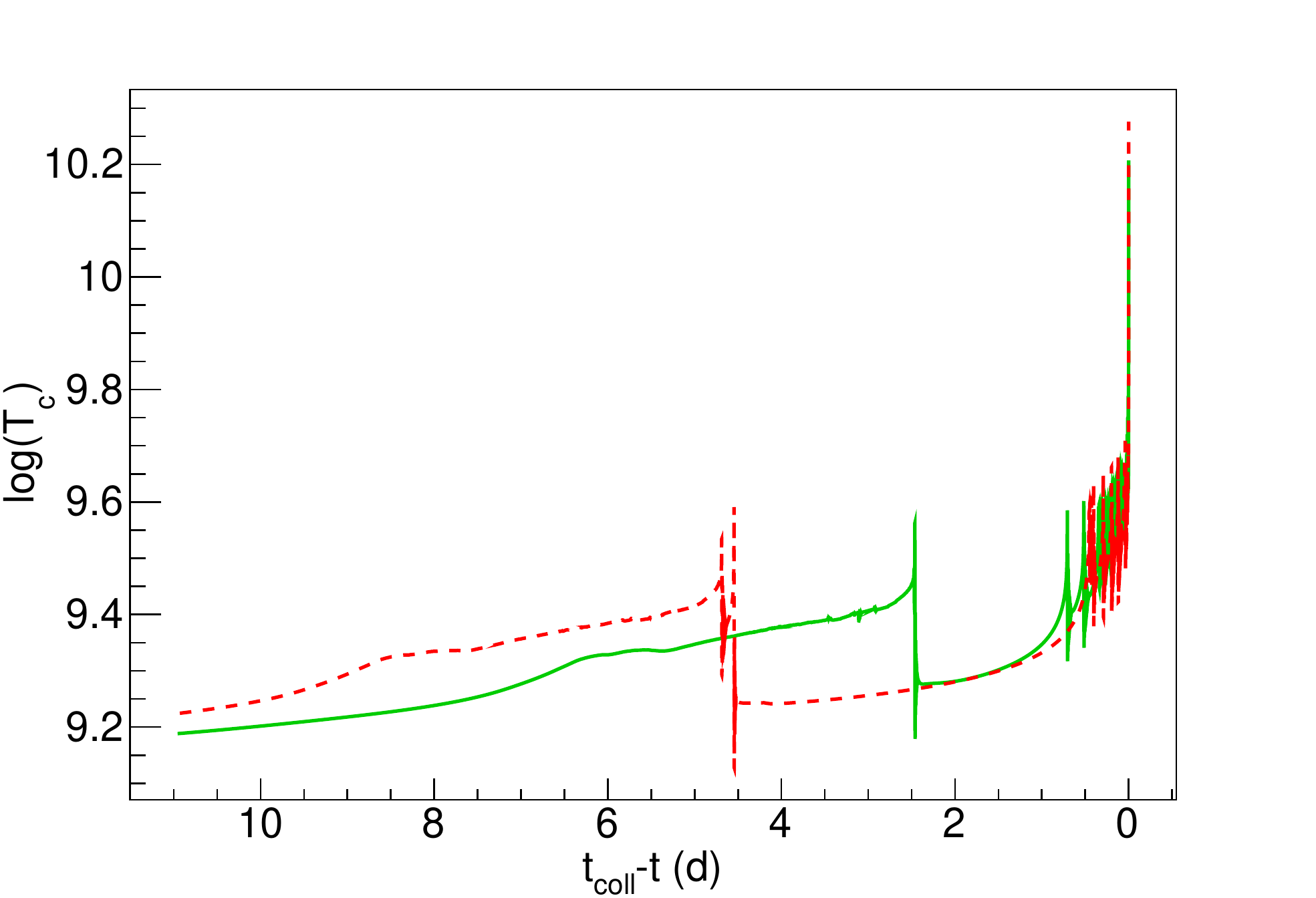}
        \includegraphics[width=0.5\textwidth]{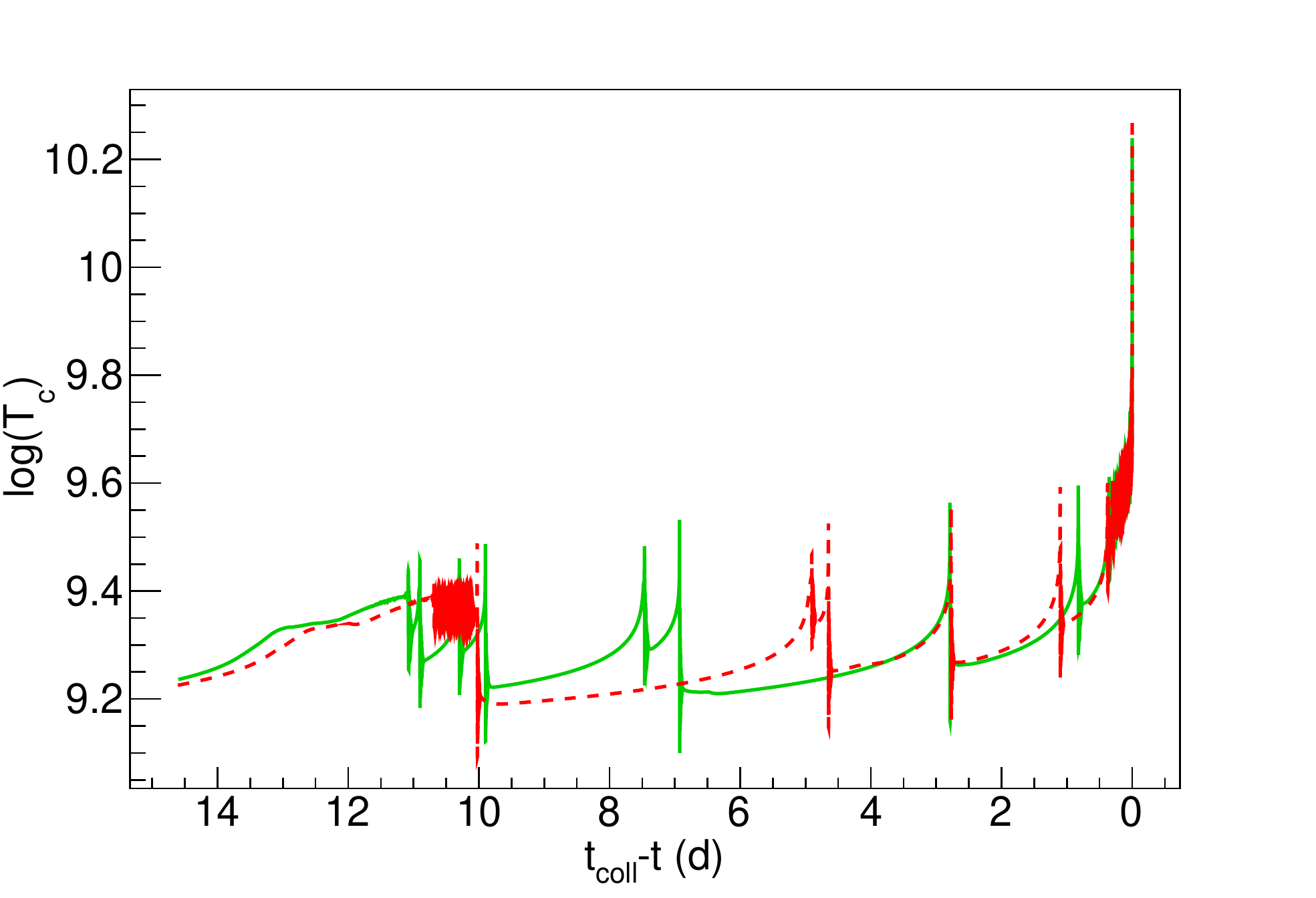}
            \includegraphics[width=0.5\textwidth]{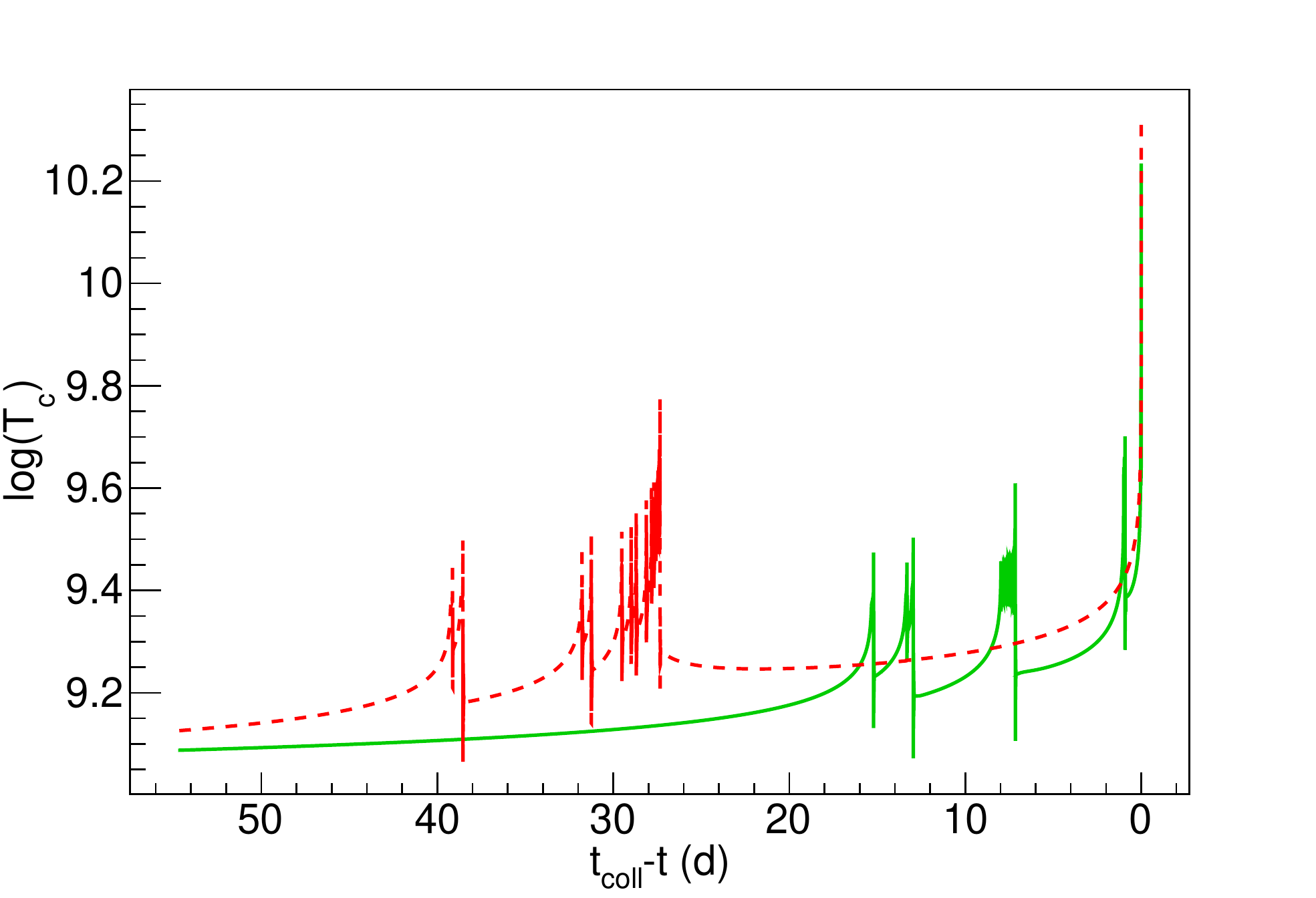}
        \includegraphics[width=0.5\textwidth]{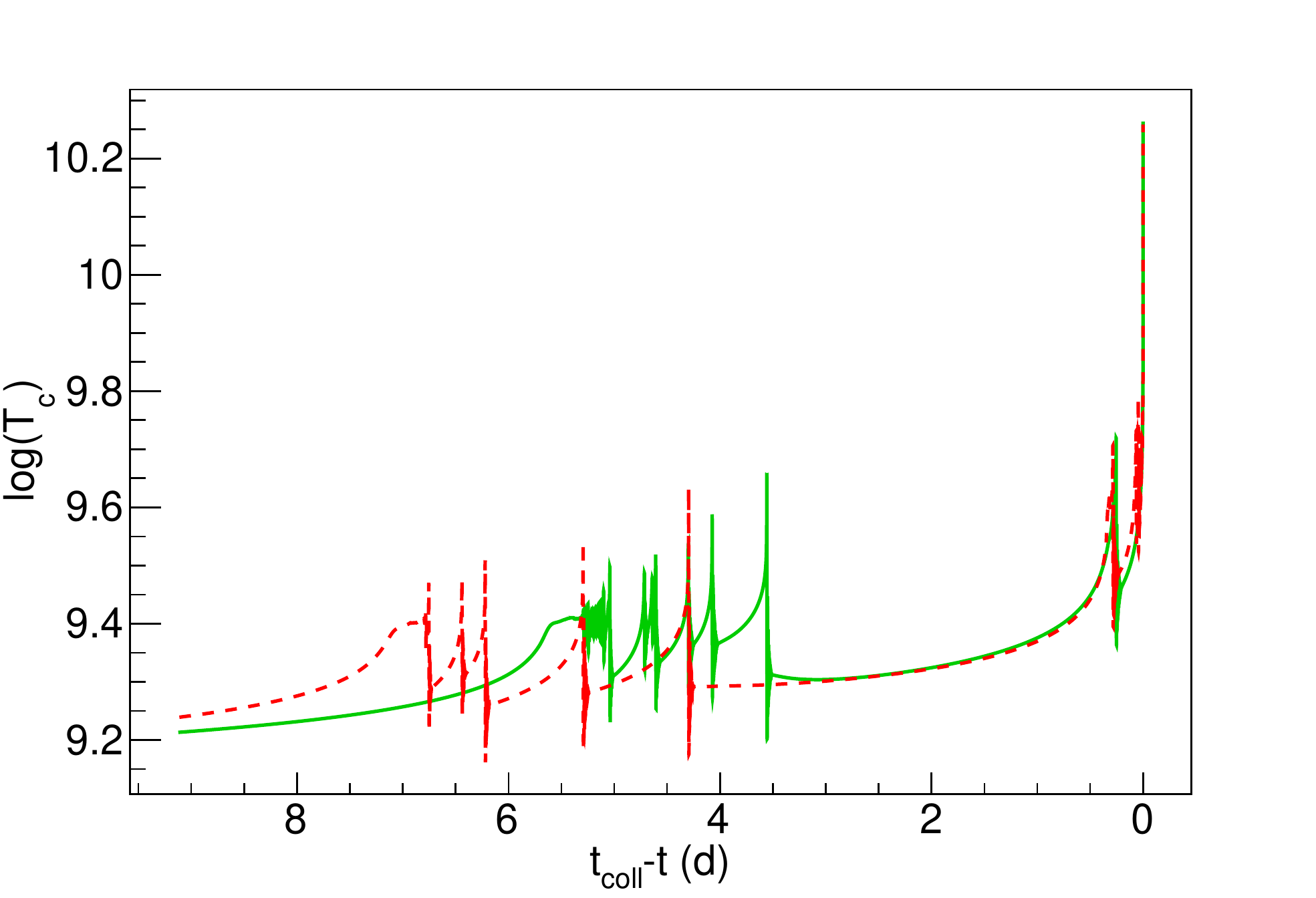}
    \caption{Detail of central temperature for three 
    progenitor models for both default (solid green lines) and relativistic (dashed red lines) screening
    rates.  
        (\textit{top left}) 52 M$_\odot$,
        (\textit{top right}) 58 M$_\odot$,
                    (\textit{bottom left}) 64 M$_\odot$, and
                        (\textit{bottom right}) 70 M$_\odot$.
    The relativistic screening case has been offset
    by 6.82 yr for the 52 M$_\odot$ model and
    by -2.79$\times$10$^4$ s for the 70 M$_\odot$ model so 
    that both evolution plots can fall within the same range.}
    \label{temp_v_time_zoom}
\end{figure*}

Time characteristics of the central temperature, which is used to define a pulse~\citep{marchant19}, are
shown for multiple representative progenitors in Figures \ref{temp_v_time} -- 
\ref{temp_v_time_zoom}.   Progenitor masses in this figure
are chosen to cover the full range of black-hole masses  resulting from PPISN as well as progenitors
in the region which produces the most massive black holes in the PPISN region. 

At the low end of the PPISN mass range, for 44.5 M$_\odot$ progenitors, stars undergo a direct collapse
after losing roughly 10 M$_\odot$ to the wind.  
At low progenitor mass, near the transition between direct collapse to PPISN, the central
temperature may exhibit an extended amount of time undergoing ``weak pulses''~\citep{woosley17} indicated by the 
oscillations in the central temperature prior to the final collapse.   
For the direct collapse BH at 44.5 M$_\odot$, the unstable pulsation for relativistic 
screening is more prominent.  However, these weak pulses do not occur in the default
screening model. 
In this borderline region, the very slightly increased reaction
rates at high temperature in the early pulsational stages results in a transition from
direct collapse to PPISN at a slightly (negligibly) lower  mass.  
While the instabilities occur at roughly the same time 
for both the default and relativistic screening models, the instabilities last for a longer
period of time and they extend the collapse to later times at low mass.  For the default screening
model, the onset of these weak pulses and unstable pulsation occurs at a slightly higher mass
in which the core temperature and density undergoes a transition from conditions that result in
direct collapse to that of a PPISN.

For masses around 70 M$_\odot$, which will be shown to produce the more massive black holes, 
the first pulses are found to occur at roughly the same time.  This is reasonable given 
that the core temperatures prior to the pulse do not exceed the threshold for 
relativistic screening until near the peak of the short time pulse.   However, because
of the slight differences in burning and composition in both models, subsequent
pulses and collapse occur at different times.  

For the 89 M$_\odot$ model, in which only one full pulse occurs prior to collapse, the final 
collapse time occurs slightly earlier for relativistic screening.  In this model, reaction rates
during the first pulse are increased by relativistic screening as the relativistic Debye length is shorter.
While the final nucleosynthetic abundances are not changed significantly, the processing is slightly faster.

In the case of the 80.92 M$_\odot$ progenitor, the default screening model proceeds to collapse, while the 
relativistic model proceeds to a PISN.  This mass is right at the PPISN/PISN boundary, and
the rates for the relativistic model produce a very slight increase in
the overall nuclear heating during the first pulse.  This region is a high-mass equivalent to
the low-mass region at which the star is close to being unstable against collapse or
explosion.  

Details of the first pulses in several models are shown in Figure \ref{temp_v_time_first}.
For the low-mass models near the boundary between direct collapse and PPISN collapse, the
weak pulses can be seen in detail.  The mass at which the weak pulses appear is lower in the
relativistic model, and the pulsation extends for a longer period of time prior to collapse
in the 45 M$_\odot$ model.   For the intermediate masses near 70 M$_\odot$, there is little
difference in the pulse morphology of the first pulses.

For the high-mass progenitors near the PPISN/PISN boundary, it can be seen that the
time to restore the star to the quiescent phase takes longer with increasing
progenitor mass.  In the case of the 89.02 M$_\odot$ model, the star can be seen to undergo
several oscillations in the central temperature before it stabilizes for the default
screening model.  However, in the case of relativistic screening, the explosion ends the stellar evolution at the first pulse.  The ultimate
effect here is to reduce the mass at which the PPISN/PISN boundary
occurs.  We note, however, that this reduction is very small and 
within the uncertainties of the current model~\citep{farmer19}.

In all cases, the evolution up to the first pulse is similar
in both screening models.   However, the subsequent evolution may be different in terms of the total time to collapse/explosion and in terms of pulse morphology.
Figure \ref{temp_v_time_zoom} shows the central temperature
evolution for the final pulses in various stellar 
progenitor models.   While the pulse morphology
looks
similar in all cases, some changes in the number of pulses,
pulse distribution and the pulse shape can be observed.
For example, in the 52 M$_\odot$ relativistic model, 
the increased rates in the final pulsation results in an 
additional period of weak pulsation near 
303 years.   The number of pulses later on
are roughly the same, though they occur at different times.
Overall, for this mass, the core spends more time
at temperatures for which relativistic screening becomes 
important.  

In the case of the 58 M$_\odot$ model, the earlier
pulse structure in the figure is dramatically different,
and the core undergoes fewer pulses at the latest times. Overall, this results in the 58 M$_\odot$ relativistic
model spending a much smaller amount of time 
at relativistic temperatures.   

In the case of the 70 M$_\odot$ model, the pulse morphology
is similar in both cases, with the number of pulses
and pulse duration similar in the default and relativistic
screening cases. The total time to collapse from the initial
pulse is only slightly longer in the relativistic model.
\begin{figure*}
        \includegraphics[width=0.5\textwidth]{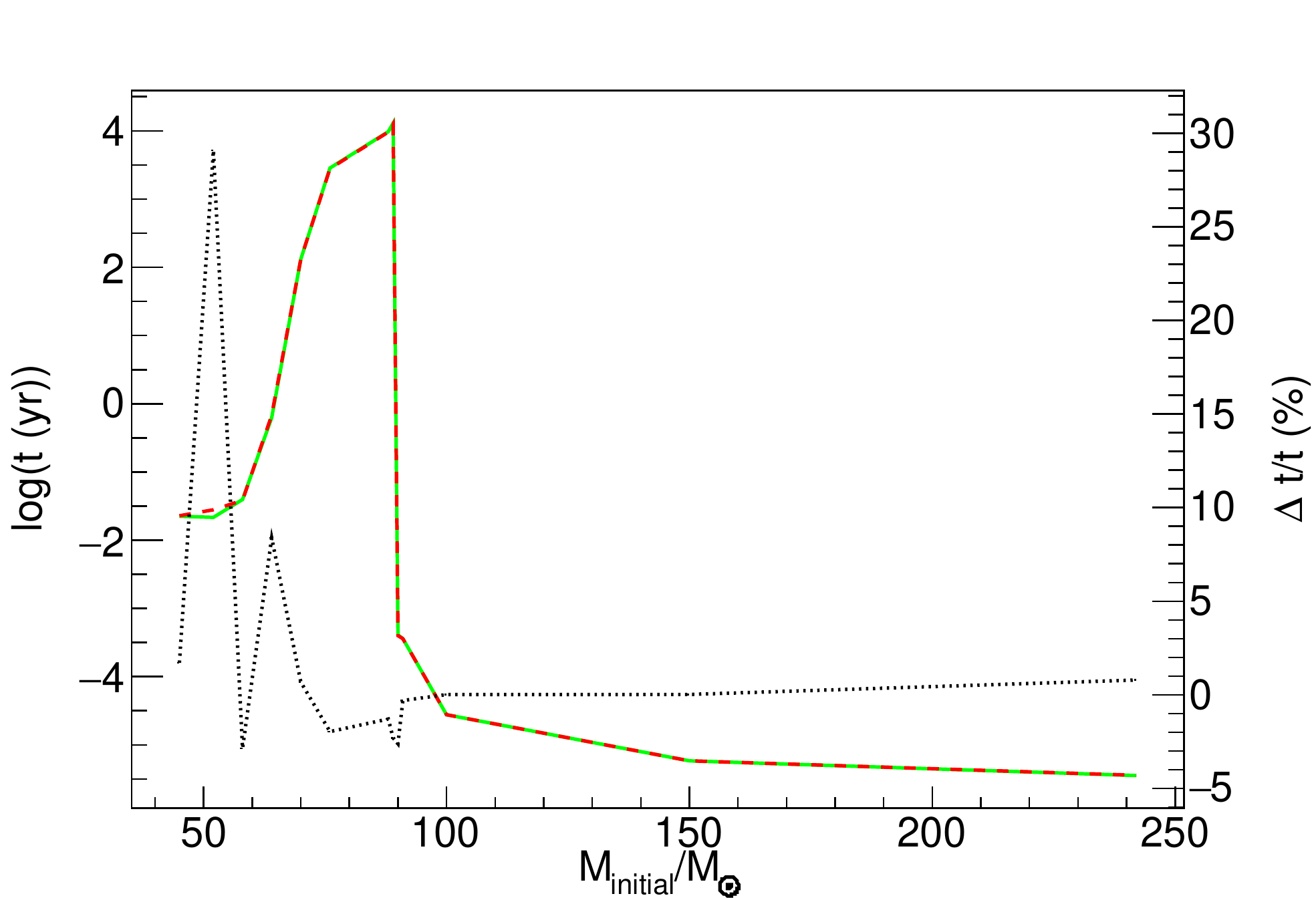}
        \includegraphics[width=0.5\textwidth]{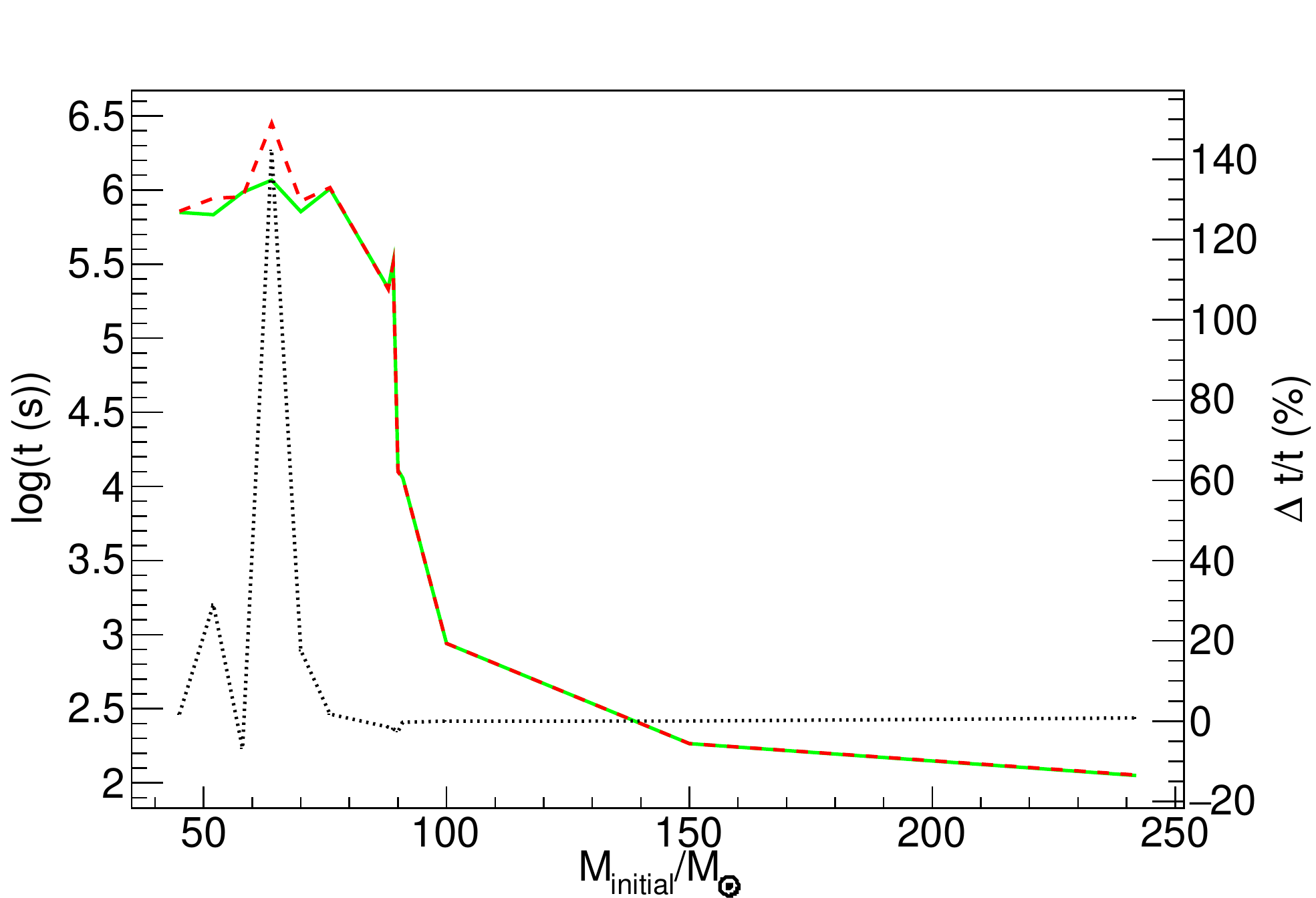}
    \caption{(\textit{left}) Time elapsed between initial pulse and final progenitor collapse or explosion as a function of initial mass.  (\textit{right}) Cumulative time for which the core temperature $kT>$ 150 keV as
    a function of initial mass.
    The solid green lines correspond to extended screening calculations.  The red dashed lines correspond to relativistic screening calculations.  The dotted black lines are the relative time difference
    between relativistic screening and default screening.}
    \label{pulse_time}
\end{figure*}

The total time from the onset of the initial pulse 
(defined as the time 
at which the central temperature first exceeds 150 keV) to final collapse or
explosion is shown in Figure \ref{pulse_time} (\textit{left}). This figure
provides a comparison of the total time spent in the 
instability region for a PPISN.  In this figure, the sharp
drop at $\sim$89 M$_\odot$ occurs at
the boundary between the PPISN and PISN.  At this mass
boundary, the PISN may undergo a single sharp rise 
in core temperature followed by an explosion as 
the entire star expands faster than the local escape 
velocity.  At the low-mass end of the PPISN region,
the total time to collapse is also small as the star 
undergoes only a single pulse.  For intermediate masses
(45 M$_\odot<$M$<$89 M$_\odot$), additional pulses
increase the time to collapse, though the time between 
the first and second pulses tends to 
dominate the evolution in this model~\citep{marchant19}.

In the intermediate mass region, the evolution is not
only dominated by the number of pulses, but
also the pulse morphology and stellar heating.  For example,
for a 52 M$_\odot$ star, the additional pulsation 
shown in Figure \ref{temp_v_time_zoom} creates
an extended period during which relativistic screening
is prominent.  This can affect the nucleosynthesis and
mass ejection during the pulse, which will be discussed later.  

Perhaps a better metric for evaluating the impact of 
relativistic effects  is the total time 
at which the star's core temperature exceeds 150 keV.
This is important as it provides a metric for mass models
for which the largest differences may be expected.
This is shown in Figure \ref{pulse_time} (\textit{right}).   Because 
of the instability of the PPISN progenitors, there
can be dramatic deviations over the entire mass range.
For example, for a 58 M$_\odot$ progenitor, the number of pulses and pulse time vary little between the first and
last pulse, resulting in little difference in times
spent at high temperature in each model.  For the 52 
M$_\odot$ progenitor, the additional instability and pulsation just prior to collapse results in 
significantly more time spent at high
temperature. The same is true for the 64 M$_\odot$ model, in which the extended 
instability and the additional time to collapse result in a significantly longer
time spent  at high temperature, where relativistic effects could
be more prominent.  The much longer time for the 64 M$_\odot$ model is shown as the large spike in the relative time difference in Figure \ref{pulse_time} (\textit{right}).
\subsection{Black Hole Masses for PPISN}
For PPISN, with progenitor core masses 44.5$\lesssim$M/M$_\odot\lesssim$89, the final
BH mass was determined in both the default and relativistic screening models.
As stated previously, the final baryonic mass is adopted as  a reasonable measure of the BH gravitational mass~\citep{fryer99}.
\begin{figure*}
        \includegraphics[width=0.5\textwidth]{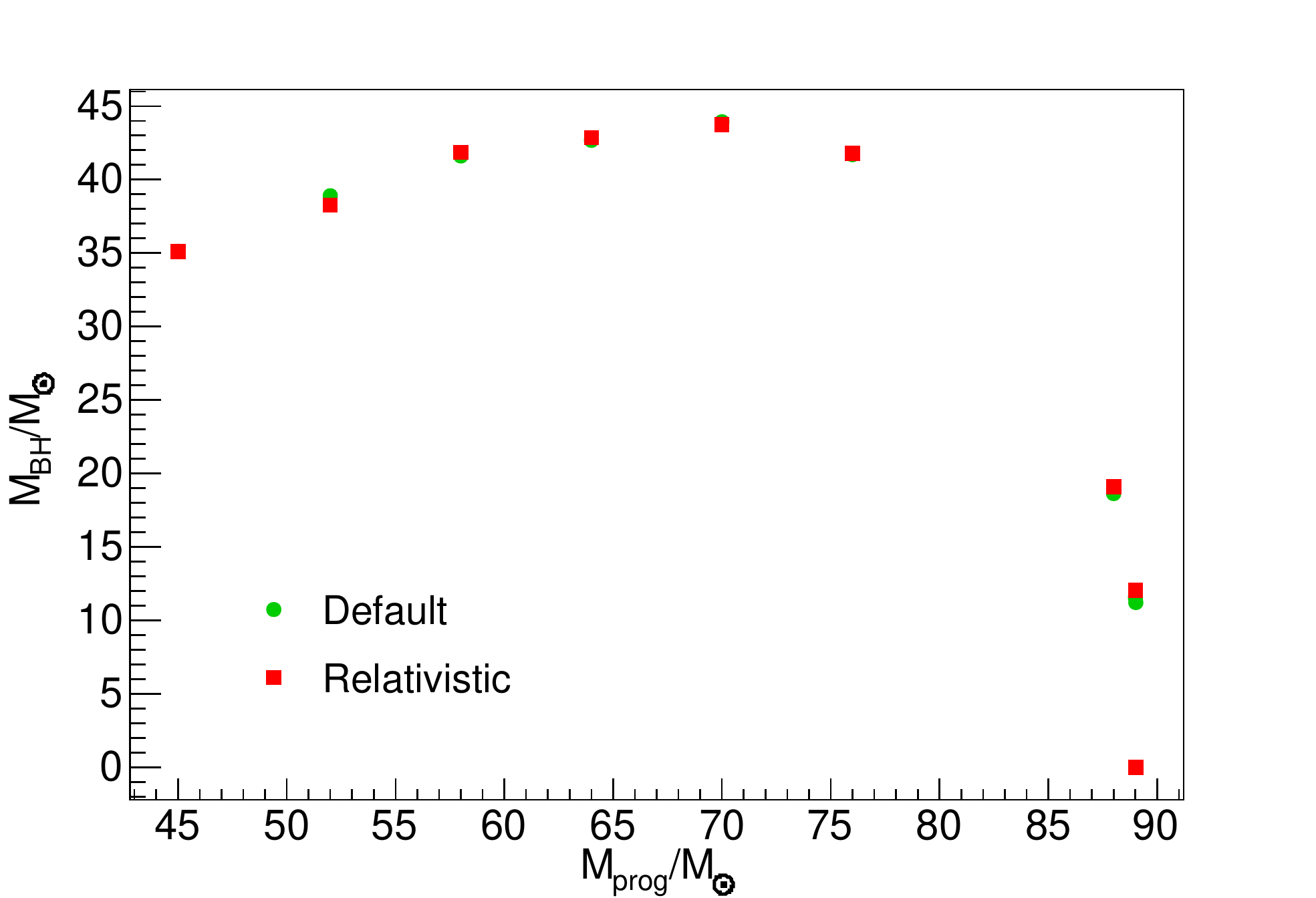}        \includegraphics[width=0.5\textwidth]{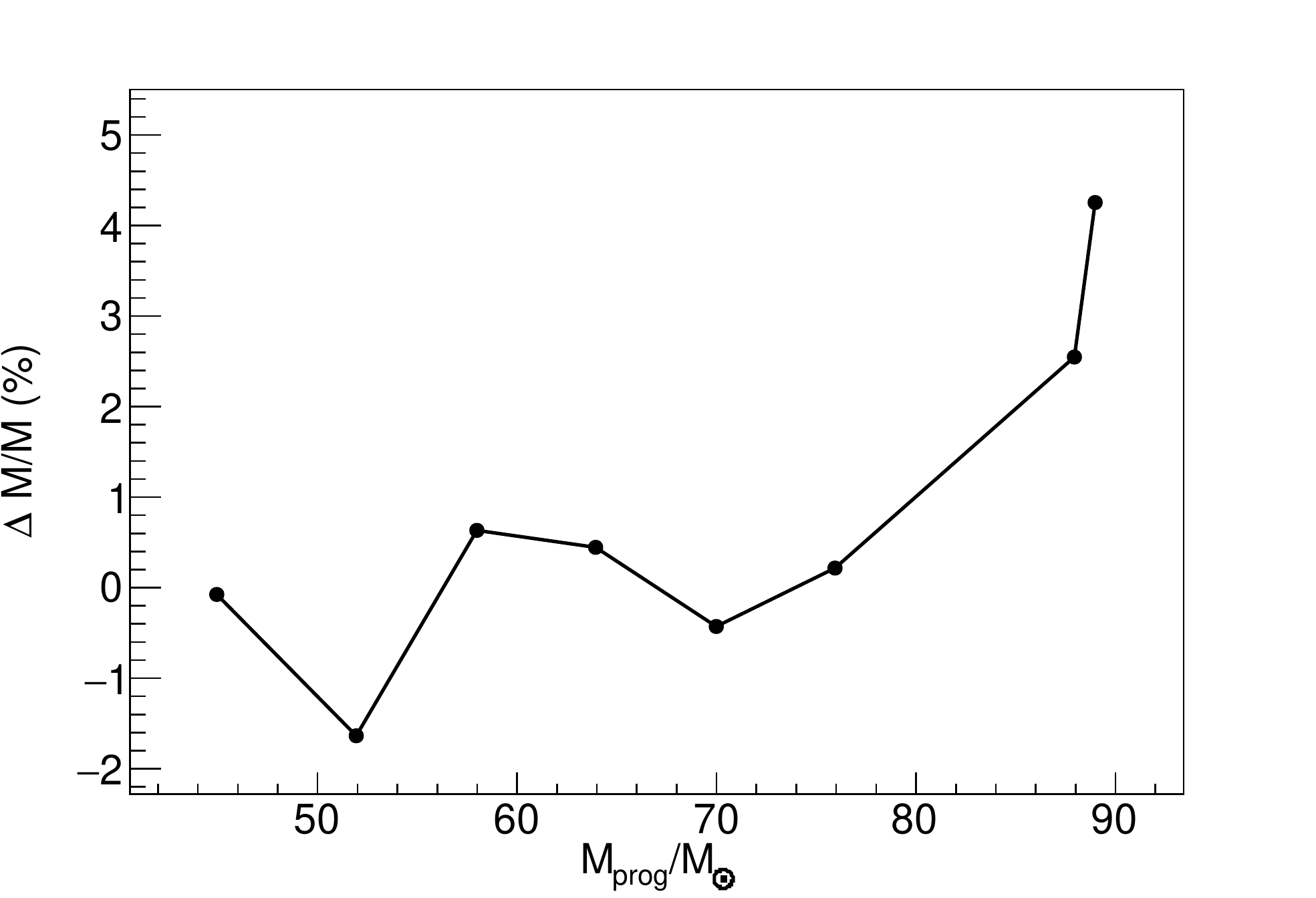}
    \caption{(\textit{left})
    Resultant BH mass versus He core progenitor mass for models including 
    the extended screening and relativistic screening. The 
    PPISN-PISN cutoff lies between 89.02M$_\odot$ and 89.05M$_\odot$ for the default screening model and 
    between 89M$_\odot$ and 89.02M$_\odot$ for the relativistic
    screening model.
    (\textit{right}) Relative shift in the resultant BH mass when 
    relativistic screening is used compared to the
    default screening model.}
    \label{BH_prog_mass}
\end{figure*}
The final BH mass as a function of progenitor mass is shown in Figure 
\ref{BH_prog_mass} (\textit{left}).  The difference in masses between both models as a function of progenitor mass is shown in Figure \ref{BH_prog_mass}(\textit{right}) where
$\Delta M/M\equiv (M_{BH,rel}-M_{BH,def})/M\times 100$\%.  In either case
it is seen that the influence of relativistic screening is negligibly small as
the change in mass is likely smaller than the numerical uncertainties of the model
\citep{farmer19}.  The BH masses shown in Figure \ref{BH_prog_mass} (\textit{left}) are in the mass range
11.22$<$M$_{BH}$/M$_\odot<$43.94.  These values are comparable to those extracted from
recent LIGO/VIRGO data for binary-black-hole merger events~\citep{abbott19}. Lighter
BH merger events, such as GW170608 contained lower-mass black holes 
which likely originated from core-collapse supernova events and not
PPISNe.  One merger event, GW170729, was found to contain a 50.2 M$_\odot$ BH, which lies
in the mass gap in which massive stars are thought to undergo PISN.  It has
been found, however, that these more massive BHs may be formed if
the $^{12}$C($\alpha$,$\gamma$)$^{16}$O lower, while still being within the 68\% 
confidence interval for this rate~\citep{farmer19}.
\begin{figure*}
        \includegraphics[width=0.5\textwidth]{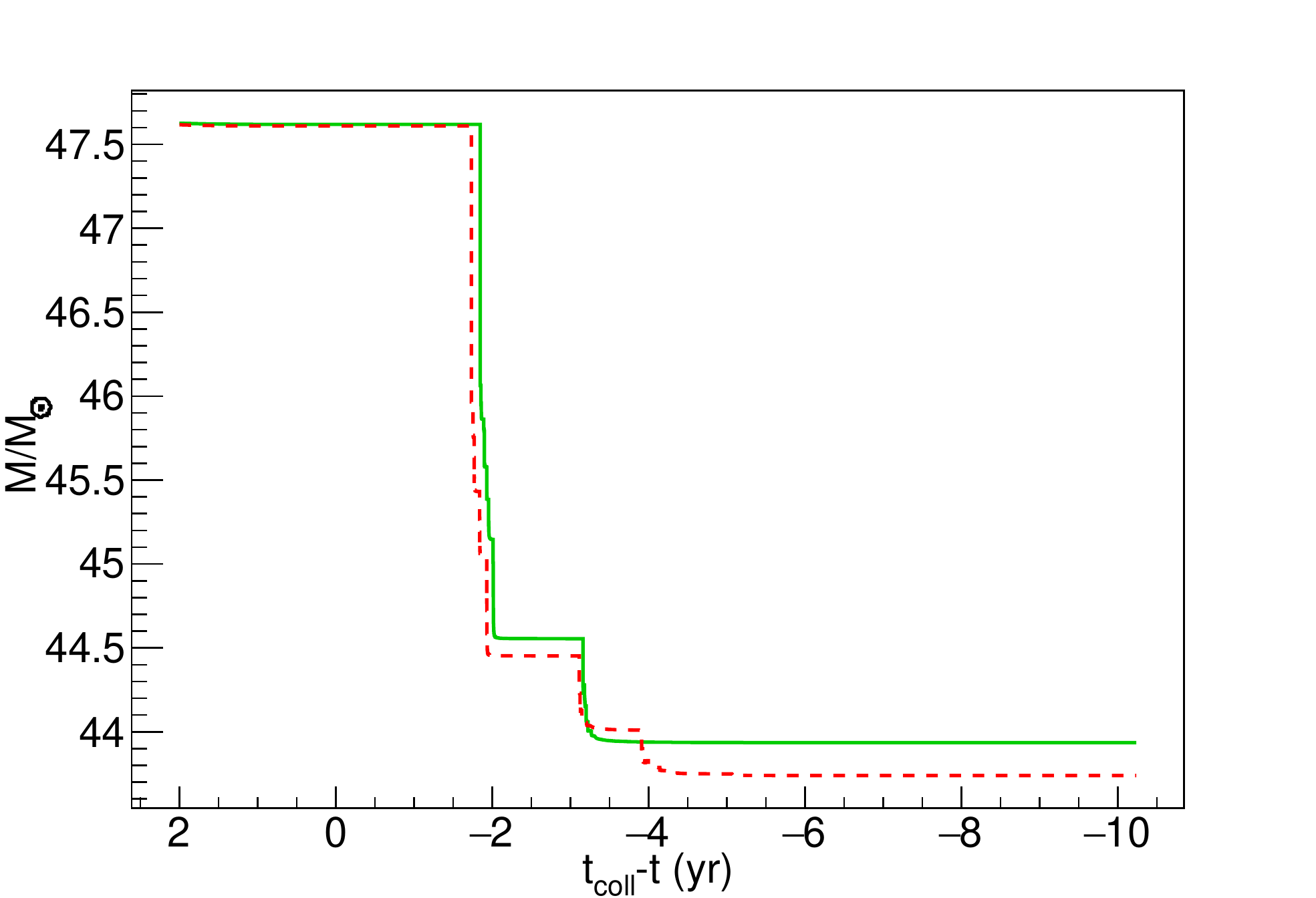}
         \includegraphics[width=0.5\textwidth]{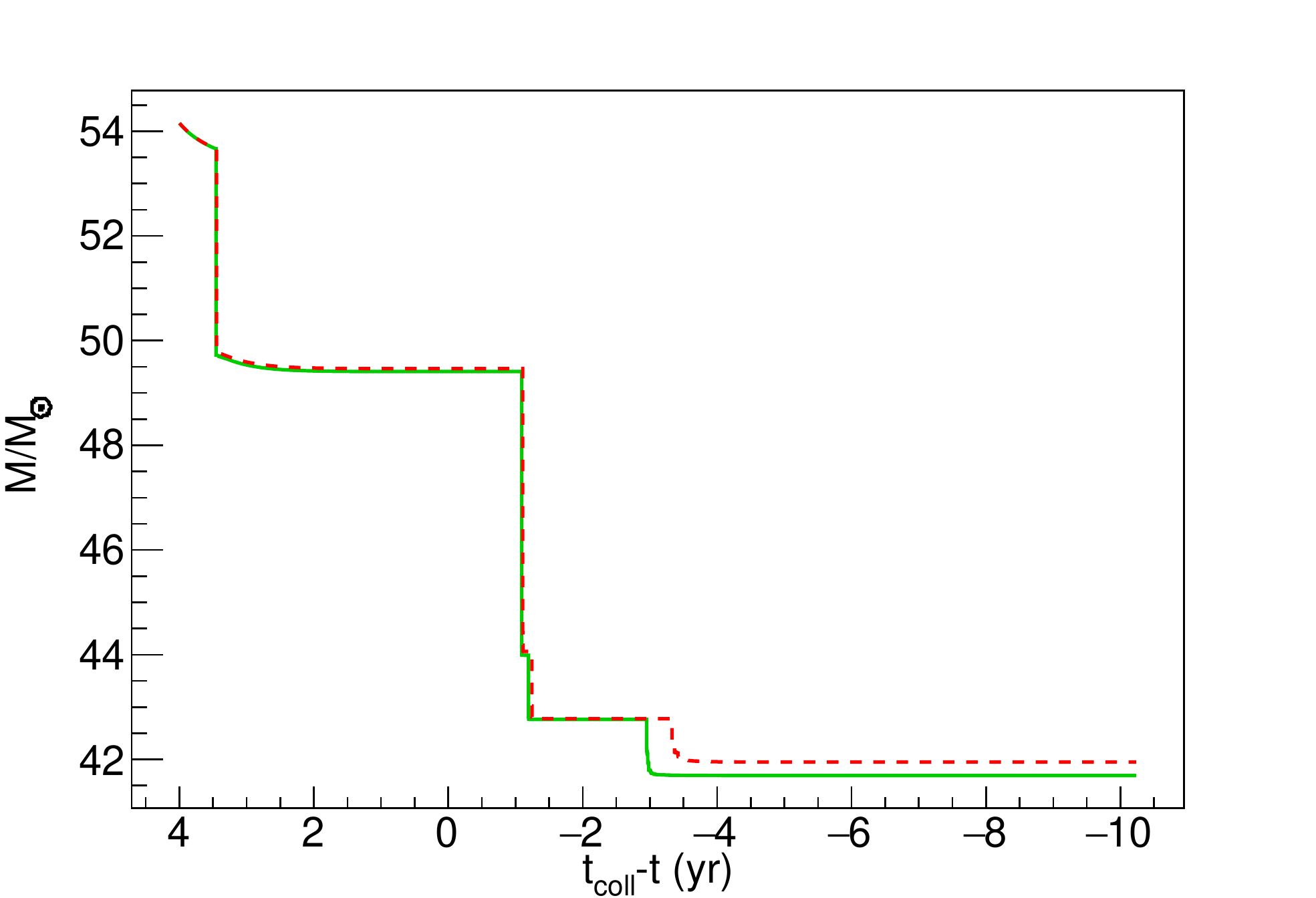}
        \includegraphics[width=0.5\textwidth]{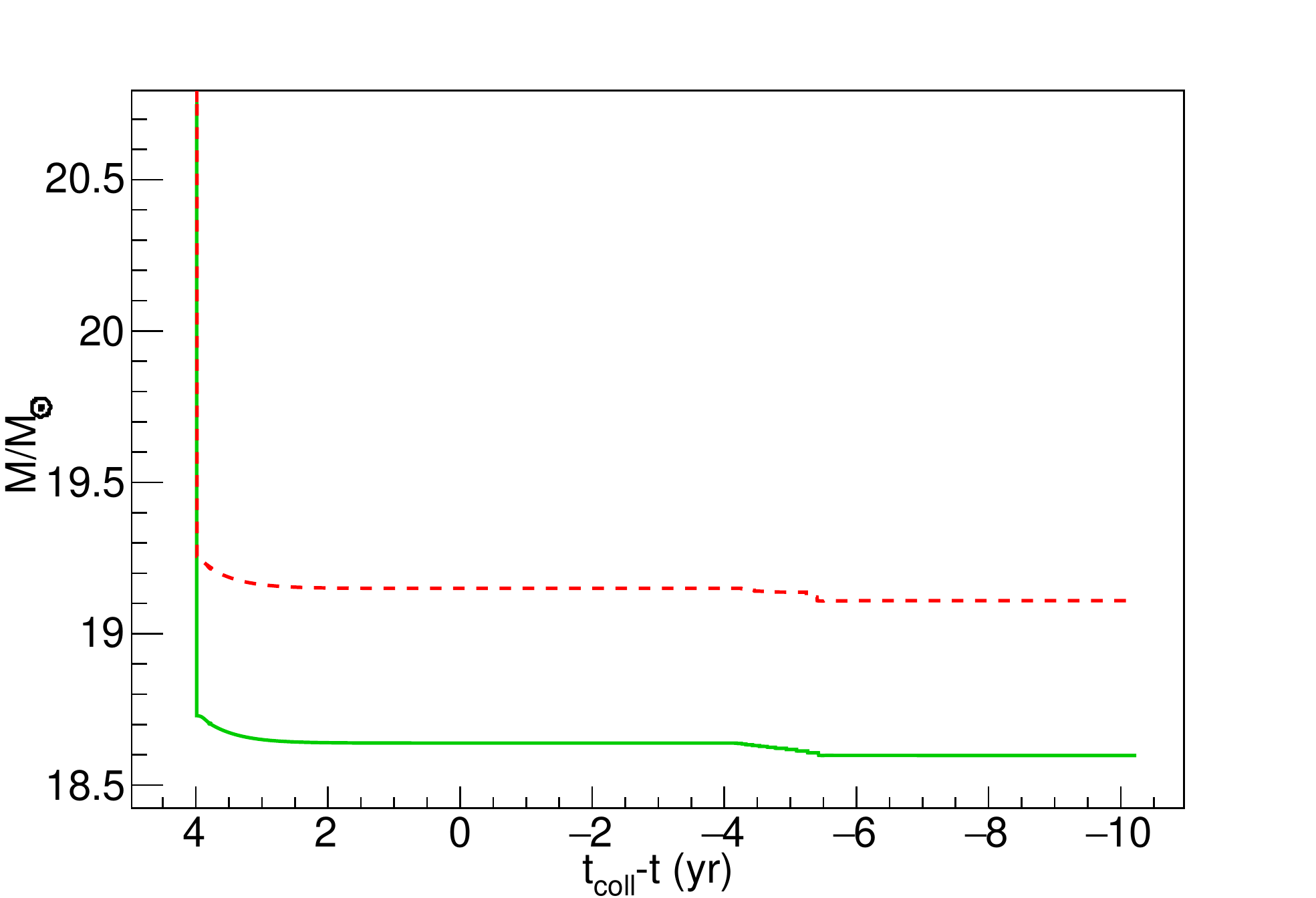}
         \includegraphics[width=0.5\textwidth]{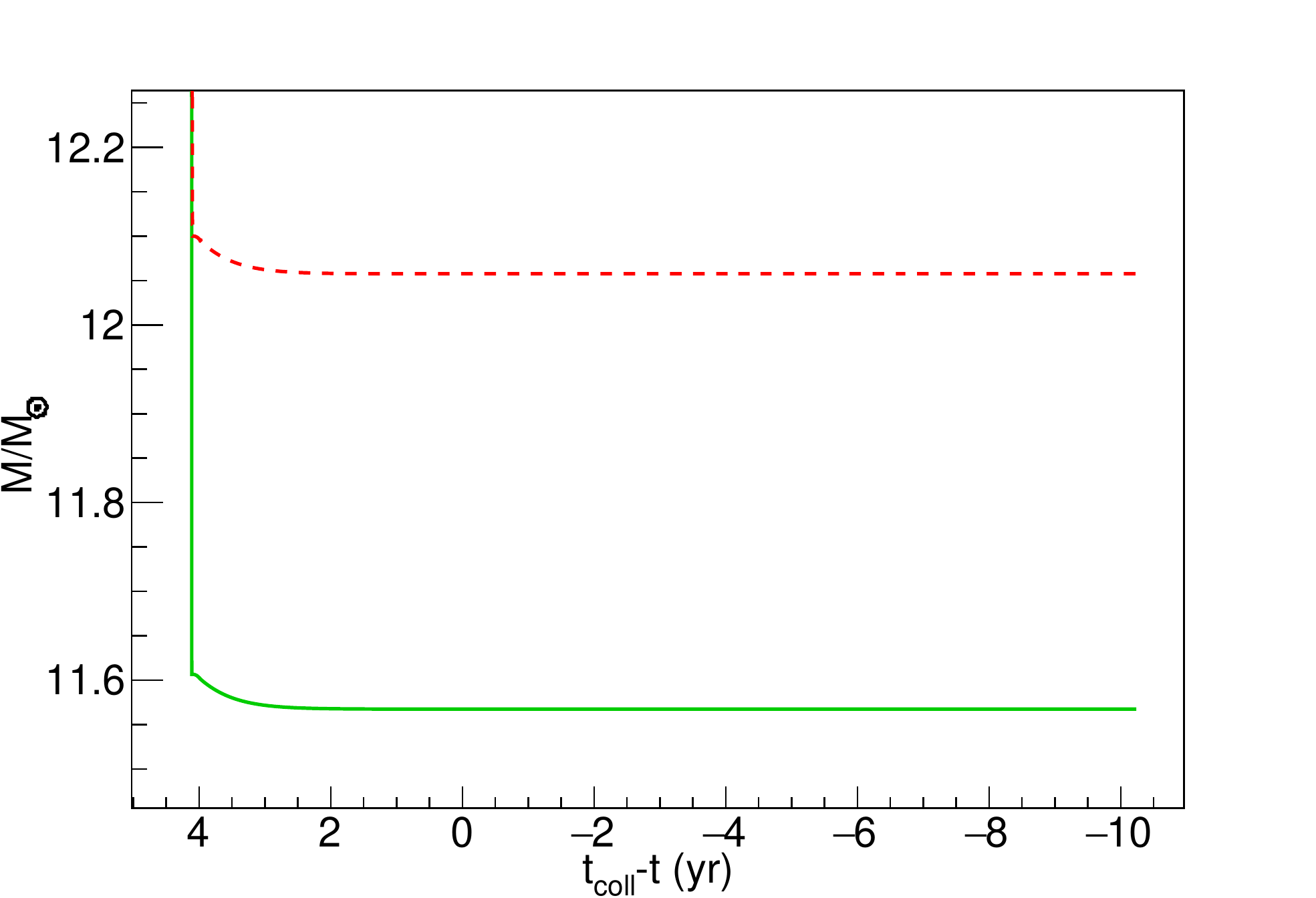}
    \caption{
    Stellar mass as a function of time.  Evolution is
    tracked as the logarithm of the time before
    final collapse and is shown after the first pulse.  Shown is the mass below the escape velocity.  The solid green line is for the default screening model, while the dashed red line is for the relativistic screening model.
            (\textit{top left}) 70 M$_\odot$,
        (\textit{top right}) 76 M$_\odot$,
                    (\textit{bottom left}) 88 M$_\odot$, and 
                        (\textit{bottom right}) 89 M$_\odot$.
    }
    \label{mass_v_time}
\end{figure*}

A significant portion of the mass loss in each case comes from the 
individual pulses themselves, with most of the pulsational mass loss from 
the first pulse.  Much of the He loss is from the wind prior to the first pulse.
This can be seen in Figure \ref{mass_v_time}, which shows the mass loss
for various progenitors at the first pulse and up to the time of collapse. Shown
in this figure is the total stellar mass below the stellar escape velocity as a function of time for various progenitor masses.  In each case, a significant amount
of mass loss occurs during the first pulse.  While subsequent pulses occur,
the mass loss may not be as pronounced.  However,  the 76 M$_\odot$ progenitor shows
significant mass loss during subsequent pulses.  
Also observable is the difference in the stellar mass for each model.
A larger resultant BH for the most massive PPISN, while the final BH for a 70 M$_\odot$ model is smaller.  
It is also noted that relativistic effects shift the PPISN/PISN boundary to lower
progenitor mass.  However, this shift is insignificant.
\subsection{Nucleosynthesis of PISN}
\begin{figure*}
        \includegraphics[width=0.33\textwidth]{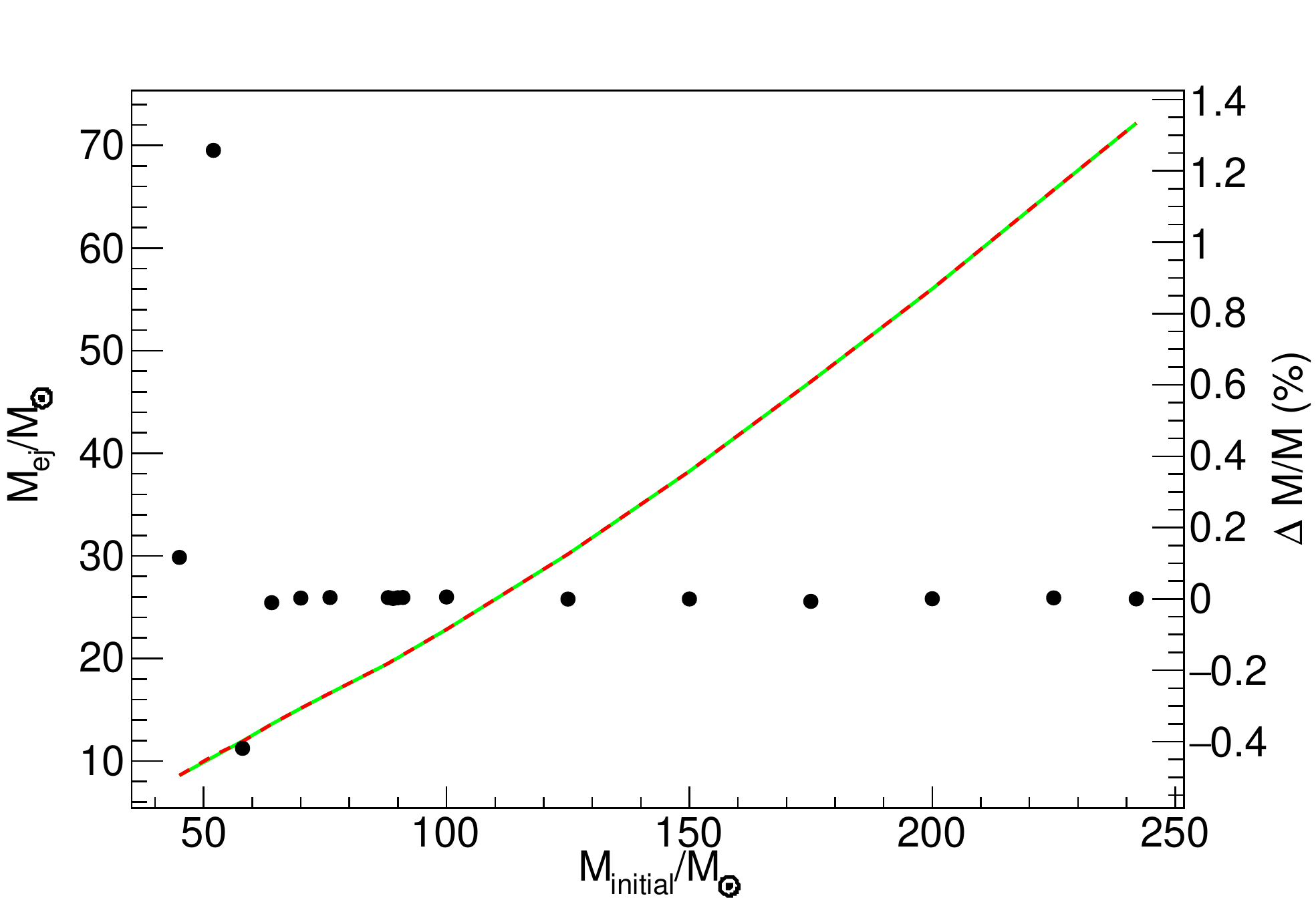}
         \includegraphics[width=0.33\textwidth]{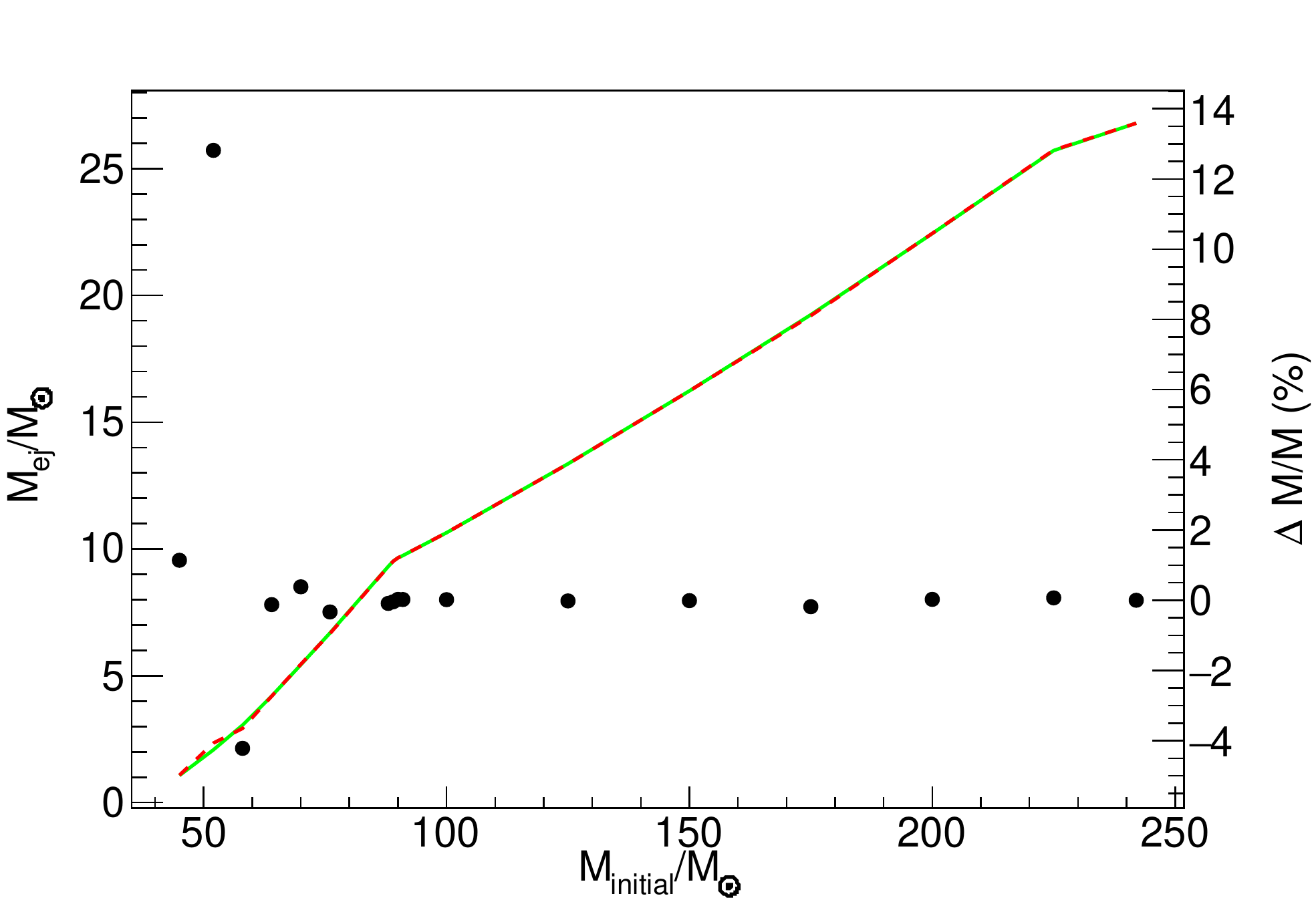}
          \includegraphics[width=0.33\textwidth]{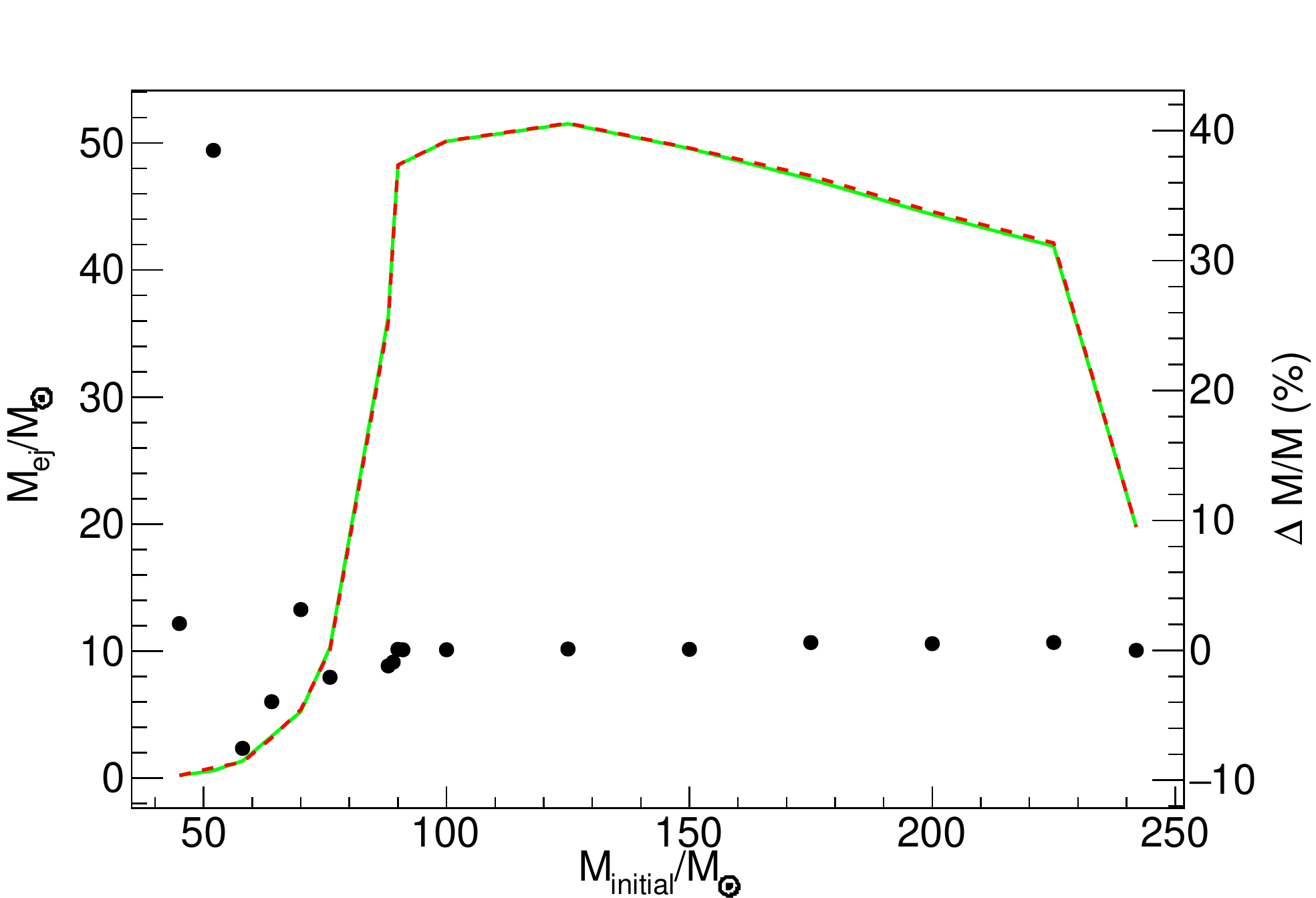}
        \includegraphics[width=0.33\textwidth]{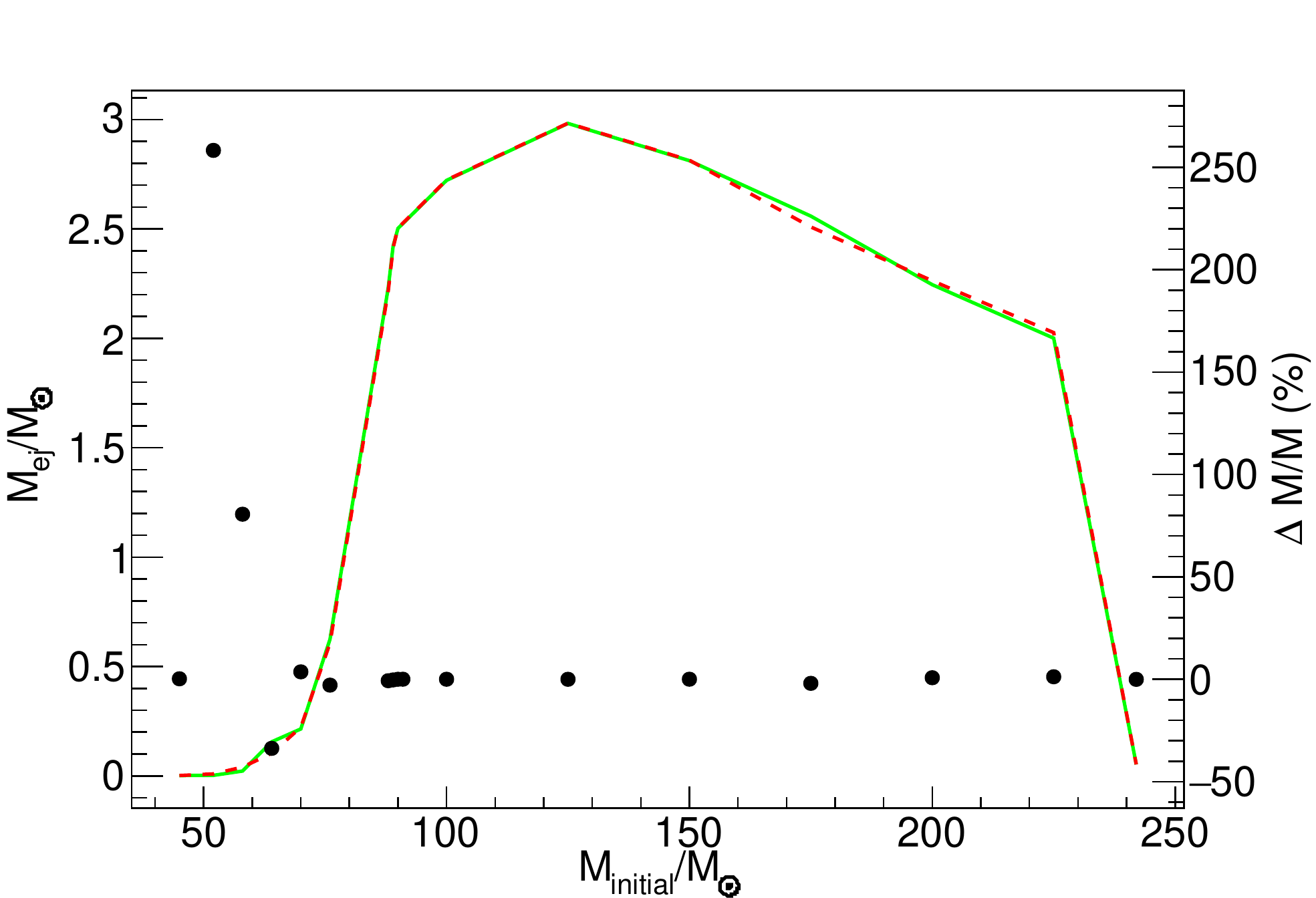}
         \includegraphics[width=0.33\textwidth]{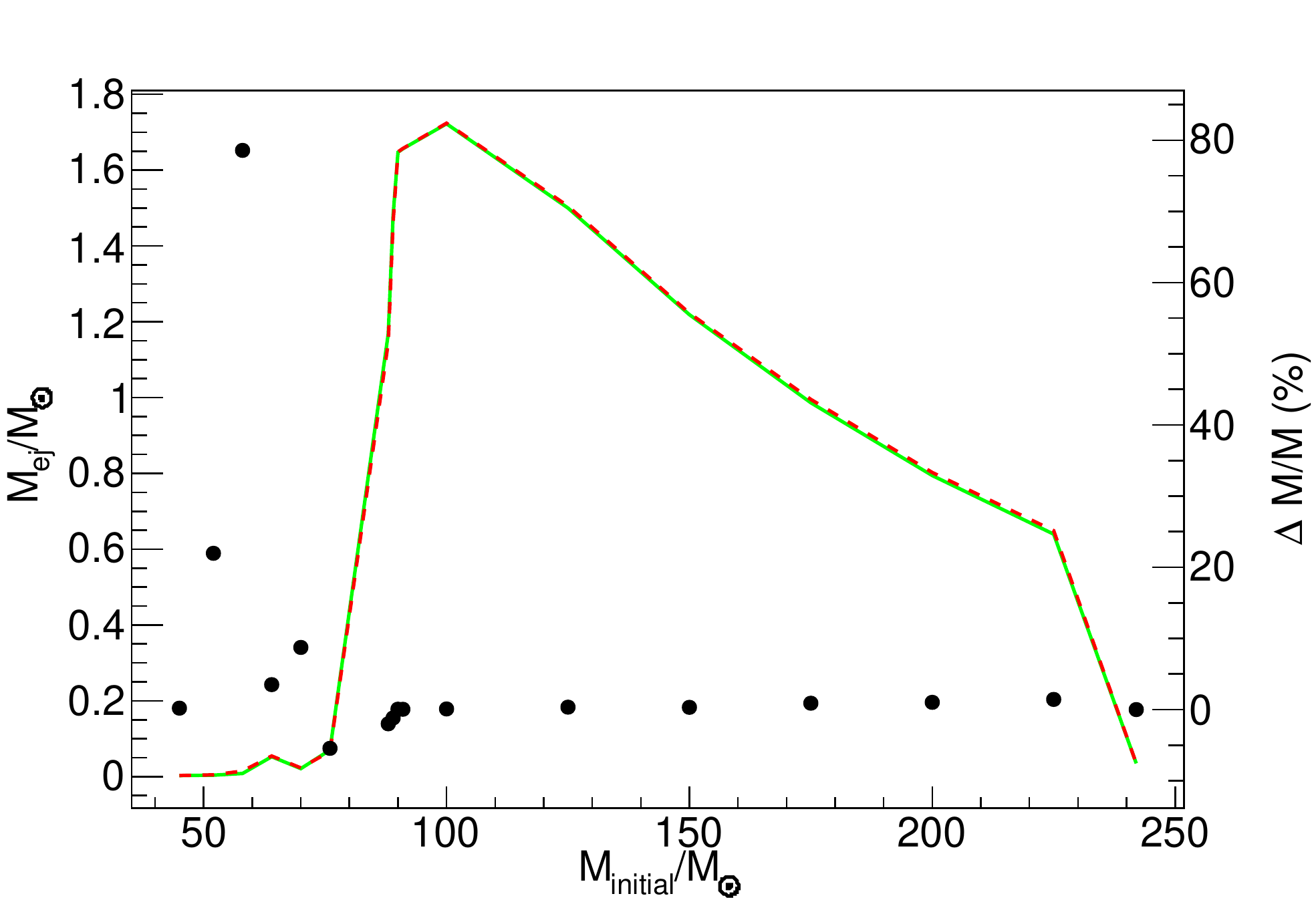}
          \includegraphics[width=0.33\textwidth]{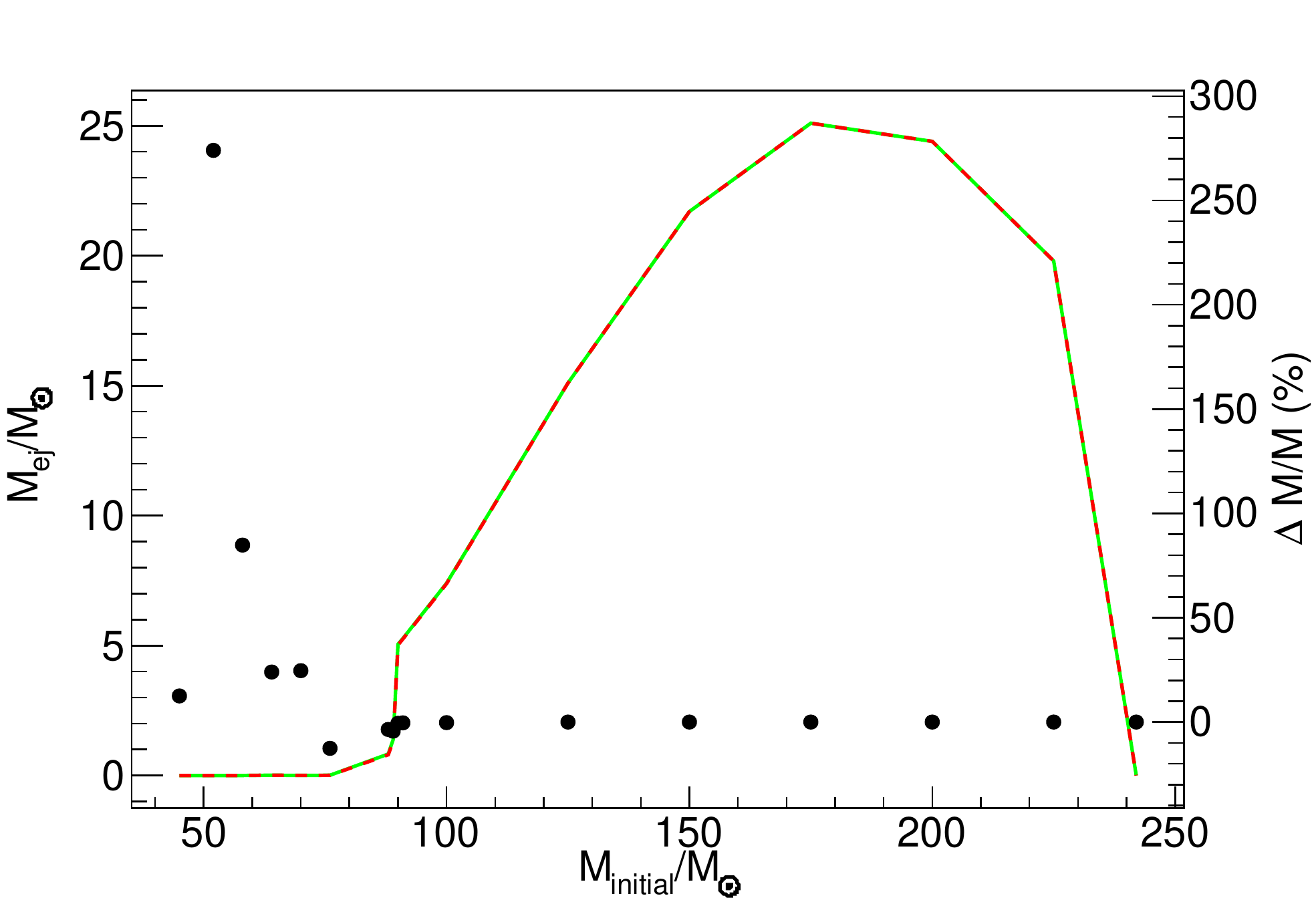}
      \caption{Cumulative ejected mass of various isotopes as a function of the initial mass for nuclei
        up to and including $^{28}$Si.  The solid greed line
    corresponds to models using extended screening, while the dashed lines correspond to models using relativistic screening.  The dots are the relative difference in ejected mass between relativistic and extended screening computations.
                (\textit{top left}) $^4$He,
        (\textit{top middle}) $^{12}$C,
                    (\textit{top right}) $^{16}$O,
                                    (\textit{bottom left}) $^{20}$Ne,
        (\textit{bottom middle}) $^{24}$Mg, and
                    (\textit{bottom right}) $^{28}$Si.
    }
    \label{ejected_mass1}
\end{figure*}
\begin{figure*}
        \includegraphics[width=0.33\textwidth]{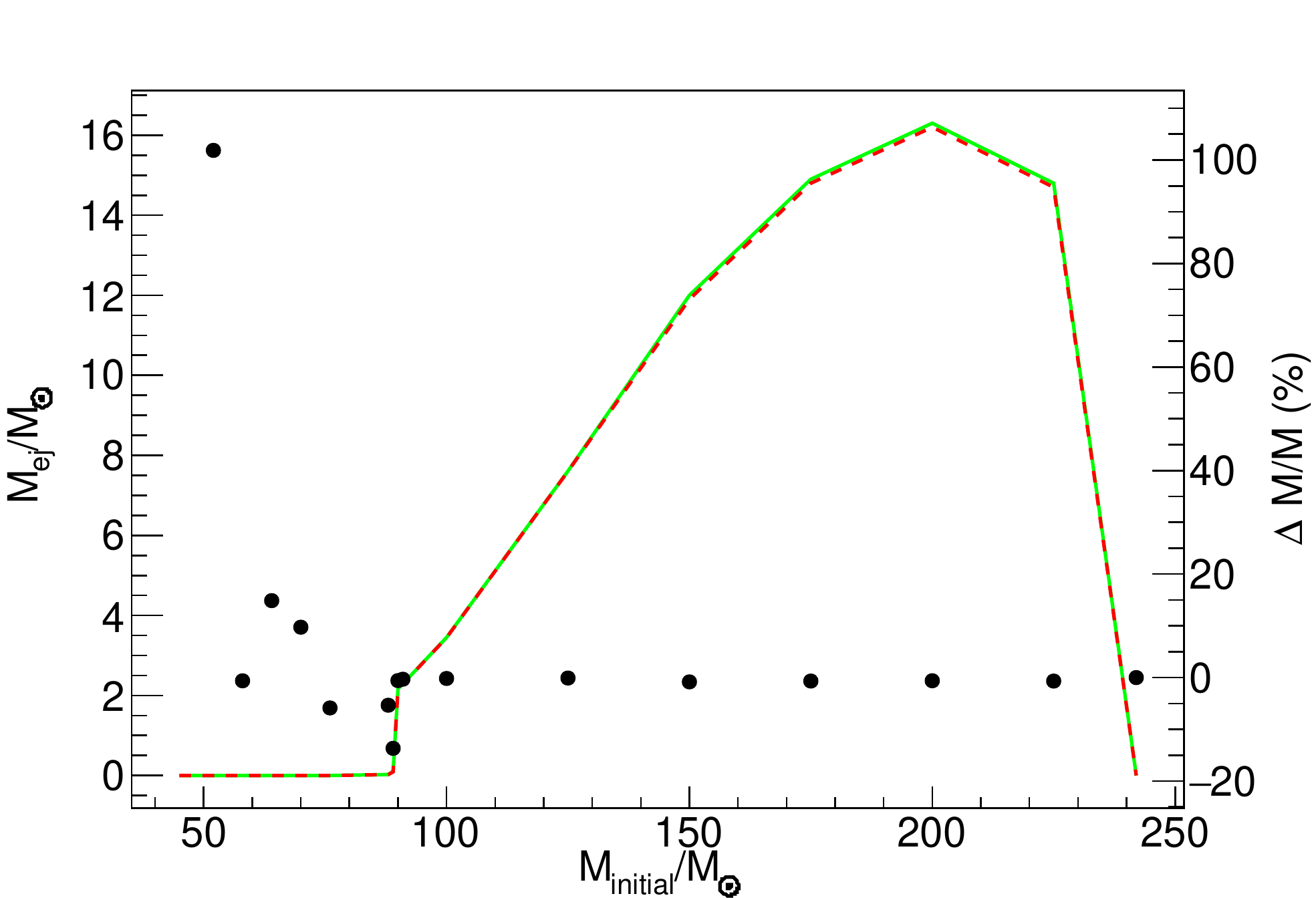}
         \includegraphics[width=0.33\textwidth]{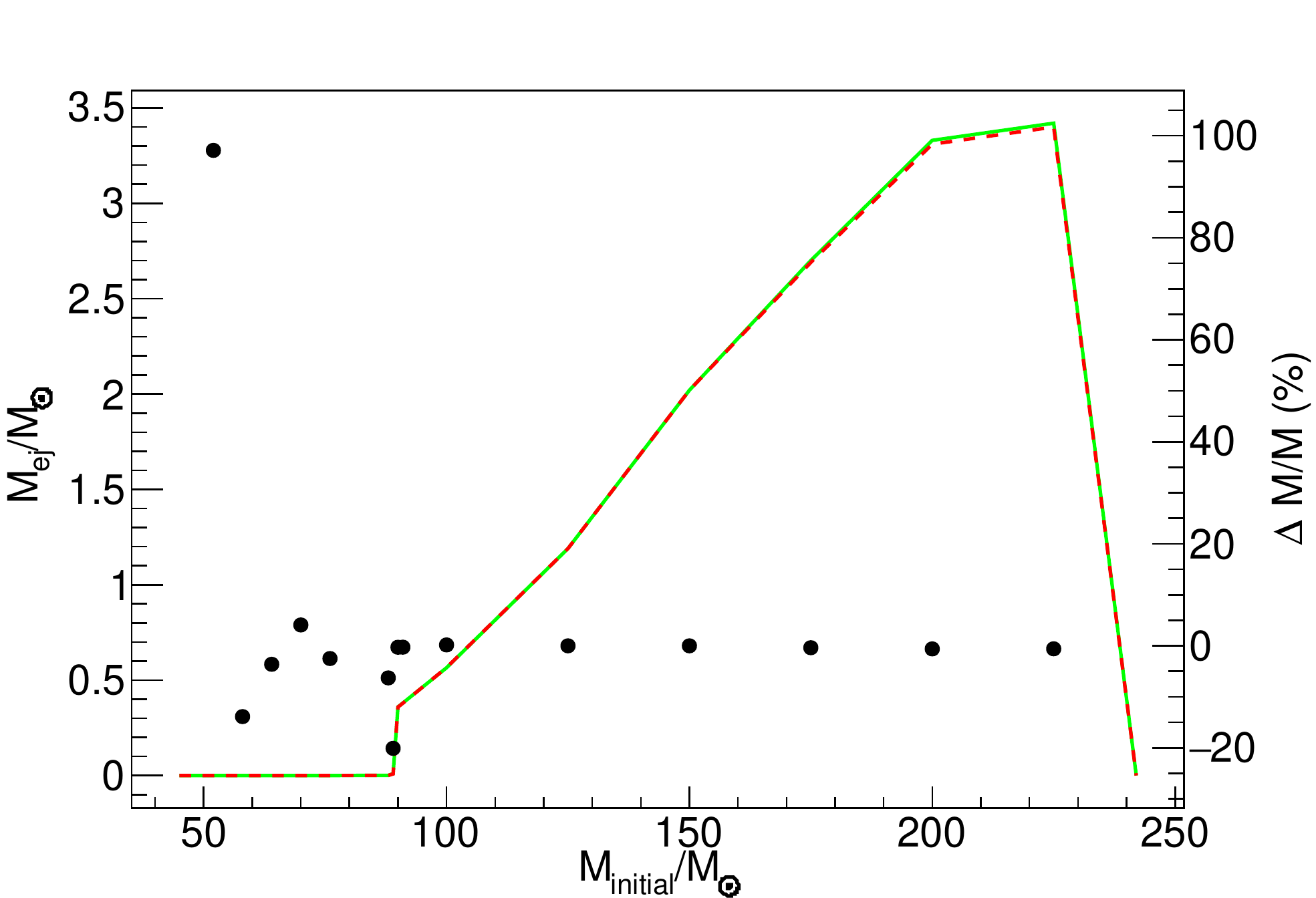}
          \includegraphics[width=0.33\textwidth]{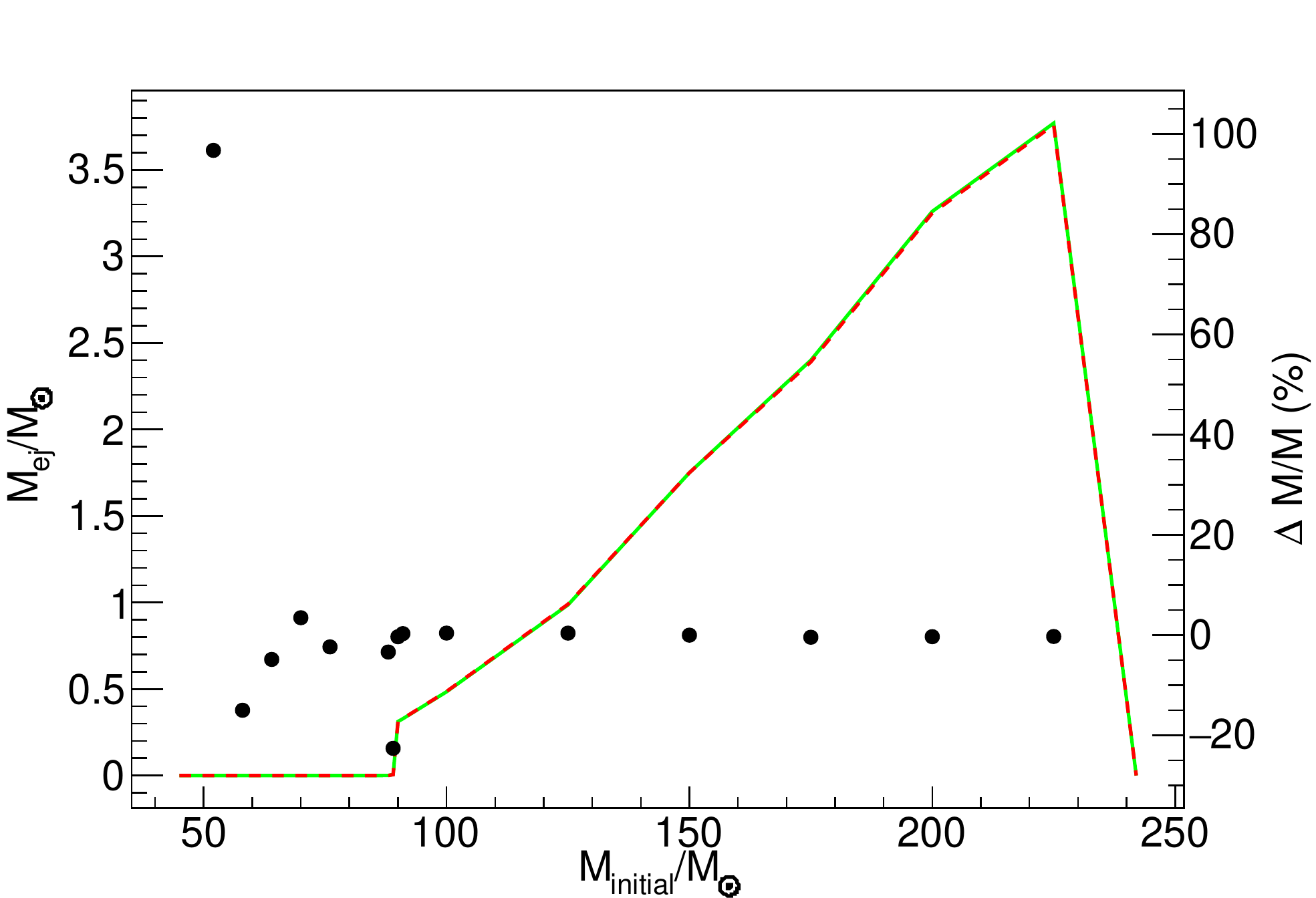}
        \includegraphics[width=0.33\textwidth]{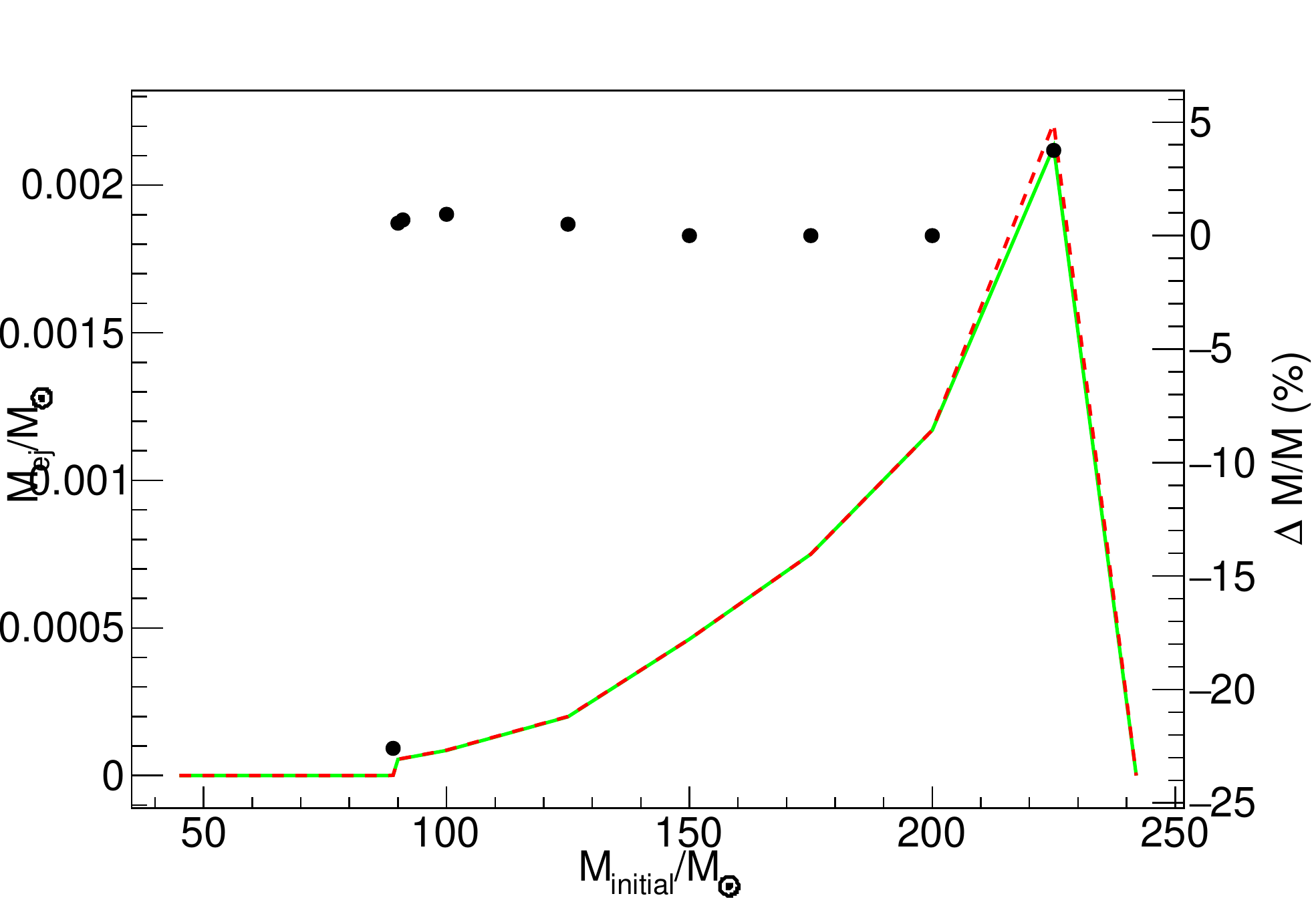}
         \includegraphics[width=0.33\textwidth]{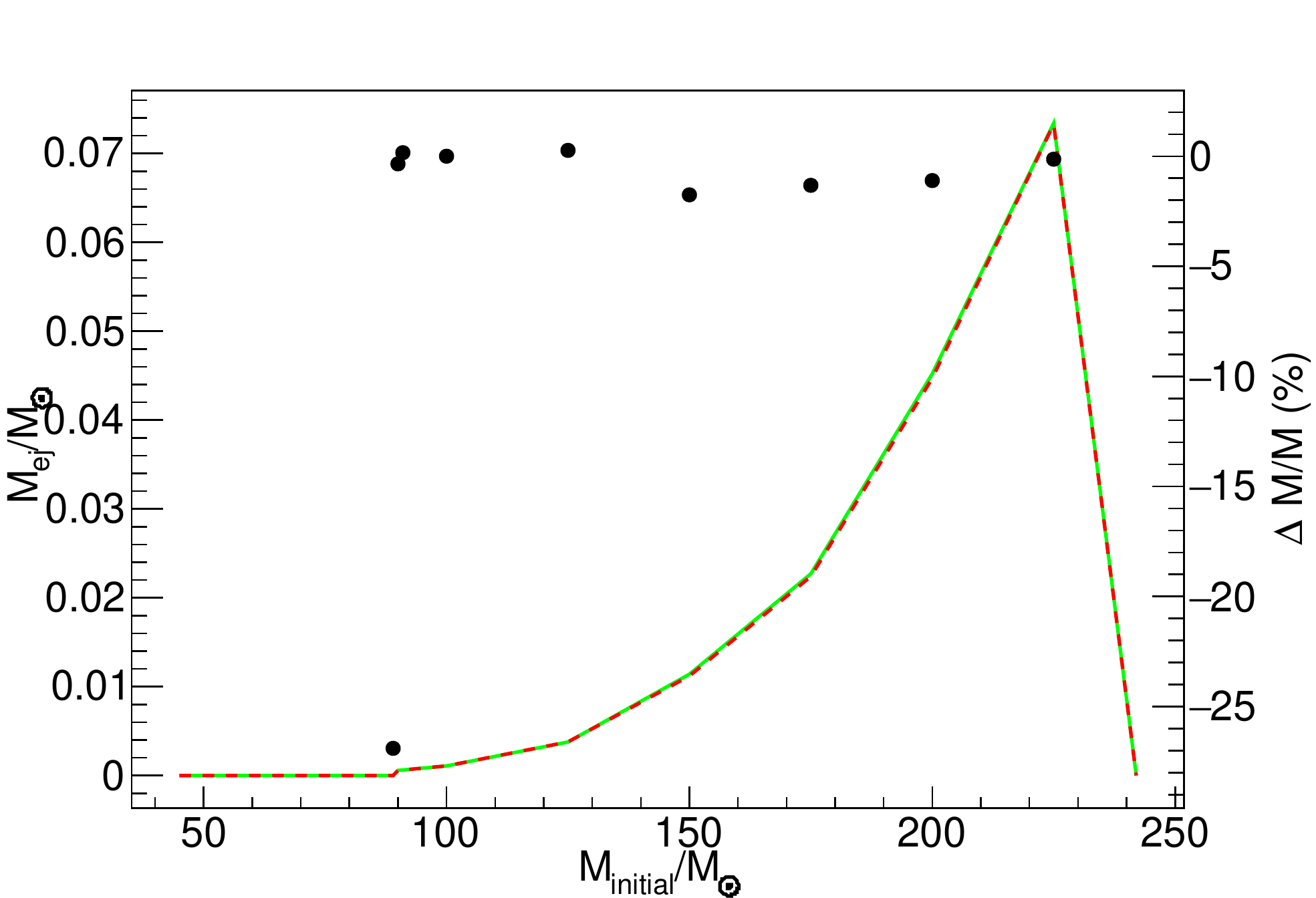}
          \includegraphics[width=0.33\textwidth]{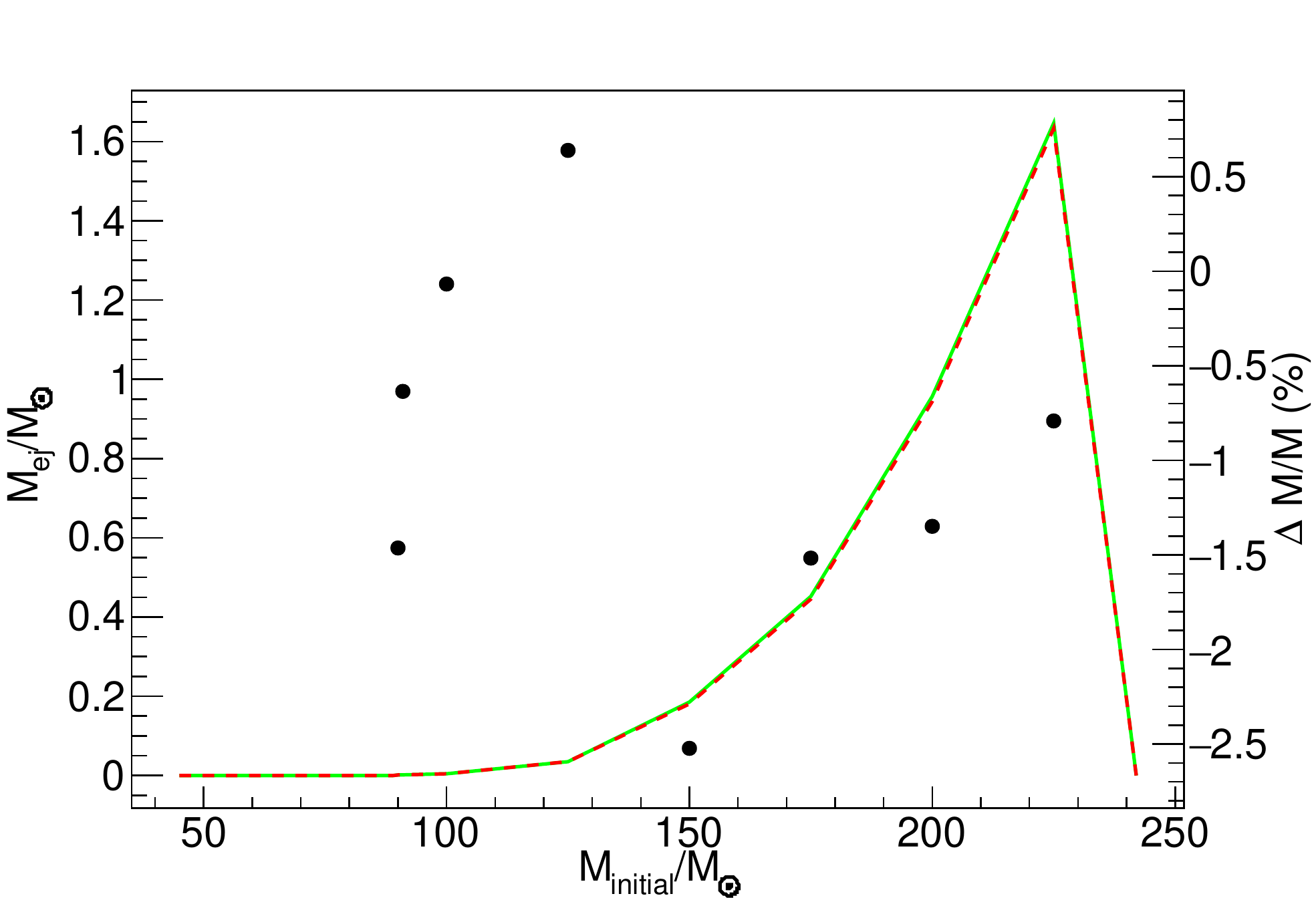}
        \includegraphics[width=0.33\textwidth]{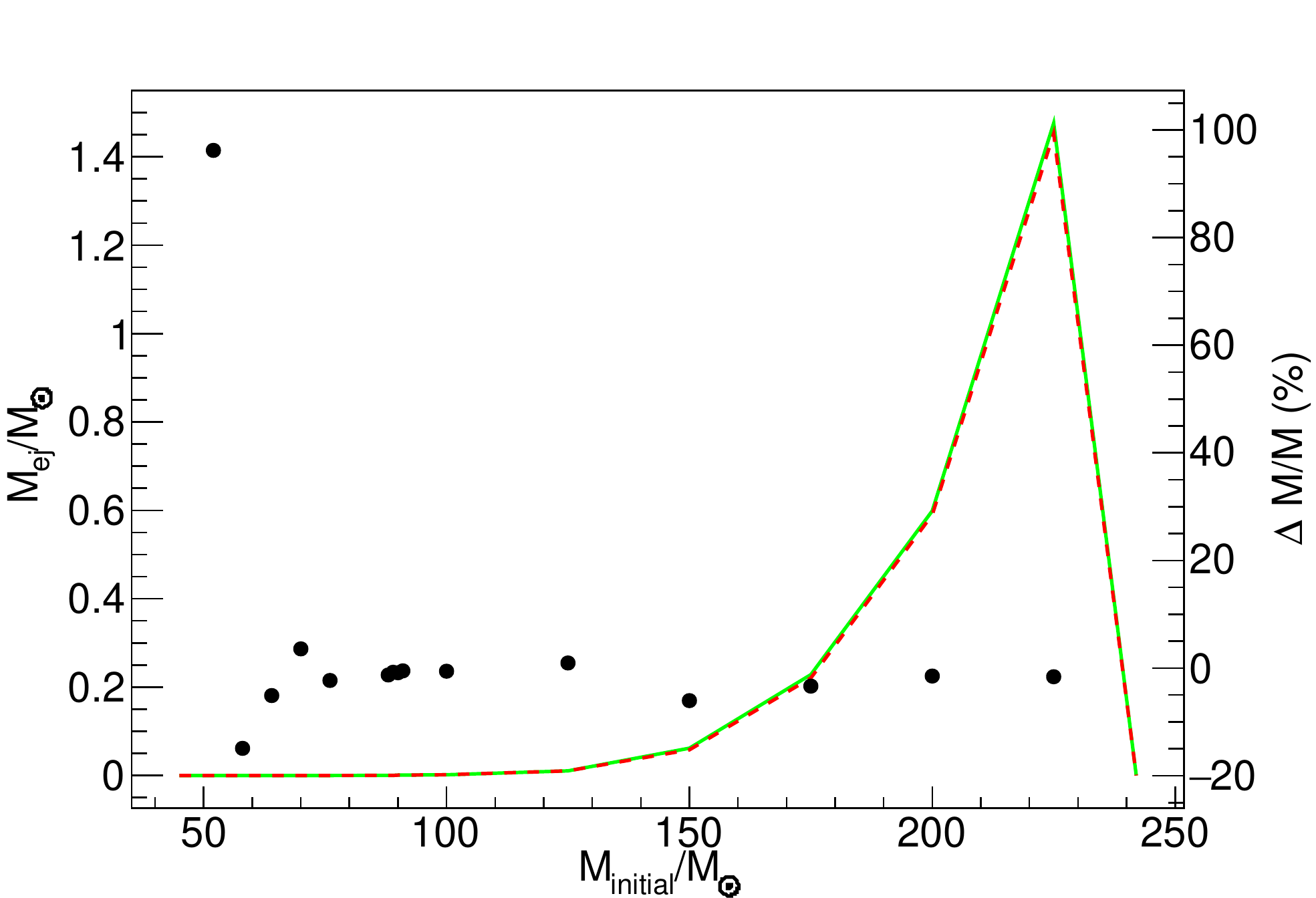}
         \includegraphics[width=0.33\textwidth]{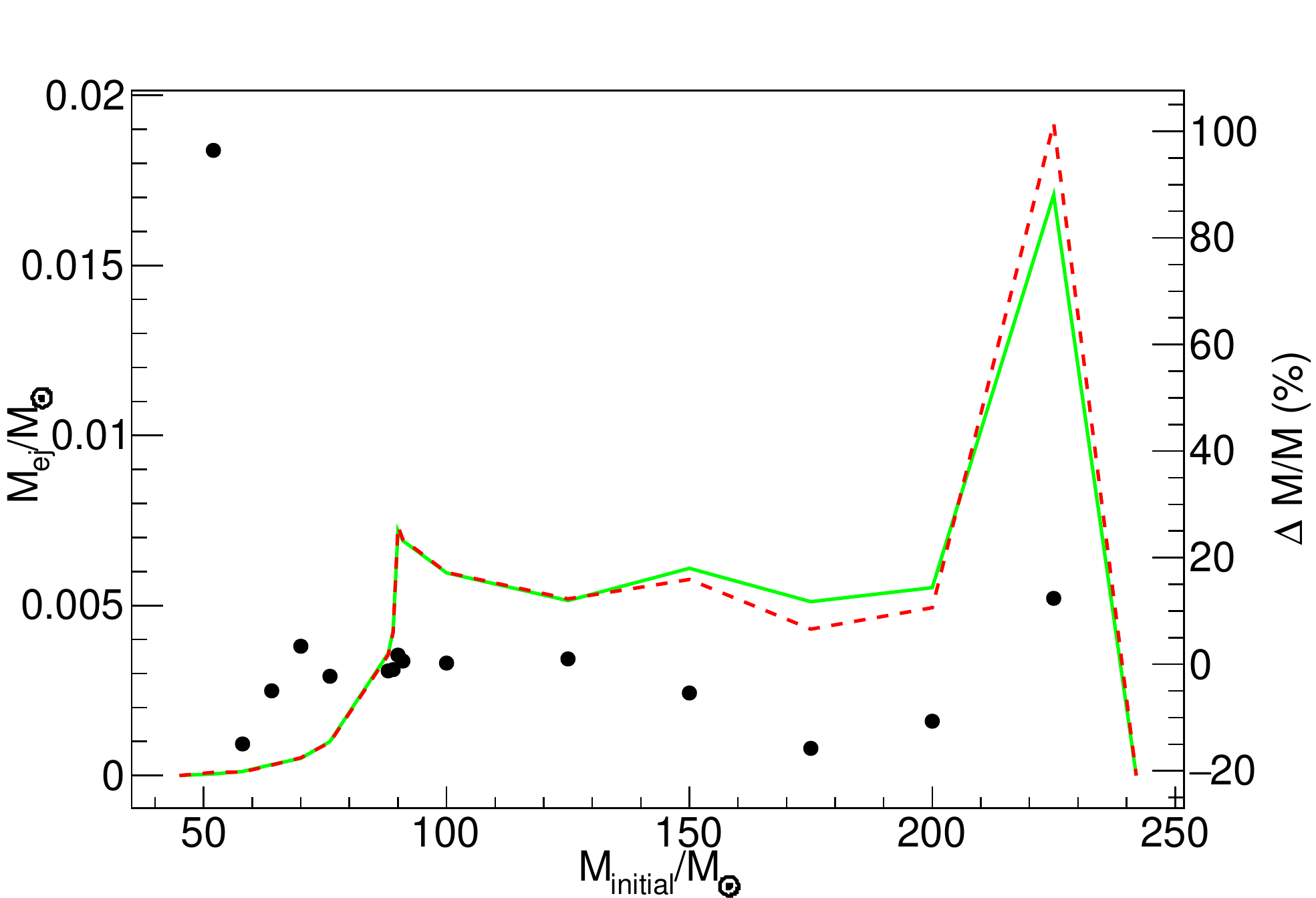}
          \includegraphics[width=0.33\textwidth]{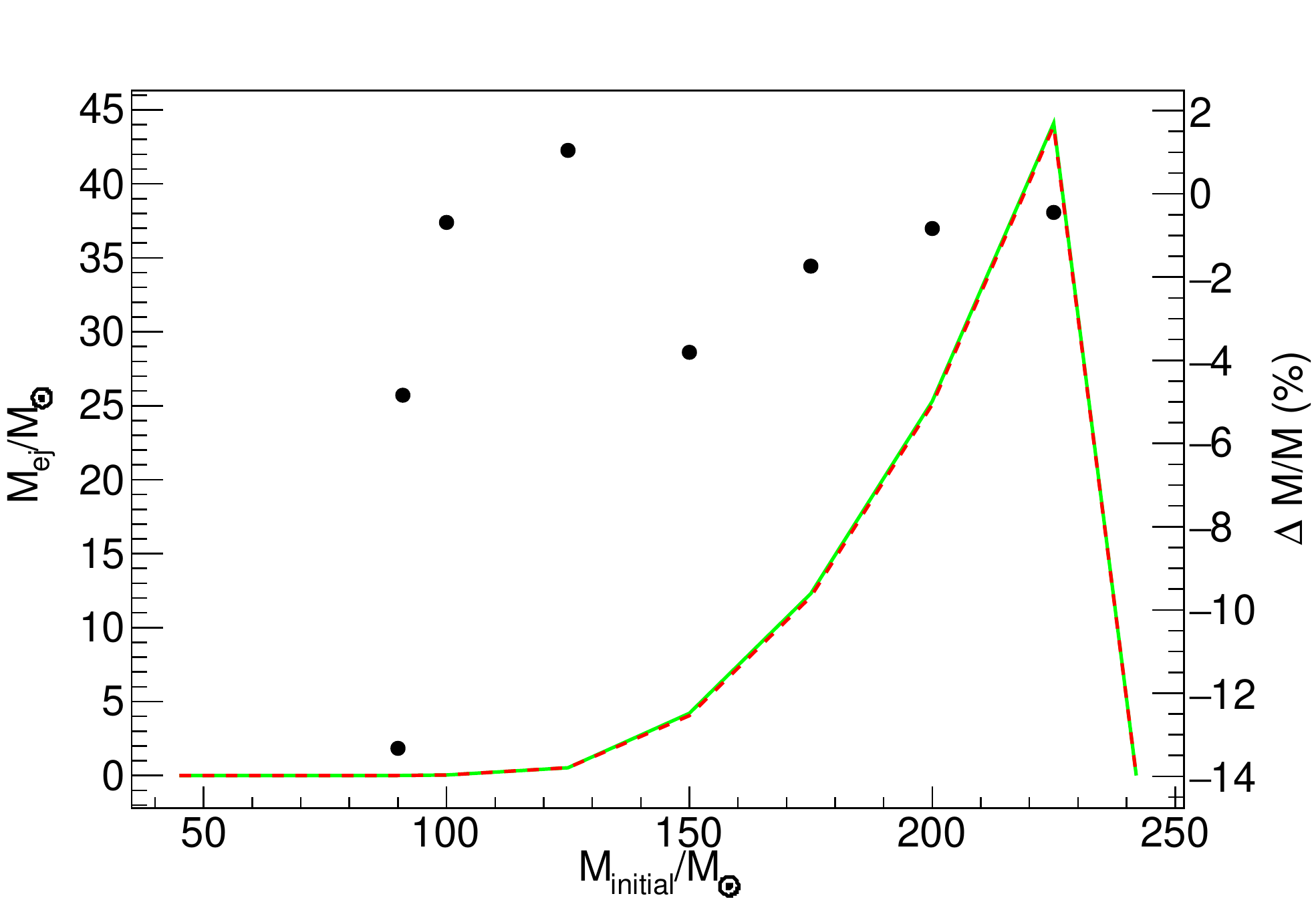}
    \caption{Same as Figure \ref{ejected_mass1} except for nuclei heavier than $^{28}$Si.
                    (\textit{top left}) $^{32}$S,
            (\textit{top middle}) $^{36}$Ar,
                    (\textit{top right}) $^{40}$Ca,
                                    (\textit{middle left}) $^{44}$Ti,
        (\textit{center}) $^{48}$Cr,
                (\textit{middle right}) $^{52}$Fe,
                                    (\textit{bottom left}) $^{54}$Fe,
        (\textit{bottom middle}) $^{56}$Fe, and 
                    (\textit{bottom right}) $^{56}$Ni.}
    \label{ejected_mass2}
\end{figure*}

\begin{figure*}
    \includegraphics[width=0.5\textwidth]{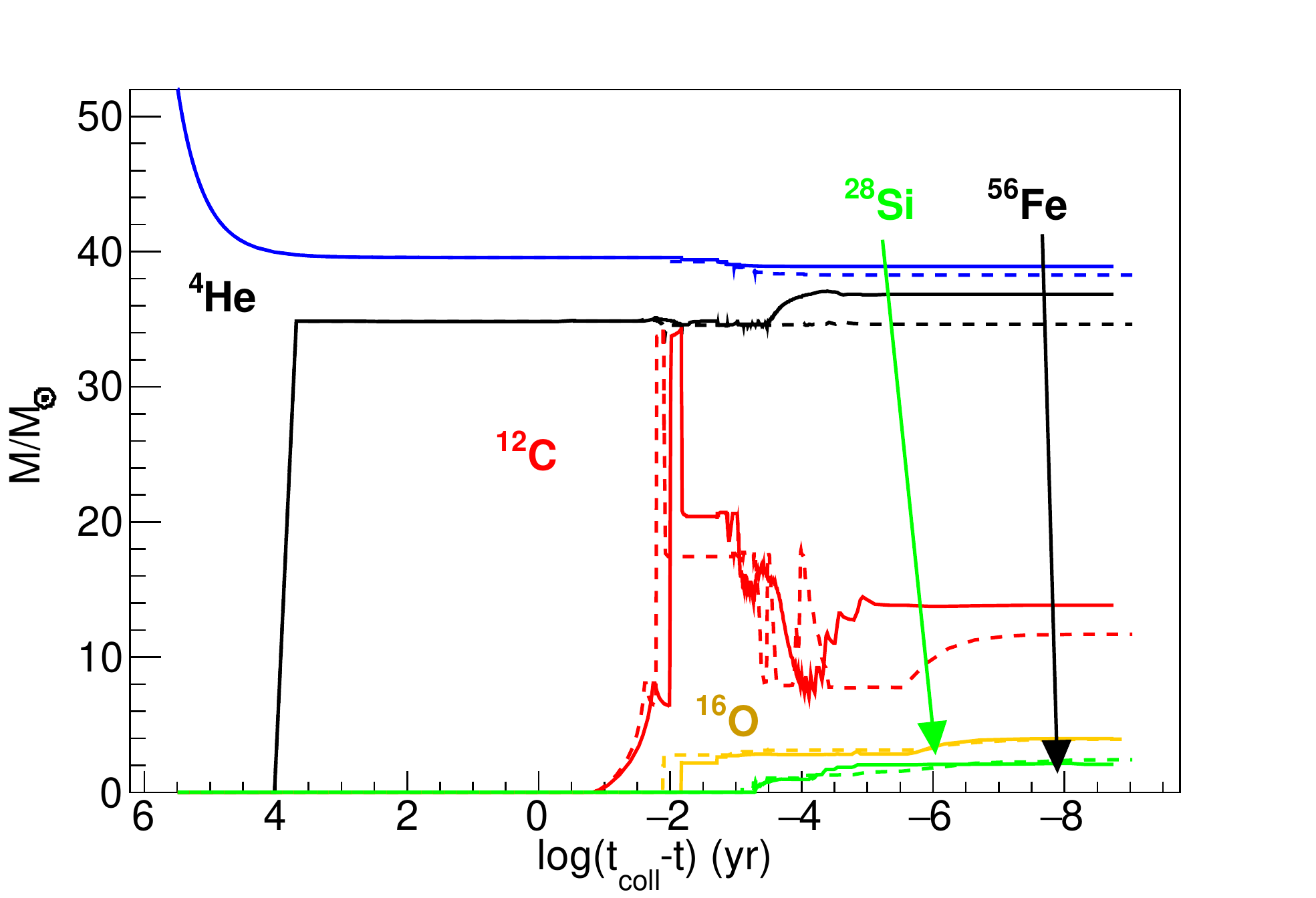}
    \includegraphics[width=0.5\textwidth]{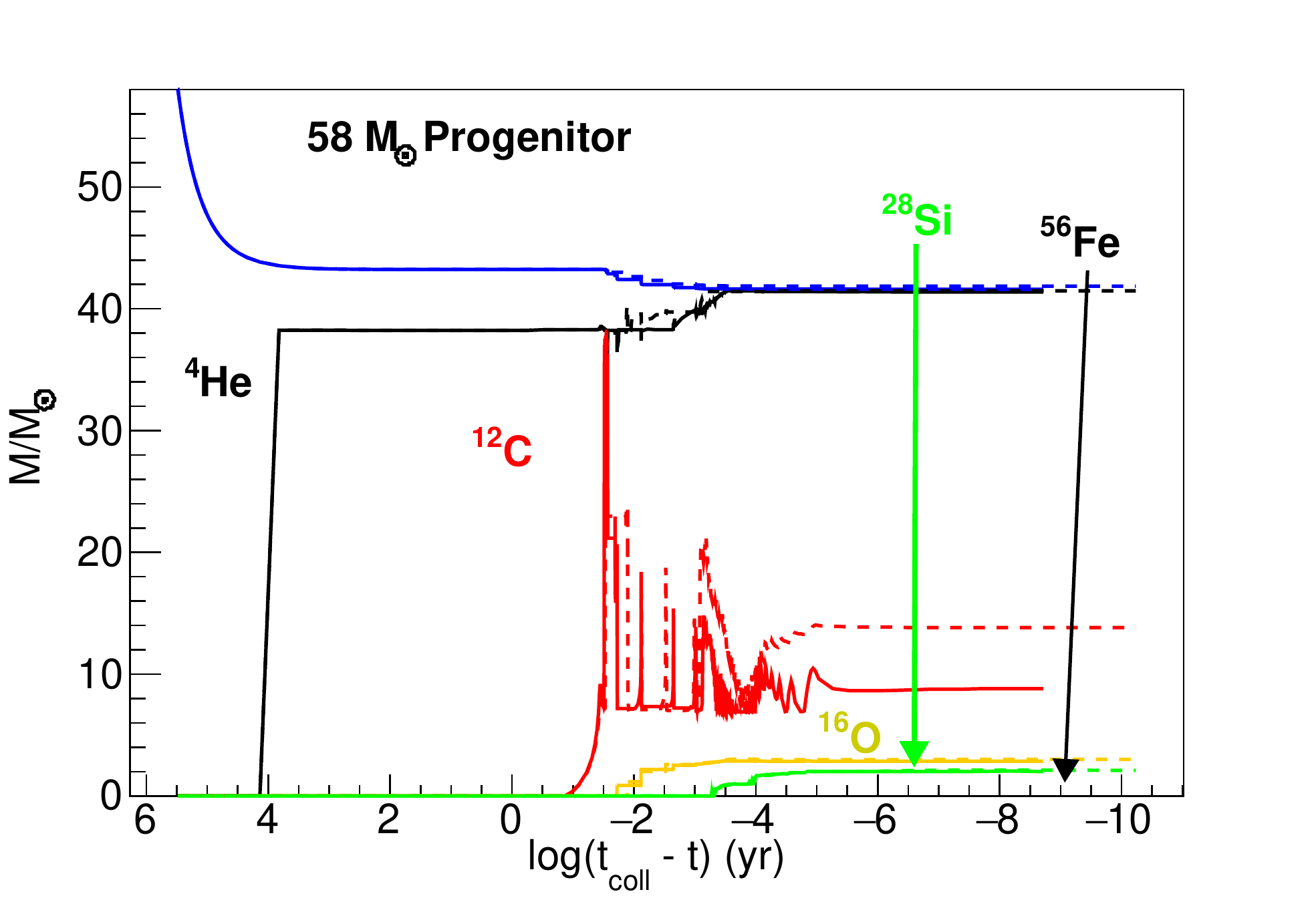}
    \includegraphics[width=0.5\textwidth]{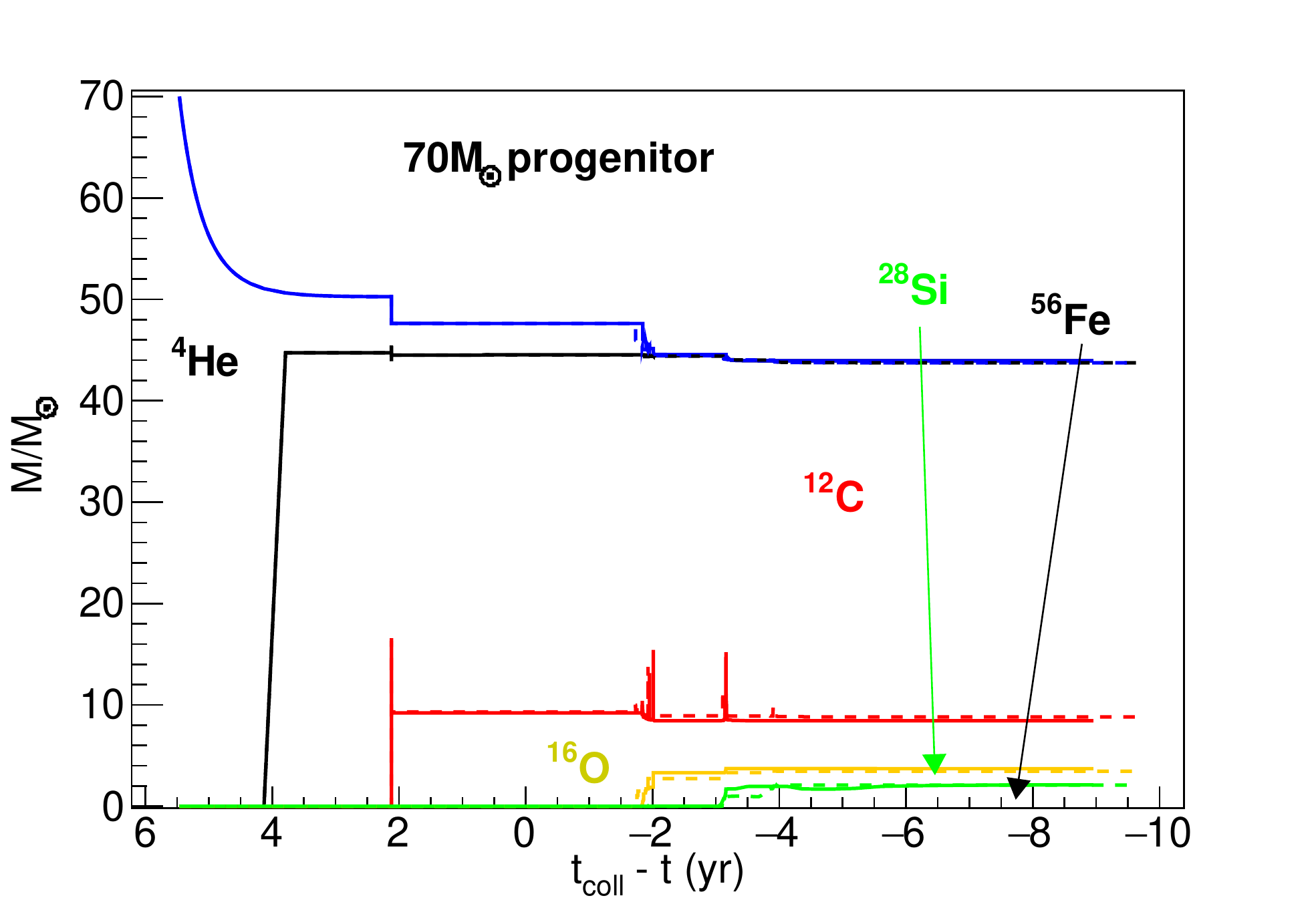}
    \includegraphics[width=0.5\textwidth]{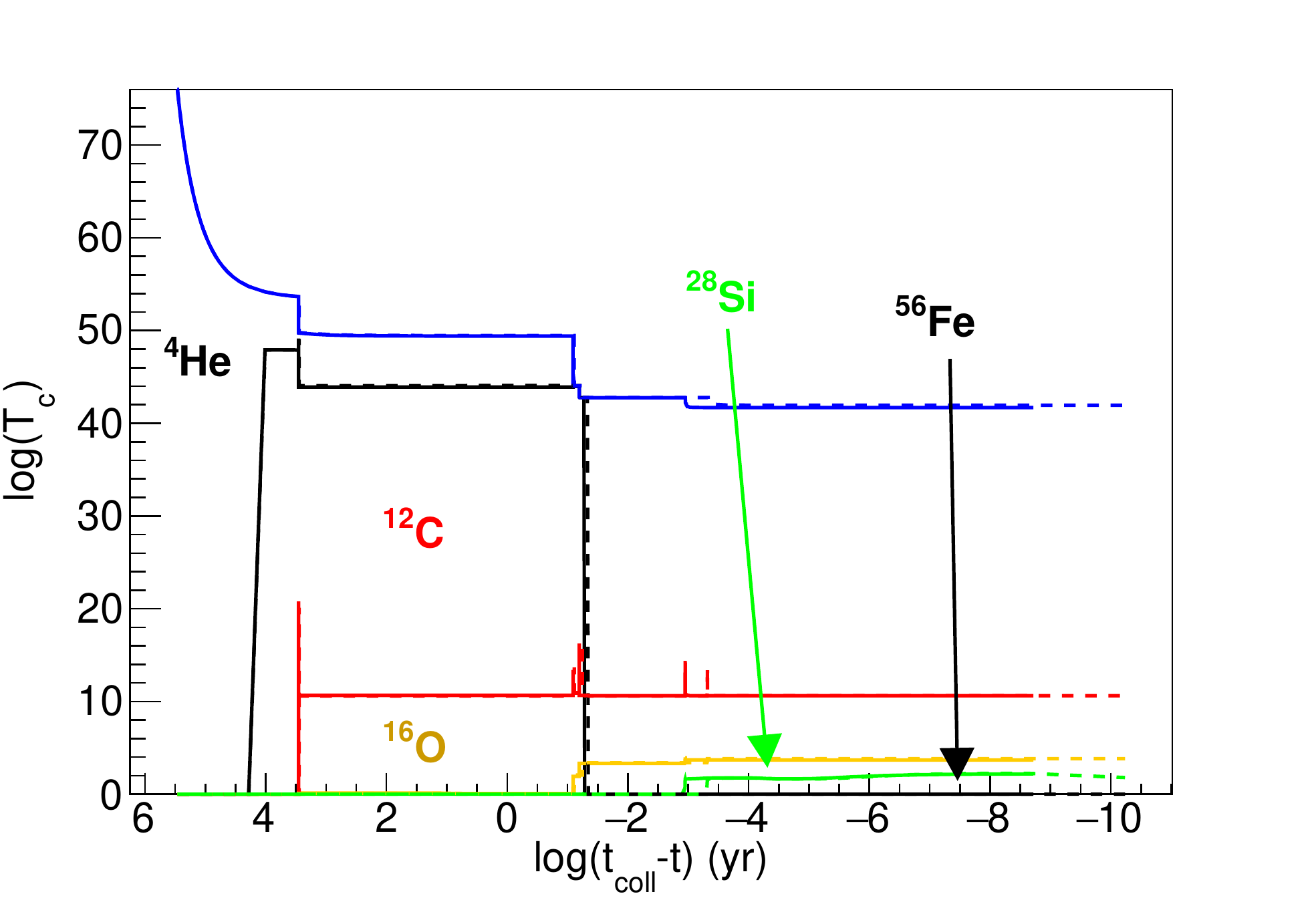}
    \caption{Kippenhahn diagram comparison for 
    four progenitors.
    Lines indicate the mass element at which the mass
    fraction of the indicated isotope falls below 10$^{-2}$.  
    The mass of each layer is indicated, along with 
    the total mass of the star below the escape velocity.
    Dashed lines are for the relativistic model and solid lines are for
    the default screening model.  Convective and burning regions have been removed
    for clarity.
                (\textit{top left}) 52 M$_\odot$,
        (\textit{top right}) 58 M$_\odot$,
                    (\textit{bottom left}) 70 M$_\odot$, and 
                        (\textit{bottom right}) 76 M$_\odot$.}
    \label{kipp_compare}
\end{figure*}

Figures \ref{ejected_mass1} and \ref{ejected_mass2} show the total ejected mass of isotopes as a function of 
the progenitor mass.   The ejected mass includes mass ejected in the wind and in the individual pulses.  Also shown in this figure is the relative difference of mass ejected between each model:
\begin{equation}
    \frac{\Delta M}{M}\equiv \frac{M_{ej,rel}-M_{ej,ext}}{M_{ej,ext}} .
\end{equation}

In the case of $^{4}$He, the difference in both models is not found to be significant.  
Nearly all of the $^{4}$He is ejected in the wind for all progenitors.  For nuclei heavier than $^{4}$He, the ejected mass is significantly affected by the subsequent pulsations 
of the star.

For many of the models, it would appear that the amount of mass ejected in the
relativistic model is largest relative to the amount of mass ejected in the extended model for a progenitor mass of $\sim$52 M$_{\odot}$.  This is probably because this progenitor undergoes the largest number
of pulses, and  more importantly, spends more time at a central temperature where
relativistic effects are important.   Here, it might not be surprising that burning
will progress more rapidly to produce more species heavier than $^{12}$C at the stellar surface for progenitors which undergo more pulses.  For progenitors with masses between 70 and 76 M$_{\odot}$, there are several pulses, and the mass ejected per pulse is 
larger.  This results in a larger amount of mass ejected as shown in Figures \ref{ejected_mass1} 
and \ref{ejected_mass2} but increased burning beyond $^{24}$Mg results in a depletion of
the lighter nuclei for higher-mass progenitors.

For nuclei heavier than $^{24}$Mg, very little mass is ejected unless the evolution results in a 
PISN, ejecting all of the mass in the star.  Because much of the mass of these nuclei is confined to the core, and because the individual pulses
proceeding a PPISN primarily eject surface layers, the heavy nuclei are maintained in the core of the star as it collapses into a BH.  This is seen in the figure as a sharp increase in the ejected mass of 
$^{28}$Si and heavier for $\sim$90 M$_{\odot}$. At a mass roughly equal to the PPISN/PISN cutoff, there is a sharp increase in the relative
difference in ejected mass.  This difference is particularly visible for the heavier nuclei and to a lesser extent for the lighter nuclei.  This sharp increase corresponds to the very narrow range of progenitor mass which results in a PISN for a the relativistic model, but a PPISN in the extended model.  For 
progenitors heavier than this cutoff in both models, the ratios are very close to zero as all of the 
stellar mass is ejected into the ISM after a very short increase in core
temperature.  Much of the production of heavier nuclei in stars undergoing PISNe is from hydrostatic burning at lower temperature.

For the Fe in the PISN region (M$_{initial}\sim$150 M$_{\odot}$), the relativistic model ejects slightly less Fe than the extended model.  However, only
a very small amount of Fe is ejected in either model, and the difference is small.  For direct collapse, progenitors with M $\ge$ 242 M$_{\odot}$ eject no Fe  as the 
star undergoes a direct collapse into a BH, and only lighter-mass nuclei are ejected in the wind prior to
collapse.  

In fact, it is seen that for the massive stars that undergo PISNe or direct collapse, there is little or no difference in the ejecta for all nuclei.  This is because the stellar interior either does
not reach a core temperature hot enough for relativistic screening to be significant, or the
core temperature is hot for a very short period of time. 

The progenitors in the mass region between 50 and 64 M$_\odot$ are particularly interesting with regard to the production of nuclei less massive than $^{24}$Mg.  Here the dynamics and interplay between the number of pulses, core
temperature, surface ejecta, and the wind can be quite complicated.  Relativistic
screening for the 52 M$_\odot$ model results in an increase in the 
ejection of lighter nuclei, with the largest enhancement for $^{20}$Ne (although it is 
important to note that only a very small amount of $^{20}$Ne is ejected
by the 52 M$_\odot$ model).  However, for the 58 M$_\odot$ model, there is a smaller
amount of He, C, and O ejected for relativistic screening.  This may be due 
to the enhanced burning of these nuclei as the star progresses to the production of
more massive nuclei.  

Figure \ref{kipp_compare} shows Kippenhahn diagrams for four
representative models.   Shown are the mass lines as a function
of log($t_{coll}-t$) where $t_{coll}$ is the final collapse time.
Each line in these diagrams corresponds to the location at which the
mass fraction of the 
labeled isotope falls below 10$^{-2}$.  Also shown is the total
stellar mass below the escape velocity. (Convective and
burning zones are not shown for simplicity.)
In the case of the 52 M$_\odot$ and 58 M$_\odot$ models,
there is a significant convection zone after the first 
pulse starting at the boundary of the C/O layer. This
is likely responsible for the rapidly changing behavior 
between these two models.  In the relativistic model,
the C layer extends to deeper in the star prior to collapse
for the 52 M$_\odot$ model, while it does not
extend as deeply as that for the default screening formulation
in the 58 M$_\odot$ model.  It can be seen in Figure 
\ref{ejected_mass1} that more C is ejected in the 52 M$_\odot$ model
if relativistic effects are considered, but less
is ejected for the 58 M$_\odot$ model. Similar results 
are seen for O.

This emphasizes the complexity of nucleosynthesis within
the PPISN model and the changes in rates between these two
formulations of electron screening.  Because the screening correction is density-dependent,
rates in the relativistic screening model do not always
exceed those of the default screening model.  Furthermore,
the pulse number, duration, shape, and quiescent period all vary
between these formulations. While fewer pulses are expected to result
in a lower ejected  mass, a longer pulsational period (including
weak pulses) can have the opposite effect.  
\section{Conclusions}
A screening model was developed for \texttt{MESA} in which 
electron-positron charge density and distribution
were treated in an environment in which pair-production
is possible.  Existing routines have been
adapted to incorporate a new screening type for all \texttt{MESA}
simulations.

As a first test of this model, relativistic screening for
pulsational pair-instability supernovae and pair instability supernovae
(PPISNe and PISNe) has been explored.  Comparisons were made to prior
models over a range of masses including low-mass He core progenitor stars, which
undergo direct BH collapse; stars which undergo direct collapse after
one or more thermal pulses; stars which explode in PISNe as every mass element 
exceeds the local escape velocity; and massive stars which undergo direct collapse.

The inclusion of relativistic screening was found to change the overall pulsational
morphology and timing characteristics of the pulsational phase of the star.
This is not surprising as relativistic screening is most prominent at higher
temperatures.  The pulsational phase of the star was found to achieve temperatures
in the core and elsewhere of well over 150 keV, the cutoff temperature
for relativistic screening.  For masses near 44.5 M$_\odot$, the unstable
temperature oscillations were found to extend for a longer period of time
prior to the final collapse of the star.  It was also found that the PPISN/PISN boundary
occurs at lower mass. However, the shift in this boundary with mass insignificant in light of the other uncertainties in this model.

Because of the unstable nature of the pulsational phase of PPISNe, slight changes in
the reaction rates and heating due to small changes in screening can
result in changes in the timing characteristics of the pulses.  A notable
case is the 64 M$_\odot$ model, for which the heating is more pronounced in the 
month prior to collapse.  For other models, shifts in pulses, the appearance
of weak pulses, and changes in the overall timing of the final pulses can
result from the inclusion of relativistic screening.

The changes induced in the temperature characteristics of the pulsational phase
can also affect the resulting nucleosynthesis slightly.  Here, we examined the
composition of material ejected in the winds and pulses.  For stars that
undergo total destruction in a PISN, little change was found in the composition
of ejected mass between the default and the relativistic model.  However,
for resulting PPISNe, there could be changes in the total masses of ejected isotopic 
species as the total surface composition could vary with each model.  For nuclei 
less massive than $^{28}$Si, differences in the ejected mass of specific
species could exceed 200\%.  However, the total mass of light nuclei
ejected by PPISNe is generally small. This reflects the effects of the steep mass
profile within the star and how small differences in mass profiles at the
surface can result in large relative changes in the total ejecta of one specie.  
For progenitors with masses 70$<$M/M$_\odot<$90, where mass loss from pulsation
becomes significant, the ejected mass of light nuclei increases.  In this region, the
relative differences in the ejected mass of a particular specie is generally less than 10\%.  

For lighter progenitors, with M$<$70 M$_\odot$, the ejected mass can be quite small, $\sim$ 5 M$_\odot$ or less for C and O.  However  the differences in 
the mass ejected can be more significant.  It is  as much as  40\%(14\%) for O(C) ejected by a 52 M$_\odot$ star.
Differences in the convection induced by heating differences in 
each model are also noted, as shown in Figure \ref{kipp_compare}.

The maximum BH mass formed in this study, which is produced by a $\sim$70 M$_\odot$ He
core, was found to vary little by the addition of relativistic screening. 
At higher progenitor masses, the resultant BH mass was found to increase
by roughly 0.5 M$_\odot$. However, this increase is probably
negligible given the uncertainties of the model used.  
Though the changes in rates induced by a relativistic treatment of screening
change the BH mass only slightly, the resultant composition of the
ejecta, being sensitive to the surface composition, was found
to vary somewhat for lighter progenitors and lighter stars.  While this
may not significantly change models for late time galactic chemical evolution, it
may be worth exploring the effects of changes in the galactic chemical evolution of
the early galaxy from the possible nucleosynthesis in Pop III stars which undergo
PPISNe.

\begin{acknowledgements}
T.K. is supported in part by Grants-in-Aid for Scientific Research of JSPS (17K05459, 20K03958).  A.B.B. is supported in part by the U.S. National Science Foundation Grants No. PHY-2020275 and PHY-2108339.  M.A.F. is supported by National Science Foundation Grant No. PHY-1712832 and by NASA Grant No. 80NSSC20K0498. K.M. is supported by Research Institute of Stellar Explosive Phenomena at Fukuoka University.  M.A.F., G.J.M.  and A.B.B. acknowledge support from the NAOJ Visiting Professor program.  Work at the University of Notre Dame (G.J.M.) supported by DOE nuclear theory grant DE-FG02-95-ER40934.
\end{acknowledgements}

\bibliographystyle{aa}
\bibliography{sample631}{}

%
%

\end{document}